\documentclass[12pt]{article}
\usepackage[T1]{fontenc}
\usepackage{lmodern}
\usepackage{microtype}
\usepackage{amsmath,amssymb,amsthm,amsfonts,mathtools}
\usepackage{setspace,graphicx,color,paralist,nicefrac,comment,booktabs,multirow,float,caption}
\usepackage[colorlinks=true, linkcolor=red, citecolor=blue, urlcolor=blue]{hyperref}
\usepackage[nameinlink,noabbrev,sort,capitalise]{cleveref} 
\usepackage[multiple]{footmisc}
\usepackage{thmtools}
\usepackage{thm-restate}
\usepackage{caption}
\usepackage{bbm}
\usepackage{needspace}
\usepackage[multiple]{footmisc}
\usepackage{natbib}
\usepackage{bibunits}
\usepackage{apptools}
\usepackage{etoolbox}
\usepackage[margin = 1in]{geometry}
\usepackage{tikz}
\usepackage{pgfplots}  
\usepgflibrary{shapes.geometric}  
\usetikzlibrary{calc,decorations.pathreplacing,patterns,arrows.meta}

\definecolor{signalRed}{HTML}{BB5566}
\definecolor{signalBlue}{HTML}{004488}
\definecolor{bertrandBlue}{rgb}{0.25,0.41,0.88}
\colorlet{regionBlue}{bertrandBlue!18}
\definecolor{figureGray}{HTML}{707070}
\tikzset{
  paper figure/.style={
    >=Stealth, font=\small, line cap=round, line join=round},
  paper axis/.style={->, thin, black!80},
  paper boundary/.style={semithick},
  paper guide/.style={densely dotted, thin, figureGray}
}

\usepackage{booktabs,array,tabularx,makecell}
\newcolumntype{L}[1]{>{\raggedright\arraybackslash}p{#1}}
\newcolumntype{Y}{>{\raggedright\arraybackslash}X}

\newcommand{\de}{\mathop{}\!\mathrm{d}}

\def\w{\omega}
\def\l{\lambda}
\def\m{\mu}
\def\t{\tau}
\def\e{\varepsilon}

\def\o{\theta}
\def\iff{\Longleftrightarrow}

\def\D{\Delta}
\def\W{\Omega}

\def\R{\mathbb{R}}
\def\E{\mathbb{E}}

\usepackage{enumitem}

\DeclareMathOperator{\Cov}{Cov}

\DeclareMathOperator{\Var}{Var}

\DeclareMathOperator{\marg}{marg}

\DeclareMathOperator*{\argmax}{argmax}
\DeclareMathOperator*{\esssup}{esssup}

\newtheorem{theorem}{Theorem}

\newtheorem{condition}[theorem]{Condition}

\newtheorem{corollary}{Corollary}

\newtheorem{example}{Example}

\newtheorem{lemma}{Lemma}

\newtheorem{proposition}{Proposition}

\theoremstyle{definition}
\newtheorem{definition}{Definition}
\newtheorem{assumption}{Assumption}

\newcommand\blfootnote[1]{%
   \begingroup
   \renewcommand\thefootnote{}\footnote{#1}%
   \addtocounter{footnote}{-1}%
   \endgroup
}
\makeatletter
\newcommand{\pushright}[1]{\ifmeasuring@#1\else\omit\hfill$\displaystyle#1$\fi\ignorespaces}
\newcommand{\pushleft}[1]{\ifmeasuring@#1\else\omit$\displaystyle#1$\hfill\fi\ignorespaces}
\patchcmd{\appendix}{\@Alph}{\@Roman}{}{}
\makeatother
\usepackage{appendix}
\usepackage{titlesec}

\titleformat{\section}
  {\normalfont\fontsize{16}{18.5}\selectfont\bfseries}
  {\thesection}{1em}{}
\titlespacing*{\section}{0pt}
  {2.75ex plus 0.90ex minus 0.18ex}
  {1.5ex plus 0.18ex}

\titleformat{\subsection}
  {\normalfont\fontsize{14}{16.5}\selectfont\bfseries}
  {\thesubsection}{1em}{}
\titlespacing*{\subsection}{0pt}
  {2.5ex plus 0.94ex minus 0.19ex}
  {1.2ex plus 0.19ex}

\title{When Is Mediated Communication Possible?}
\author{Yifan Dai\thanks{Department of Economics, MIT, \url{yfdai@mit.edu}} \\ MIT \\ {\small Job Market Paper} \\\and 
Anna Merotto\thanks{Department of Economics, MIT, \url{anna_mer@mit.edu}} \\ MIT}
\date{\vspace{-0.6em}
November 6, 2026\\[0.1em]
    {\small\href{https://www.dropbox.com/scl/fi/wk9ky9tgpbqb27alzaugu/Dai_JMP.pdf?rlkey=3zglm31jdkair0filcdcq613z&st=u4gjnn81&dl=0}
    {Click here for the latest version}}}

\BeforeBeginEnvironment{example}{\Needspace{6\baselineskip}}

\begin{document}
\noindent

\maketitle

\begin{abstract}
    We study when players can truthfully share private information through a mediator. Our analysis focuses on symmetric linear-quadratic games with strategic complements and independent private values. 
    We characterize when mediated communication is informative, meaning that it can induce an outcome different from the unique Bayes-Nash outcome without communication. 
    A sharp dichotomy emerges.
    When types that prefer higher own actions benefit less from higher opponent actions, informative communication is impossible regardless of the type distribution.
    By contrast, when those types benefit more from higher opponent actions, informative communication is possible if and only if the relevant extreme type lies sufficiently far from the mean. 
    In a linear-demand Bertrand oligopoly, informative communication is impossible under private costs but may arise under private demand.
    Finally, informative communication weakly benefits every type. But whenever information sharing can raise aggregate welfare, the maximum aggregate payoff under communication is strictly below the maximum payoff achievable without truth-telling constraints.
    
\end{abstract}

\blfootnote{We are particularly indebted to Drew Fudenberg, Stephen Morris, and Alexander Wolitzky for guidance, support, and many suggestions that substantially improved the paper. We are grateful to Charles Angelucci, Nemanja Antic, Ian Ball, Alessandro Bonatti, Roberto Corrao, Piotr Dworczak, Glenn Ellison, Ying Gao, Bob Gibbons, Andrew Koh, Haruki Kono, Dean Li, Jiangtao Li, Jonathan Libgober, Elliot Lipnowski, Moritz Meyer-ter-Vehn, Ellen Muir, Parag Pathak, Harry Pei, Sivakorn Sanguanmoo, Ludvig Sinander, Ran Spiegler, Yukun Wang, Frank Yang, Chen Zhao, Jie Zhou, as well as the audience at MIT Theory Lunch, Organizational Economics Seminar, and IO Lunch for helpful comments and conversations.}
\thispagestyle{empty} 

\clearpage
\setcounter{page}{1}

\section{Introduction}


Strategic interaction creates a tension when privately informed players communicate: although they benefit from learning what others know, they may have incentives to misrepresent their own information to influence others' decisions.
This tension has long been relevant in oligopoly markets, where firms exchange 
commercially sensitive information \citep{raith1996general,vives1999oligopoly}.
In these settings, regulators have been concerned that third-party intermediaries may facilitate such information exchange. 
Historically, information sharing has occurred through trade associations and industry consultants \citep{harrington2006cartels}. 
More recently, pricing algorithms that collect firm-specific data and provide pricing recommendations have received increasing attention from regulators. In the rental housing market, RealPage's revenue-management software combined nonpublic data from competing landlords to recommend apartment rents \citep{usdoj2024realpage,calder2024algorithmic}. Similar services have appeared in retail gasoline and hotel room pricing \citep{assad2024algorithmic,harrington2026hub}. 

Motivated by these settings, this paper asks when informative communication through an intermediary is possible among privately informed players in strategic environments.
We study this question in symmetric linear-quadratic games with strategic complementarities and independent private values, using \emph{communication equilibrium} (CE) \citep{myerson1982optimal,forges1986approach} as the solution concept. An uninformed mediator commits ex ante to a communication mechanism; after learning their types, players report to the mediator, receive possibly randomized private recommendations, and then choose actions.\footnote{The mediator need not be a literal third-party intermediary. By the revelation principle, any equilibrium outcome attainable through a communication protocol among players can be represented as a CE outcome.}  The unique Bayes-Nash equilibrium of the baseline game without communication is always a CE. Communication is \emph{informative} when it induces an equilibrium outcome different from this no-communication outcome. 

In our linear-quadratic setting, we provide a sharp characterization of when informative communication is possible. When players' types have opposite effects on their preferences over their own and opponents' actions, informative communication is impossible for every type distribution. Otherwise, informative communication is possible if and only if the relevant extreme type is sufficiently far from the mean type.

The linear-quadratic structure allows us to disentangle the preceding strategic tension cleanly, yielding a tractable representation of incentive compatibility in CE (\cref{prop:IC-characterization}). Incentive compatibility rules out profitable double deviations, in which a player first misreports their type and then disobeys the mediator's recommendation. 
After any report, the player's expected payoff from responding optimally to the recommendation depends on only two quantities: the expected opponent action and the variance of recommendations. The latter captures the value of learning about opponents' actions through recommendations, which depends on the player's report but not on their true type. 
Information can therefore serve as a screening instrument: more informative recommendations after some reports can compensate for less favorable opponent behavior. However, this information value is endogenous and constrained: the recommendations must themselves be optimal for players to follow.

Our first characterization identifies when players' preferences rule out informative communication for every type distribution (\cref{Thm: misaligned}). 
We classify preferences 
 by how players' types affect their preferences over their own and opponents' actions.
Preferences have \emph{aligned differences} when the types that prefer higher own actions also value higher opponent actions weakly more. 
Otherwise, preferences have \emph{misaligned differences}, and the opponent behavior needed to induce truthful reports is incompatible with the actions players would optimally take in response to recommendations.
We show that informative communication is impossible regardless of the prior type distribution if and only if preferences have misaligned differences.

Our second characterization shows that, when preferences have aligned differences, informative communication is possible if and only if the relevant extreme type is sufficiently far from the mean type (\cref{Thm: aligned}).
Alignment removes the conflict identified above, but players may still have an incentive to misreport to induce more favorable opponent behavior. 
Whether enough information can be generated to offset these incentives depends on the type distribution.
When the benefit from higher opponent actions increases in type and the mean type benefits from higher opponent actions, the lowest type has the weakest incentive to misreport to induce higher opponent actions. 
If the lowest type lies too close to the mean, even this type is unwilling to accept lower expected opponent actions in exchange for information. If it lies sufficiently far below the mean, 
a cutoff mechanism can sustain informative communication. When all players report below the cutoff, the mediator sometimes reveals this fact. Reporting low thus provides information about opponents' actions but induces lower expected opponent actions. By contrast, reporting above the cutoff induces higher expected opponent actions but no additional information. 
The analogous characterization applies to the highest type when the incentives are reversed. 
 
The characterizations lead to a sharp contrast in Bertrand oligopoly with private demand or cost information (\cref{Cor: Bertrand}). When firms have private information about their marginal cost, a higher-cost firm benefits less from an increase in opponents' prices, as its markup is lower. Meanwhile, higher-cost firms also prefer higher prices, so preferences have misaligned differences; no informative communication can arise by \cref{Thm: misaligned}.
When instead firms have private information about their demand intercept, a higher-demand firm benefits more both from a higher opponent price and from a higher own price, so preferences have aligned differences. Informative communication is then possible if and only if the lowest demand intercept is sufficiently lower than the mean by \cref{Thm: aligned}. In particular, a uniform upward shift in demand or a reduction in marginal cost can make informative communication impossible.

The linear-quadratic structure also delivers a sharp interim welfare implication (\cref{Prop: welfare CE}). Under every CE, every type of every player is weakly better off than under the no-communication outcome. 
Moreover, informative communication is possible if and only if some CE makes every player type strictly better off. In the Bertrand application, this characterizes when firms can increase their joint profit through mediated communication.


Next, we compare mediated communication with other forms of information sharing. 
We consider two benchmarks: Bayes correlated equilibrium, in which recommendations can condition directly on realized types, and exclusive ex-ante information sharing, in which players decide whether to give the mediator access to their information before learning their types. Both of these benchmarks relax interim incentive compatibility and attain the same maximum aggregate player payoff. Relative to this common benchmark, interim elicitation leads to a strict welfare loss unless the no-communication outcome is already welfare-maximizing (\cref{Prop: CE BCE EA}). This exception arises when preferences are sufficiently misaligned. Under weaker misalignment, information sharing can raise the aggregate payoff, but informative communication remains impossible. Interim elicitation also restricts feasibility: in private-demand Bertrand, two markets can have identical sets of outcomes under each benchmark but differ in whether informative communication is possible.

We then compare CE with public cheap talk, where players exchange simultaneous public messages before acting. Under aligned differences, informative cheap talk requires a stronger separation condition than mediation does, because a player's message can affect opponents' actions but not what the player learns from opponents' messages (\cref{prop: public_cheap_talk}). In the private-demand Bertrand environment, if equilibrium prices exceed common marginal cost, public cheap talk is always uninformative, whereas mediation can potentially sustain informative communication.
Finally, we study restricted access to mediation and show that informative communication becomes harder to sustain as fewer players can participate.\footnote{One motivation comes from the Australian gasoline market studied in \cite{byrne2025asymmetric}, where one firm lost access to a price-sharing platform while its competitors retained it.} 

\paragraph{Related literature.}
Our paper contributes to the literature on communication equilibrium and mediated communication (\cite{myerson1982optimal,forges1986approach}). Much of the more recent literature compares mediation with cheap talk in sender-receiver games where the informed sender does not control the payoff-relevant action (\cite{blume2007noisy, goltsman2009mediation, corrao2023bounds}).
Beyond those settings, much less is known about what mediation can achieve, or even when it can be informative.
We instead study mediated communication in incomplete-information games where privately informed players also choose actions, so they may misreport their information and subsequently disobey recommendations. Unlike \cite{corrao2023bounds}, we impose a linear-quadratic structure to make these double deviations tractable.
The closest paper is \cite{goltsman2014communication}, which studies a Cournot duopoly with private costs and provides a sufficient condition for a simple threshold communication mechanism to sustain informative mediated communication.\footnote{In this setting, public cheap talk is uninformative, which is also observed in \cite{ziv1993information}. In our Bertrand application, public cheap talk can likewise be uninformative while informative mediation is possible.}
In contrast, we characterize when informative mediated communication is possible in multi-player symmetric linear-quadratic games with strategic complementarities: under aligned differences, the condition implied by a similar threshold construction is also necessary, while under misaligned differences, informative communication is impossible for any type distribution.

Our leading application connects our results to the oligopoly information-sharing literature unified by \cite{raith1996general} and surveyed in \cite{vives1999oligopoly}.  
The linear-quadratic specification we adopt is standard in this literature,
which focuses on firms' ex-ante incentives to commit to an information-sharing rule before observing their private information \citep{novshek1982fulfilled,clarke1983collusion,vives1984duopoly,shapiro1986exchange}.\footnote{The precise ex-ante sharing rule differs across papers. In \cref{ssec: cost of interim}, we introduce an exclusive ex-ante sharing benchmark that preserves this timing while abstracting from the specific disclosure technology, and compare it with mediated communication.} 
It therefore ignores firms' incentives to report their private information truthfully.
In Bertrand competition, 
this literature yields a sharp contrast across forms of firm-specific information: \cite{sakai1986cournot} shows that revealing privately observed demand intercept is a dominant strategy in Bertrand duopoly, whereas \cite{gal1986information} shows that concealing private cost information is dominant. \cite{amir2010robust} generalize this contrast to oligopoly. 
We obtain a related cost-demand contrast when information must instead be elicited after firms learn their types. With private costs, interim incentives rule out informative communication. With private demand, informative communication is possible, but only when the lowest demand is sufficiently separated from the mean. Moreover, when private information affects both demand and cost, ex-ante sharing can be valuable even though interim informative communication is impossible. 

Rather than studying players' ex-ante incentives to share information, the literature on information design in games \citep{bergemann2013robust,bergemann2019information,mathevet2020information} assumes that an omniscient information designer can directly condition on the payoff-relevant state. Linear-quadratic games with Gaussian information are prominent in this literature given their tractability: \cite{yamashita2022information} characterize optimal information structures in smooth games via duality, with symmetric linear-quadratic games as main applications. \cite{miyashita2023lqg} show that, in linear-quadratic games with a Gaussian prior, it is without loss of optimality to consider Gaussian information structures, reducing the information design problem to a finite-dimensional semidefinite program.
We differ from this literature because our mediator is not omniscient and instead must elicit information from players' reports. In particular, any informative CE must be non-Gaussian: under jointly Gaussian outcomes, conditional variance of actions cannot vary across reports to provide the informational compensation needed for truthful reporting. In \cref{Sec: comparison}, we compare CE with Bayes correlated equilibrium to isolate the effect of this elicitation constraint.

Finally, our oligopoly application relates to recent work on intermediaries in collusive environments, including bidding rings \citep{kawai2026value,ortner2024mediated} and pricing algorithms \citep{sugaya2025collusion,harrington2025equilibrium,harrington2025economic,aleksenko2026third}, as well as to papers on markets for information \citep{bordoli2026selling, wu2025welfare, bonatti2025information}.  
These papers endow the intermediary with additional information or instruments, such as monetary transfers, delegated pricing, or dynamic punishments. By contrast, we focus on the pure exchange of unverifiable information through an uninformed mediator.
In a related linear-quadratic setting, \cite{bonatti2025information} consider a designer who observes a common-value state, elicits players' independent private-value types, and sells action recommendations at report-contingent prices. A key difference is that, with transfers, the truth-telling constraint is slack at the players' optimal design.


\paragraph{Outline of the paper.} \cref{sec:model} introduces the model. \cref{sec:IC}  characterizes incentive compatibility. \cref{sec: main} characterizes when informative communication is possible, derives its welfare implications, and applies the results to Bertrand oligopoly. \cref{Sec: comparison} compares mediation with other forms of information sharing. \cref{sec:general_LQ} extends the results to quadratic externalities and more general linear-best response games. \cref{sec: conclusion} concludes. 

\section{Model}\label{sec:model}
\subsection{Baseline Game}
We consider a symmetric linear-quadratic game with incomplete information. There are $n$ players, indexed by $I = \{1,\ldots,n\}$. Player $i$ has a private type $\theta_i \in \Theta_i \coloneqq [\theta_L, \theta_H]\subseteq \R$ and chooses an action $a_i \in \mathcal{A}_i \coloneqq \R$. We write $\theta = (\theta_i)_{i\in I}$ and $a = (a_i)_{i\in I}$ for the type and action profiles, respectively. Let $A_{-i} \coloneqq \sum_{j\neq i} a_j$ denote the aggregate action of players other than $i$. Types are independently drawn from a common atomless prior $p$ with full support on $[\theta_L, \theta_H]$, mean $\mu$, and CDF $F$. 

For notational convenience, let $\Theta\coloneqq\prod_i\Theta_i $ and $\mathcal{A}\coloneqq\prod_i \mathcal{A}_i$. Let $P$ denote the induced prior over $\theta$ and $P_{-i}$ player $i$'s prior over the other players' type profile $\theta_{-i}$. By independence, this prior belief is independent of $i$'s realized type.

For each type $\theta_i$, player $i$'s payoff is quadratic in the action profile $a$:
\begin{equation}\label{eq:general_payoff}
    u_i(\theta_i,a) = (\alpha + \phi \theta_i + \rho A_{-i} - \tfrac{1}{2}a_i)\,a_i + (\beta + \psi \theta_i )\,A_{-i} + g(\theta_i),
\end{equation}
where $\phi\neq 0$, $\rho \in (0,\frac{1}{n-1})$, $\alpha,\beta,\psi \in \R$ and  $g$ is a $p$-integrable function of $\theta_i$. Given $(\theta_i, A_{-i})$, player $i$'s best response is linear in $A_{-i}$:
\[
a_i = \alpha + \phi\theta_i + \rho A_{-i}.
\]
The payoff parameter $\phi$ governs the direct effect of a player's type on their optimal action: best responses are increasing in type when $\phi > 0$ and decreasing in type when $\phi<0$. 
The parameter $\rho \in (0,\frac{1}{n-1})$ quantifies the strength of strategic interaction among the players. We focus on the strategic complement case $\rho>0$, and the assumption $\rho < \frac{1}{n-1}$ guarantees that the baseline game has a unique and stable Bayes-Nash equilibrium (BNE). 
The parameter $\alpha \in \R$ shifts the baseline level of the best response. Finally, the parameters $\beta,\psi\in \R$ capture the direct payoff externality of opponents' actions, with $\psi$ measuring how this externality varies with  type.

Throughout the paper, we illustrate our main results using the following Bertrand example. 
\begin{example}[Bertrand Oligopoly]\label{eg: Bertrand}
    \emph{Firms produce differentiated goods that are imperfect substitutes and compete in prices. Firm $i$ sets price $a_i \in \R$, faces demand $D + d_i - \frac{1}{2}a_i + r A_{-i}$, and has a marginal production cost $C + c_i$. Here, $D \geq 0$ is a common demand shifter, $C \geq 0$ is a common marginal production cost, $d_i \in \R$ is a private demand shock, and $c_i \in \R$ is a private cost shock. $r \in (0,\frac{1}{2(n-1)})$ measures the strategic complementarity in prices.\footnote{We assume $r \in (0,\frac{1}{2(n-1)})$ to ensure the demand curve is micro-founded by a representative consumer.} 
    The profit of firm $i$ is 
    \begin{align*}
        &(a_i - C - c_i) (D + d_i - \tfrac{1}{2}a_i + r A_{-i})\\
        =& (D + \tfrac{1}{2}C + d_i + \tfrac{1}{2}c_i + r A_{-i} - \tfrac{1}{2} a_i) a_i - r(C+c_i) A_{-i} - (C+c_i)(D+d_i)
    \end{align*}
    We consider two distinct cases in which firms have private information of different natures.}
    
    \noindent\emph{\textbf{Private demand.} Firm $i$ faces demand uncertainty but not cost: $d_i = \theta_i$ is an i.i.d. random variable while $c_i \equiv 0$ is a constant. 
    This is a special case of \eqref{eq:general_payoff} with $\theta_i = d_i$, $\alpha = D + \tfrac{1}{2}C$, $\phi=1$, $\beta = -rC$, $\psi = 0$, $\rho = r$ and $g(\theta_i) = -CD - Cd_i$. In the BNE of this game, firms' equilibrium prices increase in their private demand. 
    }

    \noindent\emph{\textbf{Private cost.} Firm $i$ faces cost uncertainty but not demand: $c_i = \theta_i$ is an i.i.d. random variable while $d_i \equiv 0$ is a constant. 
    This is a special case of \eqref{eq:general_payoff} with  $\theta_i = c_i$, $\alpha = D + \tfrac{1}{2}C$, $\phi=\tfrac{1}{2}$, $\beta = -rC$, $\psi = -r$, $\rho = r$ and $g(\theta_i) = -CD - Dc_i$. In the BNE of this game, the markup $a_i - C - c_i$ is decreasing in $c_i$. 
    }

    \emph{The main difference between these two cases is the type-dependent welfare externality from opponents' prices: $\psi = 0$ under private demand and $\psi = -r$ under private costs.
    We will show that this difference shapes firms' incentives to influence opponents' prices and leads to a sharp contrast.}
\end{example}

\subsection{Mediated Communication}

We model mediated communication by positing that an uninformed mediator commits ex-ante to a direct communication mechanism $\t: \Theta \to \Delta(\mathcal{A})$. This mechanism induces a game among the agents, in which a strategy for agent $i$ consists of a reporting rule and an action rule, and the timing is as follows: After observing $\theta_i$, each player $i$ sends a private report $\hat\theta_i\in \Theta_i$ to the mediator; then, given the report profile $\hat{\theta}= (\hat\theta_i)_{i\in I}$, the mediator draws a profile of action recommendations $a=(a_i)_{i\in I}$ according to $\t(\cdot \mid \hat\theta)$; finally, each player $i$ privately observes $a_i$ and then takes an action.

A BNE of the communication game induced by $\t$ is called a \emph{communication equilibrium} (CE) of the baseline game \citep{myerson1982optimal,forges1986approach}. By the revelation principle, it is without loss to focus on direct incentive-compatible mechanisms: it is a BNE for all players to report their types truthfully and to obey the mediator's action recommendation. We therefore characterize communication equilibria directly in terms of the outcomes they induce.

An outcome $\pi \in \D(\Theta\times \mathcal{A})$ is a joint distribution over players' types and actions. 
Given a version $\pi(\cdot\mid \theta)$ of the regular conditional distribution of actions given type profile $\theta$, for every $i\in I$ and $\hat{\theta}_i \in \Theta_i$, define the distribution of action recommendations induced by report $\hat{\theta}_i$ as
\[
\pi_i^{\hat{\theta}_i}(B) \coloneqq \int_{\Theta_{-i}} \pi(B \mid \hat{\theta}_i, \theta_{-i}) \de P_{-i}(\theta_{-i}) \qquad \text{for every Borel } B\subseteq \mathcal{A}.
\]
This notation distinguishes the player's true type, which determines their payoff, from the reported type, which determines the continuation recommendation distribution they face.\footnote{We restrict attention to outcomes with $\E_{\pi}[\sum_{i\in I}|a_i|] < \infty$ and $\E_{\pi_i^{\hat\theta_i}}[\sum_{j\in I}a_j^2] < \infty$ for every $i,\hat\theta_i$.} This distribution only depends on the reported type since the players' types are independent. 

For player $i\in I$, $\theta_i, \hat{\theta}_i\in \Theta_i$, and measurable function $\delta_i : \mathcal{A}_i \to \mathcal{A}_i$, define 
\[
U_i^\pi(\theta_i;\hat{\theta}_i,\delta_i)
\coloneqq
\int_\mathcal{A}
u_i\bigl(\theta_i,\delta_i(a_i),a_{-i}\bigr)
\,\de \pi_i^{\hat{\theta}_i}(a).
\]
Let $\textrm{id}$ denote the identity map. We say an outcome $\pi\in \D(\Theta\times \mathcal{A})$ is a \emph{CE outcome} if it is induced by some communication equilibrium. Equivalently, $\pi$ satisfies:
\begin{itemize}
    \item[(i)] Consistency: $\marg_{\Theta} \pi = P$.
    \item[(ii)] Incentive compatibility: there exists a version of the regular conditional distribution $\pi(\cdot\mid \theta)$ such that 
    \begin{equation}\label{eq: IC}
        U_i^\pi(\theta_i; \theta_i, \textrm{id}) \geq U_i^\pi(\theta_i;\hat{\theta}_i,\delta_i) \tag{IC}
    \end{equation}
    for every $i \in I$, $\theta_i, \hat{\theta}_i \in \Theta_i$ and every measurable $\delta_i: \mathcal{A}_i \to \mathcal{A}_i$.
\end{itemize}
Condition \eqref{eq: IC} is a \emph{double-deviation} constraint: given a direct mechanism, no player with type $\theta_i$ strictly benefits from unilaterally misreporting type $\hat{\theta}_i$ and then taking an action $\delta_i(a_i)$ different from the recommended action $a_i$.

\subsection{Benchmark Outcomes}

A necessary condition of \eqref{eq: IC} is \emph{obedience}: conditional on truthful reporting, a player cannot benefit from deviating from the recommendation. Formally, there exists a version of the regular conditional distribution $\pi(\cdot\mid\theta)$ such that 
\begin{equation*}\label{eq: OB general}
        U_i^\pi(\theta_i; \theta_i, \textrm{id}) \geq U_i^\pi(\theta_i;\theta_i,\delta_i)
\end{equation*}
for every $i\in I$, $\theta_i \in \Theta_i$, and every measurable $\delta_i: \mathcal{A}_i\to \mathcal{A}_i$.

We say that an outcome is a \emph{Bayes correlated equilibrium (BCE) outcome} if it is consistent and obedient. This is the relevant benchmark when the mediator can condition recommendations on the players' types without needing to elicit them, for example, because reports are directly verifiable.

Let $\pi^{NI}$ denote the outcome induced by the unique BNE of the baseline game, which is also a CE outcome induced by an uninformative mechanism. This is the \emph{no-information benchmark}: players choose actions based only on their private types with equilibrium actions
\begin{equation}\label{eq: NI}
    a_i^{NI}(\theta_i) = \alpha + \phi\theta_i + \rho(n-1) \bar{a}, \qquad \bar a = \frac{\alpha+\phi\mu}{1-\rho(n-1)} \tag{NI}
\end{equation}
where $\bar a$ is the expected action in equilibrium. In this case, $\Var_{\pi^{NI}}(a_i\mid \theta_i)=0$ for every $i\in I$.

Finally, let $\pi^{FI}$ denote the \emph{full-information benchmark}: the realized type profile is publicly disclosed, and players then choose the corresponding
complete-information Nash equilibrium actions,
\begin{equation}\label{eq: FI}
a_i^{FI}(\theta)
=
\bar a
+\frac{\phi}{1+\rho}(\theta_i-\mu)
+\frac{\rho\phi}{(1+\rho)(1-\rho(n-1))}
\sum_{j\in I}(\theta_j-\mu).
\tag{FI}
\end{equation}
Since the recommended actions are
best responses to the realized type profile, $\pi^{FI}$ is a BCE outcome. As we discuss in \cref{ssec: cost of interim}, however, it is never a CE outcome.

Throughout the analysis below, whenever we consider a CE (respectively, BCE) outcome $\pi \in \D(\Theta\times \mathcal{A})$, we fix an incentive-compatible (obedient) version of $\pi(\cdot \mid \theta)$, and all report-contingent objects are understood relative to this version.

\section{Incentive Compatibility in Linear-Quadratic Games}\label{sec:IC}
Communication affects a player's payoff through two distinct channels. By changing their report, a player might alter the actions their opponents are induced to take. The report may also change what the player learns about those actions before choosing their own. In general, these effects are intertwined as both can depend on the entire continuation distribution. The linear-quadratic structure allows us to disentangle them. 

For every consistent outcome $\pi \in \D(\Theta \times \mathcal{A})$, suppose player $i\in I$ reports $\hat{\theta}_i\in \Theta_i$. Player $i$'s expectations of the aggregate opponent action $A_{-i}$ before and after observing recommendation $a_i$ are, respectively,
\[
\bar A_{-i} (\hat{\theta}_i) \coloneqq \E_{\pi_i^{\hat{\theta}_i}} \left[ A_{-i}\right], \qquad \text{and} \qquad \bar A_{-i} (a_i;\hat{\theta}_i) \coloneqq \E_{\pi_i^{\hat{\theta}_i}} \left[ A_{-i} \middle| \, a_i \right].
\]

If player $i$ reports type $\theta_i$ truthfully, conditional on recommendation $a_i$, only the posterior mean $\bar A_{-i}(a_i;\theta_i)$ matters for the player's decision. Obedience is therefore equivalent to: for every $i \in I$ and every $\theta_i \in \Theta_i$,
\begin{equation}\label{eq: OB}
    a_i = \alpha + \phi \theta_i + \rho \bar A_{-i}(a_i;\theta_i) \qquad \text{for } \pi^{\theta_i}_i\text{-almost every } a_i \in \mathcal{A}_i. \tag{OB}
\end{equation}

To characterize the continuation payoff following an arbitrary report, we first define the indirect payoff from responding optimally to a given $A_{-i}$, net of $g(\theta_i)$:
\begin{align*}
    V(\theta_i, A_{-i}) \coloneqq \frac{1}{2}(\alpha + \phi\theta_i + \rho A_{-i})^2 + (\beta + \psi \theta_i) A_{-i}. 
\end{align*}

Now consider a player of true type $\theta_i$ who reports $\hat\theta_i$ and responds optimally after observing the recommendation. Under obedience, the optimal action differs from the recommendation by $\phi(\theta_i - \hat{\theta_i})$. Writing $\sigma_i^2(\hat\theta_i) \coloneqq \Var_{\pi_i^{\hat\theta_i}}(a_i)$, the optimized continuation payoff net of $g(\theta_i)$ is
\begin{align*}
U_i^\pi(\theta_i;\hat\theta_i)
\coloneqq
\E_{\pi_i^{\hat\theta_i}}
\left[
V\bigl(\theta_i,\bar A_{-i}(a_i;\hat\theta_i)\bigr)
\right] 
=&\;
V(\theta_i, \bar A_{-i}(\hat{\theta}_i)) + \tfrac{1}{2} \Var_{\pi_i^{\hat\theta_i}}(\rho \bar A_{-i}(a_i;\hat{\theta_i}))\\
=&
\underbrace{V(\theta_i, \bar A_{-i}(\hat{\theta}_i))}_{\text{Manipulation value}}
+
\underbrace{\tfrac{1}{2}\sigma_i^2(\hat{\theta}_i)}_{\text{Information value}}.
\end{align*}
The two terms capture the two channels described above. The first is the \emph{manipulation value} of report $\hat\theta_i$ for type $\theta_i$: the payoff from optimally responding to the expected opponent actions $\bar A_{-i}(\hat\theta_i)$ before observing the recommendation. The second is the \emph{information value}: the additional payoff from conditioning the player's action on the recommendation. Importantly, this value depends on the report but not on the player's true type.

The double-deviation constraint \eqref{eq: IC} consists of obedience after truthful reporting and \emph{truth-telling}: no player can strictly benefit from misreporting their type and then responding optimally to the recommendation. In our linear-quadratic setting, truth-telling reduces to an information-manipulation trade-off. 

\begin{proposition} \label{prop:IC-characterization}
A consistent outcome $\pi \in \D(\Theta \times \mathcal{A})$ is a CE outcome if and
only if it admits a version of regular conditional distribution satisfying \eqref{eq: OB} and, for every $i\in I$ and $\theta_i, \hat{\theta}_i\in \Theta_i$,
\begin{equation}
\label{eq: IC simplified}
\underbrace{
\frac{1}{2} \sigma_i^2(\theta_i)
-
\frac{1}{2} \sigma_i^2(\hat{\theta}_i)
}_{\text{informational difference } \hat{\theta}_i \to \theta_i}
\ge
\underbrace{
V\bigl(\theta_i, \bar A_{-i}(\hat{\theta}_i)\bigr)
-
V\bigl(\theta_i, \bar A_{-i}(\theta_i)\bigr)
}_{\text{manipulation difference } \theta_i \to \hat{\theta}_i}. \tag{TT}
\end{equation}
\end{proposition}

In other words, the mediator is designing a menu of information values and expected opponent actions $(\sigma_i^2(\hat{\theta}_i), \bar A_{-i}(\hat{\theta}_i))$ to screen players' private information. This is reminiscent of a mechanism-design problem where $\bar A_{-i}(\hat\theta_i)$ plays the role of an allocation and $\sigma_i^2(\hat\theta_i)$ plays the role of a transfer. Unlike in a standard screening problem, however, the mediator cannot choose them independently: both must arise from the same communication mechanism and be linked by obedience.  
In particular, the opponent behavior used for screening must be induced by recommendations they are willing to follow.
This interaction between \eqref{eq: OB} and \eqref{eq: IC simplified} is exactly what determines whether informative communication is feasible.

\section{Informative Communication}\label{sec: main}
A CE outcome is \emph{informative} if it differs from the no-information benchmark $\pi^{NI}$; otherwise, it is \emph{uninformative}. Informative communication is possible if an informative CE outcome exists. 

\subsection{Impossibility under Misaligned Differences}

Whether informative communication is possible depends on how the players' types affect their preferences over their own and opponents' actions. The first effect is captured by the cross-partial $\phi = \frac{\partial^2}{\partial a_i \partial \theta_i} u_i$. Its sign determines whether payoffs have increasing or decreasing differences in type and own action, and hence how a player’s preferred action varies with type. The second is captured by 
\[
V_{\theta A} \coloneqq \frac{\partial^2 V(\theta_i, A_{-i})}{\partial \theta_i \partial  A_{-i}} = \rho \phi + \psi.
\]
The sign determines whether the manipulation value has increasing or decreasing differences between type and expected opponent actions, and hence which types benefit more from inducing higher opponent actions.


\begin{definition}\label{def:misaligned}
    Preferences have aligned differences if $\phi V_{\theta A} \geq 0$. Preferences have misaligned differences if $\phi V_{\theta A} < 0$.
\end{definition}

Preferences have aligned differences 
when the effect of type on the player’s preference for a higher own action is aligned with its effect on the preference for higher opponent actions. 
As the next theorem shows, misaligned differences rule out informative communication for every prior type distribution.

\begin{theorem} \label{Thm: misaligned}
    Informative communication is impossible regardless of the prior type distribution if and only if preferences have misaligned differences.\footnote{Here, we allow any atomless prior supported on a bounded interval.}
\end{theorem}

To see the intuition, consider a symmetric CE outcome and suppose that $\phi > 0$; the case $\phi < 0$ is symmetric. Holding aggregate opponent actions fixed, a higher-type player prefers taking a higher action. Since actions are strategic complements, obedience \eqref{eq: OB} requires a nonnegative covariance between types $\theta_i$ and expected opponent actions $\bar A_{-i}(\theta_i)$. 
When preferences have misaligned differences, $V_{\theta A} < 0$, so lower types value an increase in opponent actions relatively more than higher types. To ensure truthful reporting \eqref{eq: IC simplified}, higher opponent actions should be allocated to types that value them relatively more, so $\bar A_{-i}(\theta_i)$ is weakly decreasing in type. Therefore, obedience and truth-telling are incompatible unless $\bar A_{-i}$ is constant. In this case, players are not responding to any information beyond their own types, yielding precisely the uninformative outcome.

This result also clarifies the role of the type-dependent welfare externality term $\psi \theta_i A_{-i}$. Absent this term, $V_{\theta A} = \rho\phi$, which has the same sign as $\phi$ because $\rho >0$. Thus, misaligned differences cannot arise from the strategic channel alone; they require the type-dependent welfare externality to be strong enough to overturn the strategic effect.

The impossibility result extends to richer environments.
For instance, the mediator could also use report-dependent transfers, or the players' payoffs have additional second-order pure welfare externality terms that do not affect the players' best responses. As long as those additional terms do not interact with the players' true type, they behave like the informational value in the current information-manipulation decomposition, and hence do not affect the tension between \eqref{eq: IC simplified} and \eqref{eq: OB} under misaligned differences. In \cref{sec:general_LQ}, we show an analogue of this impossibility result in a class of general linear best-response games with strategic complementarities.

\subsection{Possibility under Aligned Differences}
When preferences have aligned differences, informative communication is possible if and only if the informational benefit of truthful reporting exceeds the manipulation gain from inducing more favorable opponent actions.

Let 
\[
\overline{V}_{A} \coloneqq \frac{\partial}{\partial  A_{-i}}[V(\theta_i,A_{-i})] \vert_{(\theta_i, A_{-i}) = (\mu, (n-1) \bar a)}=\rho \bar a + \beta + \psi \mu,
\]
be the marginal payoff gain of the mean type $\mu$ from an increase in aggregate opponent action at the no-information benchmark $\pi^{NI}$. 

\begin{definition}
    We say that the \emph{extreme-type separation condition} holds if 
    \begin{equation}\label{eq: main cutoff}
    | \mu - \theta_{ext}| > \frac{|\overline{V}_{A}|}{\frac{\rho |\phi|}{2(1-\rho(n-1))} + |V_{\theta A}|} \tag{SEP}
    \end{equation}
    where $\theta_{ext}=\theta_L$ if $\phi\overline{V}_{A} \geq 0$ and $\theta_{ext}=\theta_H$ otherwise. 
\end{definition}

\begin{theorem}\label{Thm: aligned}
    If preferences have aligned differences, then there exists an informative CE outcome if and only if 
    the extreme-type separation condition holds.
\end{theorem}

To see the intuition, suppose $\phi > 0$ and $\overline{V}_A > 0$. Under the no-information outcome $\pi^{NI}$, every type expects the same aggregate opponent action. Obedience implies that the prior average of expected opponent action across types remains at the no-information level under any CE, so increases for some types must be offset by decreases for others. Consider an informative perturbation from $\pi^{NI}$ that raises expected opponent actions for higher types and lowers them for lower types. Since $\overline{V}_A > 0$ and preferences have aligned differences, types above the mean benefit from the increases. Types facing reductions, however, might have an incentive to misreport to induce higher opponent actions. For a given increase in opponent actions, this incentive is weakest for the lowest type $\theta_L$, which values higher opponent actions the least. Therefore, $\theta_L$ is the easiest to incentivize to accept lower expected opponent actions in exchange for information. The extreme-type separation condition captures the first-order information-manipulation tradeoff at $\pi^{NI}$. 

The necessity of this local condition follows from the tension between truth-telling and obedience. In any informative CE outcome, the sum of covariances between players' actions and their own types is strictly higher than the no-information level. 
For a fixed covariance, truth-telling requires sufficient information to deter players from misreporting to induce more favorable opponent actions. Since information value is proportional to the conditional variance of action recommendations, this imposes a lower bound on the variance in actions. Obedience imposes an upper bound: greater action variance requires stronger correlation across players' actions by \eqref{eq: OB}, while their independent private types prevent them from being arbitrarily correlated. Both bounds coincide at $\pi^{NI}$, and the extreme-type separation condition determines their relative slopes there. When the condition fails, the variance required by truth-telling exceeds that permitted by obedience for every positive increase in covariance, ruling out all informative CE outcomes.

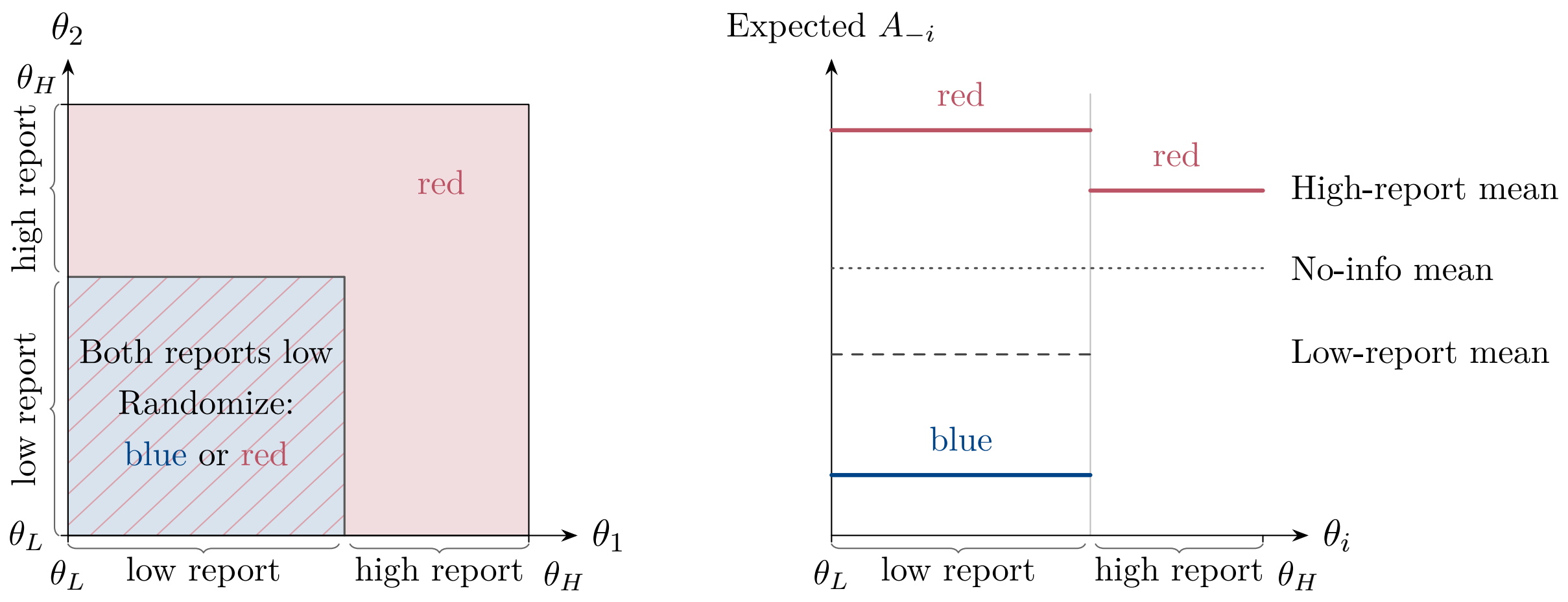
\begin{figure}[H]
\begin{minipage}{\textwidth}
\centering
\definecolor{signalRed}{HTML}{BB5566}
\definecolor{signalBlue}{HTML}{004488}
\begin{minipage}[t]{0.44\linewidth}
  \centering\vspace{0pt}
  \begin{tikzpicture}[
      x=0.94cm,y=0.88cm, >=Stealth,
      font=\small, line cap=round, line join=round]
    \def\T{5}
    \def\W{3}

    \fill[signalRed!20] (0,0) rectangle (\T,\T);
    \fill[signalBlue!15] (0,0) rectangle (\W,\W);
    \begin{scope}
      \clip (0,0) rectangle (\W,\W);
      \foreach \hatch in {-3,-2.64,...,3}{
        \draw[signalRed!55,line width=0.4pt]
          (\hatch,0) -- ++(3,3);
      }
    \end{scope}
    \draw[thin] (0,0) rectangle (\T,\T);
    \draw[semithick,black!65] (\W,0)--(\W,\W)--(0,\W);

    \node[font=\footnotesize] at (1.5,2.10) {Both reports low};
    \node[align=center,font=\footnotesize] at (1.5,1.25)
      {Randomize:\\
       \textcolor{signalBlue}{blue} or \textcolor{signalRed}{red}};
    \node[text=signalRed,font=\footnotesize] at (4.05,4.10) {red};

    \draw[->] (0,0) -- (5.53,0) node[right] {$\theta_1$};
    \draw[->] (0,0) -- (0,5.53) node[above] {$\theta_2$};
    \foreach \x/\lab in {0/\theta_L,5/\theta_H}{
      \draw (\x,0) -- ++(0,-0.075);
      \draw (0,\x) -- ++(-0.075,0);
    }
    \node[left=2.5pt,font=\footnotesize] at (0,0) {$\theta_L$};
    \node[anchor=south east,xshift=-2.5pt,yshift=3pt,
      font=\footnotesize,inner sep=0pt] at (0,\T) {$\theta_H$};

    \draw[thin,black!60,decorate,decoration={brace,amplitude=2.5pt}]
      (-0.10,0)--(-0.10,2.94);
    \draw[thin,black!60,decorate,decoration={brace,amplitude=2.5pt}]
      (-0.10,3.06)--(-0.10,\T);
    \node[rotate=90,font=\footnotesize,inner sep=0pt] at (-0.45,1.47) {low report};
    \node[rotate=90,font=\footnotesize,inner sep=0pt] at (-0.45,4.03) {high report};

    \draw[thin,black!60,decorate,decoration={brace,mirror,amplitude=2.5pt}]
      (0,-0.10)--(2.94,-0.10);
    \draw[thin,black!60,decorate,decoration={brace,mirror,amplitude=2.5pt}]
      (3.06,-0.10)--(\T,-0.10);
    \node[below=1pt,font=\footnotesize,inner ysep=0pt] at (1.47,-0.20) {low report};
    \node[below=1pt,font=\footnotesize,inner ysep=0pt] at (4.03,-0.20) {high report};
    \node[below=1pt,font=\footnotesize,inner ysep=0pt] at (0,-0.24) {$\theta_L$};
    \node[right=3pt,font=\footnotesize,inner sep=1pt] at (\T,-0.47) {$\theta_H$};
    \path (0,-0.80) rectangle (0,6.2);
    \node[anchor=north,font=\small,inner sep=0pt] at (2.5,-1.12)
      {(a) Communication mechanism};
  \end{tikzpicture}
\end{minipage}\hfill
\begin{minipage}[t]{0.54\linewidth}
  \centering\vspace{0pt}
  \begin{tikzpicture}[
      x=0.88cm,y=0.88cm, >=Stealth,
      font=\small, line cap=round, line join=round]
    \def\T{5}
    \def\W{3}
    \def\zBlue{0.70}
    \def\zRedLow{4.70}
    \def\zLowMean{2.10}
    \def\zNI{3.10}
    \def\zRedHigh{4.00}

    \draw[thin,black!25] (\W,0) -- (\W,5.12);
    \draw[->] (0,0) -- (5.53,0) node[right] {$\theta_i$};
    \draw[->] (0,0) -- (0,5.53);
    \node[above,font=\footnotesize] at (0,5.53) {Expected $A_{-i}$};

    \draw[dotted,semithick,black!70] (0,\zNI)--(\T,\zNI);
    \node[right=4pt,font=\footnotesize] at (\T,\zNI) {No-info mean};
    \draw[dashed,semithick,black!75] (0,\zLowMean)--(\W,\zLowMean);
    \node[right=4pt,font=\footnotesize] at (\T,\zLowMean) {Low-report mean};

    \draw[very thick,signalRed] (0,\zRedLow)--(\W,\zRedLow);
    \draw[very thick,signalBlue]   (0,\zBlue)--(\W,\zBlue);
    \draw[very thick,signalRed] (\W,\zRedHigh)--(\T,\zRedHigh);
    \node[right=4pt,font=\footnotesize] at (\T,\zRedHigh) {High-report mean};
    \node[above=3pt,text=signalRed,font=\footnotesize] at (1.5,\zRedLow) {red};
    \node[above=3pt,text=signalBlue,font=\footnotesize] at (1.5,\zBlue) {blue};
    \node[above=3pt,text=signalRed,font=\footnotesize] at (4,\zRedHigh) {red};

    \foreach \x in {0,5}{
      \draw (\x,0) -- ++(0,-0.075);
    }
    \draw[thin,black!60,decorate,decoration={brace,mirror,amplitude=2.5pt}]
      (0,-0.10)--(2.94,-0.10);
    \draw[thin,black!60,decorate,decoration={brace,mirror,amplitude=2.5pt}]
      (3.06,-0.10)--(\T,-0.10);
    \node[below=1pt,font=\footnotesize,inner ysep=0pt] at (1.47,-0.20) {low report};
    \node[below=1pt,font=\footnotesize,inner ysep=0pt] at (4.03,-0.20) {high report};
    \node[below=1pt,font=\footnotesize,inner ysep=0pt] at (0,-0.24) {$\theta_L$};
    \node[right=3pt,font=\footnotesize,inner sep=1pt] at (\T,-0.47) {$\theta_H$};
    \path (0,-0.80) rectangle (0,6.2);
    \node[anchor=north,font=\small,inner sep=0pt] at (2.5,-1.12)
      {(b) Posterior means};
  \end{tikzpicture}
\end{minipage}

\captionof{figure}{Cutoff construction ($\phi>0$ and $\overline V_A>0$).
Panel (a) illustrates the mechanism for two players.
In panel (b), colored segments show posterior means of opponent actions conditional on the public message, the dashed line averages over messages following a low report, and the dotted line is the no-information benchmark.}
\label{fig:cutoff-construction}
\end{minipage}
\end{figure}

We show the sufficiency of the separation condition by constructing a cutoff mechanism illustrated in \cref{fig:cutoff-construction}. Players report whether their types are above or below a cutoff. If all reports are low, the mediator randomizes between two public messages, blue and red; otherwise it sends red for sure (panel (a)). The blue message thus reveals that all reports are low. As panel (b) shows, reporting low allows a player to distinguish between two possible posterior means of opponent actions, while a high report provides no additional information. A low report thus provides information but induces lower expected opponent actions. Under the separation condition, there exists an interior cutoff type that is indifferent between two reports, and aligned differences ensure truthful reporting. 

\subsection{Interim Pareto Improvement}
The incentive compatibility constraint has a sharp welfare consequence: every informative CE outcome is an interim Pareto improvement over the no-information one.
\begin{proposition}\label{Prop: welfare CE}
Under any CE outcome, every type of every player obtains an interim expected payoff weakly higher than under the no-information benchmark. Moreover, if one type of a player is strictly better off, then every type of that player is strictly better off. 
\end{proposition}

Therefore, in our linear-quadratic setting, the existence characterizations \cref{Thm: misaligned,Thm: aligned} determine exactly when mediated communication can generate a strong interim Pareto improvement relative to the no-information outcome. Whenever an informative CE exists, it can be symmetrized so that every player type strictly benefits from communication.

The result follows from the information-manipulation decomposition \eqref{eq: IC simplified} and the convexity of the manipulation value in expected opponent actions. Truthful reporting dominates a deviation where a player draws a random report from the prior and then responds optimally to the recommendation. By obedience, the expected aggregate opponent action under this deviation equals the no-information level. The convexity of manipulation value then implies that the deviation yields at least the no-information payoff, so every type weakly benefits from communication.

\subsection{Illustration: Bertrand Oligopoly}\label{sec: illustration}
\cref{Thm: misaligned,Thm: aligned} imply a sharp contrast between the two Bertrand environments in \cref{eg: Bertrand}. For the private-demand environment, we maintain the following \emph{nonnegative-markup condition}
\[
D + \theta_L \geq \left(\frac{1}{2} - r(n-1) \right) C,
\]
which ensures that the demand is sufficiently higher than the common cost, and thereby every obedient price recommendation is weakly above marginal cost.\footnote{The condition says that the lowest-demand firm's best response is weakly higher than the marginal cost when all opponents price at the marginal cost. By \cref{lem:bounded-actions-BCE}, this ensures that in any BCE outcome, the recommended price is almost surely weakly higher than the marginal cost.} 

\begin{corollary}
\label{Cor: Bertrand}
Consider the Bertrand environments in Example \ref{eg: Bertrand}.

\begin{enumerate}
    \item[(i)] In the private-cost environment, the only CE outcome is the no-information benchmark.
    
    \item[(ii)] In the private-demand environment, under the nonnegative-markup condition, informative communication is possible if and only if
    \begin{equation*}
        \mu-\theta_L > \frac{2D+(2r(n-1)-1)C+2\mu}{3-2r(n-1)}.
    \end{equation*}
\end{enumerate}
\end{corollary}

In a Bertrand oligopoly, a firm benefits from higher opponent prices because they increase its demand. In the private-cost environment, a higher-cost firm benefits less from an increase in opponent prices than a lower-cost firm does, because the extra demand diverted by higher opponent prices is less profitable when its own cost is high. Therefore, preferences have misaligned differences: higher reported costs are associated with higher prices in equilibrium, while the lower-cost firms benefit more from an increase in opponents' prices, leading to a pooling outcome.


In the private-demand environment, by contrast, preferences have aligned differences: a higher-demand firm values higher opponent prices more than a lower-demand firm does. Under the nonnegative-markup condition, every obedient price is weakly above marginal cost, so all firms benefit from inducing higher opponent prices. The lowest-demand firm has the weakest incentive to induce higher opponent prices, so informative communication is feasible if and only if information can screen the lowest-demand firm, which occurs when the lowest demand is sufficiently far below the mean demand.


The cutoff in \cref{Cor: Bertrand} also yields comparative statics for the feasibility of informative communication.
\begin{proposition}
\label{Prop: Bertrand demand CS}
In the private-demand Bertrand environment, under the nonnegative-markup condition, the extreme-type separation condition becomes harder to satisfy when the common demand shifter $D$ rises, the private demand shocks shift upward uniformly, the common cost $C$ decreases, the strategic complementarity $r$ becomes stronger, or the market size $n$ grows.
\end{proposition}

\cref{fig1Bertrand} illustrates how the characterization in  \cref{Cor: Bertrand} varies with the common demand shifter $D$, common cost $C$, and a first-order stochastic shift in the demand shock distribution. 

\begin{figure}[t]
    \centering
    \begin{tikzpicture}[paper figure, x=1.42cm, y=4.2cm]
      \pgfmathsetmacro{\boundaryangle}{atan(0.3*4.2/1.42)}
      \foreach \separation/\panelshift in
          {0.5/0,1.0/5.25,1.5/10.50} {
        \begin{scope}[xshift=\panelshift cm]
          \pgfmathsetmacro{\ceintercept}{0.3*\separation-0.2}
          \begin{scope}
            \clip (0,0) rectangle (3,1.2);
            \fill[regionBlue]
              (0,\ceintercept)--(3,{0.9+\ceintercept})
              --(3,0.7)--(0,-0.2)--cycle;
            \fill[pattern=north east lines, pattern color=black!25]
              (0,-0.2)--(3,0.7)--(3,-0.2)--cycle;
            \draw[paper boundary, figureGray] (0,-0.2)--(3,0.7);
            \draw[paper boundary, bertrandBlue]
              (0,\ceintercept)--(3,{0.9+\ceintercept});
          \end{scope}
          \node[font=\footnotesize] at (1.5,1.34){$\mu-\theta_L=\separation$};

          \node[align=center, font=\footnotesize, text=black!80]
            at (0.92,1.01) {Only\\no-information};
          \node[text=bertrandBlue, font=\footnotesize,
                rotate=\boundaryangle, inner sep=1pt]
            at (1.92,{0.3*1.92+0.15*\separation-0.2})
            {Informative CE exists};
          \node[align=center, font=\scriptsize, text=black!70]
            at (2.16,0.14) {Markup\\condition fails};

          \draw[paper axis] (0,0)--(3.10,0) node[right] {$C$};
          \draw[paper axis] (0,0)--(0,1.24);
          \foreach \x in {0,1.5,3}
            \draw[thin, black!65] (\x,0)--++(0,-0.012)
              node[below=2pt, font=\scriptsize] {\x};
          \foreach \y in {0,0.4,0.8,1.2} {
            \draw[thin, black!65] (0,\y)--++(-0.035,0);
            \ifdim\panelshift pt=0pt
              \node[left=3pt, font=\scriptsize] at (0,\y) {\y};
            \fi
          }
          \ifdim\panelshift pt=0pt
            \node[above=2pt] at (0,1.24) {$D$};
            \node[left=3pt, text=figureGray, font=\scriptsize]
              at (0,0.27) {$D_0$};
          \fi

          \draw[paper guide] (0,0.27)--(0.8,0.27)--(0.8,0);
          \draw[thin, figureGray] (0.8,0)--++(0,-0.012)
            node[below=2pt, font=\scriptsize] {$C_0$};
          \fill[black, draw=white, line width=0.5pt]
            (0.8,0.27) circle[radius=2pt];
        \end{scope}
      }
      \node[font=\footnotesize, text=figureGray] at (5.197,-0.22)
        {Fixed: $n=5$, $\theta_L=0.2$, $r=0.05$;
         $(C_0,D_0)=(0.8,0.27)$ in all three panels.};
    \end{tikzpicture}
    \caption{Informative communication in the private-demand
    Bertrand environment.}
    \begin{minipage}{0.93\textwidth}
\small \emph{Note.} Each panel plots the $(C,D)$-plane for a different
    value of $\mu-\theta_L$. The hatched region violates the
    nonnegative-markup condition; the blue-shaded region admits an informative CE;
    in the remaining region, the no-information outcome is the unique CE outcome.
\end{minipage}
    \label{fig1Bertrand}
\end{figure}

Within each panel, sufficiently low demand relative to cost violates the nonnegative-markup condition. Among primitives satisfying this condition, a higher common demand shifter $D$ or a lower common marginal cost $C$ makes informative communication harder to sustain: moving upward or to the left eventually crosses the blue boundary into the region where the no-information outcome is the unique CE outcome. Both changes raise firms' markups, increasing their gains from higher opponent prices without creating additional information value to compensate low-demand firms. 

Across panels, an upward first-order stochastic shift in the demand distribution with fixed support increases $L = \mu - \theta_L$, expanding the set of  $(C,D)$ where informative communication is possible. For shifter values fixed at $ (C_0, D_0)$, only the no-information outcome can arise when $L = 0.5$, whereas informative communication is possible when $L = 1$ or $1.5$. 

Stronger price complementarity $r$ or a larger number of firms $n$ affects both information and manipulation incentives. A stronger $r$ makes information about opponents more valuable, but it also strengthens firms' incentives to induce higher opponent prices. Similarly, since the marginal effect of each opponent's price is $r$ regardless of the market size, a larger $n$ increases both the value of learning about opponents and the incentive to influence their prices. In the Bertrand environment, the manipulation effect dominates, so the extreme-type separation condition becomes more demanding. These comparative statics depend on the particular form of strategic interaction. If demand instead depends on $r$ times the average opponent price, the aggregate interaction would not grow with $n$, and the cutoff in \cref{Cor: Bertrand} would be independent of $n$. In \cref{app:investment}, we provide an application in which stronger strategic complementarity can instead make the extreme-type separation condition less demanding.

\section{The Limits of Communication} \label{Sec: comparison}
The preceding analysis characterizes how interim incentive compatibility limits informative communication. We now explore the economic implications of these limits by comparing mediated communication with several alternative benchmarks.

\subsection{The Costs of Truth-telling} \label{ssec: cost of interim}
In this subsection, we isolate the restriction imposed by the truth-telling constraint. BCE relaxes this requirement by allowing the mediator to condition recommendations directly on players' realized types, so there is no need to elicit private information. In many applications, however, participation in information sharing is voluntary: before learning their types, players can decide whether to grant a mediator or algorithm access to their private information. This motivates the second benchmark, which we call \emph{exclusive ex-ante information sharing}.

Under exclusive ex-ante information sharing, the mediator first commits to a family of mechanisms, one for each set of participating players. Before observing private types, players decide whether to participate. Participants grant the mediator access to their realized types and receive action recommendations that depend on participants' information. Non-participants neither reveal their types nor receive any information from the mediator, and choose their actions based only on their private types. Hence, participation is determined ex-ante, and no private information needs to be elicited once participation has been chosen.


We consider BNEs of the induced games and call the corresponding outcomes \emph{ex-ante sharing} (EA) outcomes. 
EA outcomes are BCE outcomes subject to an ex-ante participation constraint: each participant's ex-ante expected payoff is at least as high as the payoff from opting out. In our linear-quadratic setting, non-participation yields the no-information payoff, so this constraint reduces to a comparison between the payoff from participation and the no-information benchmark. By \cref{Prop: welfare CE}, the incentive compatibility constraint of CE implies this participation constraint, so every CE outcome is also an EA outcome.\footnote{In general games, EA outcomes are nested in BCE outcomes, while CE outcomes are not necessarily nested in EA outcomes. EA and CE are different restrictions of BCE: EA adds an endogenous participation constraint, while CE requires elicitation of private information.}

We compare the maximum aggregate player payoff achievable under CE, BCE, and EA. Under this objective, the EA participation constraint does not restrict the optimum, so EA and BCE attain the same maximum welfare. The comparison with CE then isolates the welfare cost of truth-telling. 

\begin{samepage}
\begin{proposition}\label{Prop: CE BCE EA}
    BCE and EA attain the same maximum aggregate player payoff. Moreover, the following are equivalent:
    \begin{enumerate}
        \item[(i)] CE attains this common maximum;
        \item[(ii)] The no-information outcome is welfare-maximizing among BCE outcomes;
        \item[(iii)] Preferences have \emph{strongly misaligned differences}:
        \begin{equation}
            \frac{V_{\theta A}}{\phi} \leq  -\frac{\rho}{2(1-\rho(n-1))}.
        \end{equation}
    \end{enumerate}
\end{proposition}
\end{samepage}

Therefore, interim elicitation leads to a strict welfare loss unless no information is already optimal even when the mediator can condition directly on players' types. Whenever some BCE outcome strictly improves total player payoff over no information, the welfare-maximizing CE cannot attain the common BCE and EA optimum. 

Revealing more information allows players to adapt better to the realized types. Under misaligned differences, the only CE outcome is uninformative. 
Under aligned differences, the adjustment in opponent actions also benefits the types that value it more, so full information maximizes total welfare. However, revealing types fully is never feasible as a CE: it makes information value identical across reports, leaving only the manipulation incentive. Therefore, CE must distort away from full information to create differences in information value for truth-telling.

\begin{figure}[H]
\centering
\pgfmathsetlengthmacro{\regionwidth}{(\linewidth-5.2cm)/3}
\begin{tikzpicture}[
    paper figure, x=\regionwidth, y=1cm
]
\coordinate (c) at (1,0);
\coordinate (zero) at (2,0);

\draw[paper axis]
    (0,0) -- (3.04,0)
    node[right=3pt] {$V_{\theta A}/\phi$};

\draw[figureGray, thin, dashed] (c) -- ++(0,-2.7);
\draw[signalBlue, thin, dashed] (zero) -- ++(0,-2.7);
\fill[signalRed] (c) circle (2.2pt);
\fill[signalBlue] (zero) circle (2.2pt);

\node[above=4pt, text=signalRed, align=center] at (c)
    {$-\dfrac{\rho}{2[1-\rho(n-1)]}$};
\node[above=4pt, text=signalBlue] at (zero) {$0$};

\draw[
    decorate,
    decoration={brace,amplitude=5pt},
    signalRed,
    semithick
]
(0.02,1.25) -- (1.98,1.25)
node[midway, above=7pt, text=signalRed]
    {\textbf{Misaligned differences}};

\draw[
    decorate,
    decoration={brace,amplitude=5pt},
    signalBlue,
    semithick
]
(2.02,1.25) -- (2.98,1.25)
node[midway, above=7pt, text=signalBlue]
    {\textbf{Aligned differences}};

\node[anchor=east, xshift=-8pt] at (0,-0.8) {\textbf{BCE and EA}};
\node[anchor=east, xshift=-8pt] at (0,-1.95) {\textbf{CE}};

\node at (0.5,-0.8) {NI optimal};
\node at (1.5,-0.8) {\textbf{NI suboptimal}};
\node at (2.5,-0.8) {FI optimal};

\node at (0.5,-1.95) {Only NI};
\node at (1.5,-1.95) {\textbf{Only NI}};
\node[align=center] at (2.5,-1.95)
    {\textbf{Informative CE}\\[-1pt]
     \footnotesize iff (SEP)};

\end{tikzpicture}
\caption{Communication and welfare across parameter regions.}
\label{fig:three-regions}
\end{figure}
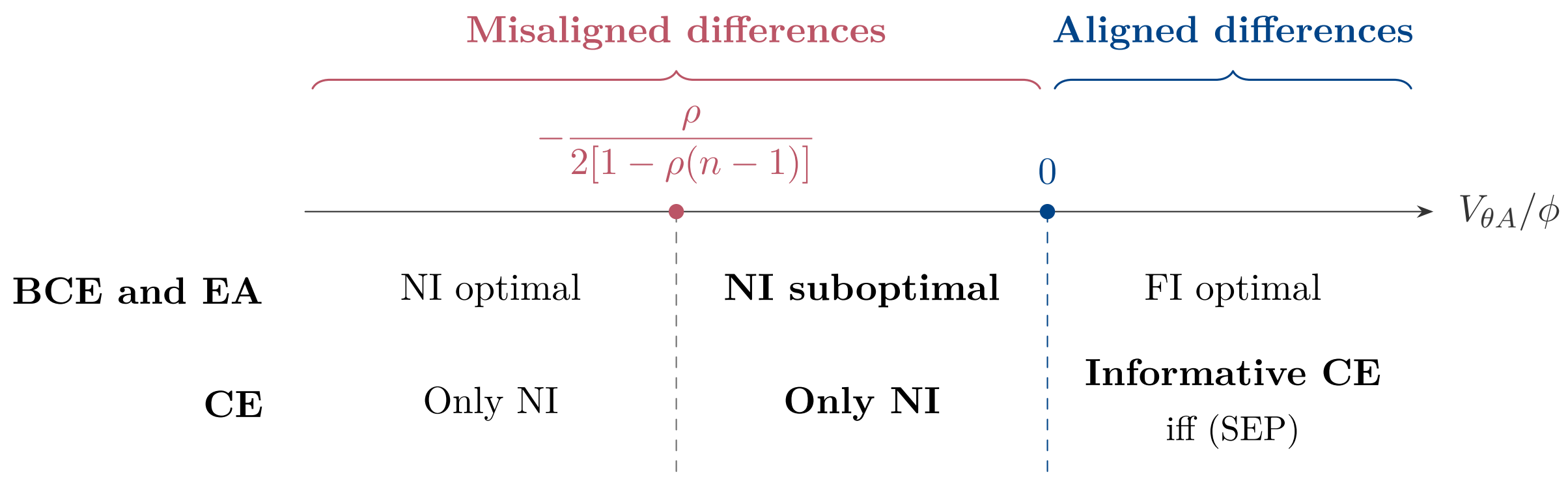

Part (iii) further provides a primitive characterization of when no information is welfare-maximizing. Under misaligned differences, information sharing has two opposing welfare effects. While it allows players to adapt their actions more closely to their types, the induced adjustments in opponent actions can reduce welfare relative to no information. To see the latter effect, suppose $\phi > 0$. Obedience requires types to be positively correlated with expected opponent actions, while misaligned differences imply that higher types value a change in opponent actions less than lower types do. Since the average expected opponent action remains at its no-information level, information only reallocates opponent actions towards types that value them less, thereby reducing expected total welfare. Under strongly misaligned differences, this effect is strong enough to outweigh the adaptation benefit, so that no information is optimal among BCE outcomes. When preferences are misaligned but not strongly misaligned, by contrast, information is valuable ex ante but cannot be elicited truthfully, so interim elicitation leads to a strict welfare loss. Combining this with our main characterizations, Figure \ref{fig:three-regions} summarizes the resulting pattern across parameter regions.

\begin{example}[Bertrand Oligopoly with Private Quality]
    Consider a variant of \cref{eg: Bertrand} in which $\theta_i$ captures privately observed product quality, which both raises consumers' willingness to pay and increases marginal production cost. Let $d_i=w\theta_i$ and $c_i=(1-w)\theta_i$, where $w\in[0,1]$. The cases $w=0$ and $w=1$ reduce to the private-cost and private-demand specifications in
    \cref{eg: Bertrand}. Firm $i$'s profit is
    \[
        \bigl(a_i-C-(1-w)\theta_i\bigr)
        \left(D+w\theta_i-\tfrac12 a_i+r A_{-i}\right).
    \]
    This is a special case of \eqref{eq:general_payoff} with $\phi=(1+w)/2$, $\psi=-r(1-w)$, and $\rho=r$. Preferences have strongly misaligned differences when $w\leq (1-2r(n-1))/(7-6r(n-1))$ and become aligned when $w \geq 1/3$. 
\end{example}


The private-quality example illustrates the difference between ex-ante and interim incentives to share information, extending the classic comparison between cost and demand information \citep{raith1996general}. In our Bertrand environment, pure cost information generates strongly misaligned differences, whereas pure demand information generates aligned differences. When a private characteristic such as quality affects both demand and cost, preferences can instead fall in the intermediate region: ex-ante information sharing is valuable, but informed firms cannot be induced to communicate it truthfully.


Interim elicitation not only restricts the maximum attainable payoff, but also the set of feasible outcomes. Primitives that are irrelevant for both obedience and ex-ante participation can limit informative communication once information must be elicited interim.

\begin{samepage}
\begin{corollary}
    Consider the private-demand Bertrand environment. Fix $r$, $n$, and the distribution of private demand shocks. For every sufficiently large $K$, consider markets satisfying $D + C/2 = K$ and the nonnegative-markup assumption. 

    The set of BCE outcomes is identical across all such markets, and so is the set of EA outcomes. However, there exists $C^* \in (0, 2K)$ such that, within this family, the only CE outcome is uninformative when $C \leq C^*$, while an informative CE exists otherwise.
\end{corollary}
\end{samepage}
To see the distinction, vary the common marginal cost $C$ while adjusting the demand shifter $D$ so $D+C/2$ remains fixed. This leaves the firms' pricing incentives unchanged, as \eqref{eq: OB} is unaffected, so the set of BCE outcomes stays the same. The set of EA outcomes is also unchanged: although this variation affects firms' expected profits, it does not change their ex-ante gain from participation, since the expected prices remain the same.
By \cref{Prop: Bertrand demand CS}, the extreme-type separation condition is more demanding when costs are lower and demands are higher, as it increases firms' interim manipulation incentives. No informative CE can be sustained when the cost is sufficiently low. Therefore, markets that are indistinguishable under both BCE and EA can differ in whether informative communication is possible.

\subsection{Public Cheap Talk}
The previous subsection shows how the truth-telling constraint restricts communication. We now compare CE with public cheap talk, a canonical communication protocol that retains interim incentive constraints but removes the mediator's commitment power.

Under public cheap talk, players simultaneously send public messages after observing their types. All players observe the entire message profile and then choose actions. A public cheap-talk outcome is the joint distribution of types and actions induced by a perfect Bayesian equilibrium of this game.\footnote{The sets of equilibrium outcomes are the same under BNE or perfect Bayesian equilibrium.} We characterize when an informative public cheap-talk outcome exists. 

\begin{definition}
    We say that the \emph{cheap-talk separation condition} holds if 
    \begin{equation*}\label{eq: CT cutoff}
    | \mu - \theta_{ext}| > \frac{|\overline{V}_{A}|}{\frac{\rho^3 (n-1)|\phi|}{2(1+\rho)(1-\rho(n-1))}+ |V_{\theta A}|}
    \end{equation*}
    where $\theta_{ext}=\theta_L$ if $\phi\overline{V}_{A} \geq 0$ and $\theta_{ext}=\theta_H$ otherwise. 
\end{definition}

\begin{proposition}
\label{prop: public_cheap_talk}
The following hold for public cheap talk:

\begin{enumerate}
    \item[(i)] Public cheap talk is uninformative regardless of the prior type distribution if and only if preferences have misaligned differences.
    
    \item[(ii)] If preferences have aligned differences, then there exists an informative public cheap-talk outcome if and only if the cheap-talk separation condition holds.
\end{enumerate}
\end{proposition}

The distinction between mediation and public cheap talk arises under aligned differences. Comparing the cheap-talk separation condition with \eqref{eq: main cutoff}, informative public cheap talk requires greater separation of the relevant extreme type from the mean. With simultaneous public messages, a player's message can affect how opponents respond but not what the player learns from opponents' messages. Therefore, public cheap talk can induce informative communication only through its effects on opponents' actions. Under mediation, however, the information a player receives can depend on what they report to the mediator.

\begin{corollary}
    In the private-demand Bertrand environment of \cref{eg: Bertrand}, if the nonnegative-markup and extreme-type separation conditions hold, then public cheap talk is uninformative, whereas mediation can sustain informative communication. 
\end{corollary}

Under the nonnegative-markup condition, all firms benefit from inducing higher opponent prices. Public cheap talk is hence uninformative: all firms prefer the message that induces the highest opponent prices. By contrast, mediation can still sustain informative communication by compensating low-demand firms for truthful reporting with more informative recommendations.


\subsection{Restricted Access to Communication} In the main model, all players participate in communication. In practice, however, participation may be restricted by regulators. 
For instance, in an Australian retail gasoline case studied in \cite{byrne2025asymmetric}, one of the five major gasoline retailers exited a price-sharing platform while the others retained full access after case settlement. 

Motivated by this, we consider a setting with $n$ players in total, but only a subset of $k < n$ players communicate. We focus on the nontrivial case $k \geq 2$. The mediator commits to a mechanism that receives type reports from participants and provides action recommendations to participants. Non-participants neither send reports nor receive recommendations, and choose actions based only on their private types. 

We consider BNEs of the induced communication games and the corresponding equilibrium outcomes, which are called CE outcomes with $k$ participants. The following result shows how restricting participation affects the possibility of informative communication.

\begin{proposition} \label{Prop: comm size}
    If preferences have aligned differences, there exists an informative CE outcome with $k$ participants if and only if 
    \[
    |\mu - \theta_{ext}| > \frac{|\overline{V}_{A}|}{\frac{\rho|\phi|}{2(1-\rho(k-1))} + |V_{\theta A}|},
    \]
    where the corresponding cutoff is weakly decreasing in $k$.
\end{proposition}

Restricting access to communication reduces the information value available to reward truthful reporting, while leaving the incentive to manipulate others' actions unchanged. When there are $k < n$ participants, the non-participants' equilibrium strategies are precisely their no-information strategies in the $n$-player game. Therefore, restricting the group does not change how much a participant values a favorable change in expected opponent action: they have the same benchmark manipulation incentive as in the $n$-player problem. With fewer participants, the mediator's recommendations can only influence fewer players' actions and therefore provide less information value for the participants. 

Therefore, informative communication may be possible with full participation but impossible when some players are excluded. 
This static mechanism is distinct from the dynamic effects of asymmetric access studied by \cite{byrne2025asymmetric}, where exiting the platform also changes how quickly a firm can observe and respond to opponents' prices.



\section{Extensions}\label{sec:general_LQ}


Our main characterizations exploit the linear-quadratic structure of the baseline model in \cref{sec:model}. We examine their robustness in two complementary directions. We first enrich the quadratic model with second-order welfare externalities, and then move beyond quadratic payoffs to identify the conditions that drive the impossibility result.

\subsection{Second-order Welfare Externality}

The baseline model assumes that the opponents' actions enter the players' payoffs linearly. We now consider a departure from this structure by allowing the payoff to depend quadratically on the aggregate opponent action:
\begin{equation}\label{eq:kappa}
    u_i(\theta_i,a) = (\alpha + \phi \theta_i + \rho A_{-i} - \tfrac{1}{2}a_i)a_i + (\beta + \psi \theta_i ) A_{-i} + \kappa A_{-i}^2 + g(\theta_i).
\end{equation}
We focus on the case $\kappa>0$. The extra term $\kappa A_{-i}^2$ changes the mean type's marginal value of a change in opponents' actions at the no-information benchmark $\pi^{NI}$, which we now denote by
\[\overline V^\kappa_A \coloneqq \rho \bar a + \beta + \psi \mu+2\kappa (n-1) \bar a.\]
The cross-partial derivative $V_{\theta A}$ is unaffected, and we extend our main results as follows.
\begin{proposition}\label{prop: second-order_effect}
\begin{enumerate}
\item[(i)] Informative communication is impossible regardless of the prior type distribution if and only if preferences have misaligned differences.
    \item[(ii)] Suppose preferences have aligned differences. There exists an informative CE outcome if
    \begin{equation} \label{eq: cutoff kappa}
        | \mu - \theta_{ext}| > \frac{|\overline{V}^\kappa_{A}|}{\frac{\rho |\phi|}{2(1-\rho(n-1))}+\kappa(n-1) |\phi|\frac{2-\rho(n-1)}{1-\rho(n-1)} + |V_{\theta A}|}
    \end{equation}
     where $\theta_{ext}=\theta_L$ if $\phi\overline{V}^\kappa_A \geq 0$ and $\theta_{ext}=\theta_H$ otherwise. Moreover, this condition is also necessary if $n=2$.
\end{enumerate}
\end{proposition}

Part (i) shows that the impossibility argument under misaligned differences extends to the case with positive second-order externalities. It is particularly robust and does not rely on the sign of $\kappa$, because the additional term neither affects the best responses nor interacts with the players' types. Under aligned differences, the conclusion is more sensitive to the payoff structure.\footnote{Negative quadratic externalities also arise naturally in classic beauty contest and coordination environments \citep{keynes1936general,dessein2006adaptive, dessein2016rational}. $\kappa < 0$ can lead to qualitatively different conclusions under aligned differences. With sufficiently negative $\kappa$, informative public cheap talk can be sustained.  The interim Pareto improvement in \cref{Prop: welfare CE} can also fail; see \cref{sec: non_lq_improvement}.} Part (ii) extends the baseline construction and provides a sufficient condition for informative communication, which is an extreme-type separation condition with a modified threshold. In the two-player case, this condition is also necessary.




In the two-player case, the cutoff remains exact, but the second-order externality affects it through two channels. First, it changes the incentive to manipulate opponents' behavior through misreporting, which is captured by $\overline V^\kappa_A$ in the numerator. Second, the variance of opponents' actions becomes payoff-relevant. Differences in this variance across reports can provide additional incentive for truthful reporting, beyond the information value of recommendations. This effect appears as an additional positive term in the denominator.



Positive second-order externalities arise, for instance, in R\&D investment and research joint ventures, where firms choose cost-reducing investments before competing in the product market \citep{d1988cooperative,suzumura1992cooperative, kamien1992research}. We illustrate this with a two-stage Bertrand model in which firms have private R\&D costs and communicate before investing. Since the second-stage competition payoff is quadratic, while R\&D shifts production costs linearly, the first-stage payoff contains a positive quadratic externality in opponents' R\&D.
\begin{example}[R\&D with Positive Spillovers]
  Two firms simultaneously choose R\&D investments $a_i, a_j\in \mathbb{R}$, which are publicly observed, and the firms then compete in prices $p_i,p_j\in \mathbb{R}$.
  
  The firms are identical except in their R\&D costs. Firm $i$ has a private cost parameter $\theta_i \in [\theta_L, \theta_H] \subseteq \R_+$ and incurs an R\&D cost $\theta_i a_i+\frac{1}{2}\gamma a_i^2$. It faces demand $D-\frac{1}{2}p_i+rp_j$ and has a marginal production cost $C-a_i-\xi a_j$, with $r\in (0,1/2)$, $\xi\in (\tfrac{r}{1-2r^2},1]$, and $\gamma > 1/2$. Thus, higher R\&D lowers the production cost of both the investing firm and its opponent.

\end{example}
Substituting the second-stage equilibrium prices, the first-stage payoffs take the form in \eqref{eq:kappa} after normalization. Under the stated assumptions, this game has strategic complementarities, positive second-order externalities, and aligned differences. \cref{prop: second-order_effect}(ii) then characterizes the parameter values for which informative communication lets the duopolists coordinate their R\&D investments.

\subsection{Beyond Linear-Quadratic Games}
The impossibility result in \cref{Thm: misaligned} does not rely on quadratic payoffs. Consider a symmetric game in which the utility of player $i$ with type $\theta_i\in \Theta_i$ at the action profile $a$ is
\[u_i(\theta_i,a)=f(\theta_i,a_i)+h(\theta_i,a_i)A_{-i}.\]
The function $f(\theta_i,a_i)$ governs the direct effect of own action on payoff, possibly influenced by the agent's type, generalizing $(\alpha + \phi \theta_i - \tfrac{1}{2}a_i)a_i+g(\theta_i)$ in the baseline payoff of \cref{sec:model}. 
The function $h(\theta_i,a_i)$ represents the marginal payoff from the total opponent action, holding player $i$'s own action fixed. It generalizes the $(\beta + \psi \theta_i + \rho a_i)$ component in the baseline payoff. 
Suppose that $f(\theta_i, \cdot)$ and $h(\theta_i, \cdot)$ are continuously differentiable, and that for every type $\theta_i$, $f(\theta_i,\cdot)$ has a unique maximizer $a^0(\theta_i)$. Assume that $a^0(\theta_i)$ is bounded and $\Var_p(a^0(\theta_i))>0$.\footnote{We restrict attention to outcomes with finite truthful interim payoffs.}

We impose two restrictions that preserve the features of the linear-quadratic model used by the impossibility result. First, the best response is linear in the posterior mean of aggregate opponent action with a common slope, while the intercept can differ across types.
\begin{assumption}\label{ass:linearBR}
    There exists $\rho\in (0,1/(n-1))$ such that, for any $\theta_i\in \Theta_i$ and $A_{-i}\in \mathbb{R}$,
    \[\argmax_{a_i\in \mathbb{R}}f(\theta_i,a_i)+h(\theta_i,a_i)A_{-i}=a^0(\theta_i)+\rho A_{-i}. \]
\end{assumption}

Second, after centering own action around $a^0(\theta_i)$, the effect of own action on the value of opponent actions has a common shape across types.
\begin{assumption}\label{ass:value_learning}
    There exists a function $H: \R \to \R$ with $H(0) = 0$, such that for every $\theta_i\in \Theta_i$ and $a_i \in \R$,
    \[
    h(\theta_i, a_i)  = h(\theta_i, a^0(\theta_i) ) + H(a_i - a^0(\theta_i)).
    \]
\end{assumption}


Together, these assumptions preserve the type-independence of the information value in the baseline model. For any $\theta_i$ and $A_{-i}$, let $v$ be the indirect utility:
\[
v(\theta_i,A_{-i})\coloneqq \max_{a_i\in \mathbb{R}} \{f(\theta_i,a_i)+h(\theta_i,a_i)A_{-i}\}.
\]
By the envelope theorem,
\[
v(\theta_i, A_{-i}) = f(\theta_i, a^0(\theta_i)) + h(\theta_i, a^0(\theta_i)) A_{-i} + K(A_{-i})
\]
where $K(A_{-i}) = \int_0^{A_{-i}} H(\rho t)\de t$ is common across types. Since the type-dependent terms are affine in $A_{-i}$, the information value is independent of the player's true type.


Under these restrictions, the impossibility argument of \cref{Thm: misaligned} extends beyond the linear-quadratic model.

\begin{proposition}\label{prop:beyond_lq}
   Suppose that \cref{ass:linearBR} and \cref{ass:value_learning} hold. If for every $\theta_i, \theta_i' \in \Theta_i$ such that $a^0(\theta_i) \neq a^0(\theta_i')$, 
   $$[a^0(\theta_i)-a^0(\theta_i')][h(\theta_i,a^0(\theta_i))-h(\theta_i',a^0(\theta_i'))]<0,$$
   then the only CE outcome is uninformative.
\end{proposition}

This condition is a natural analogue of misaligned differences: types that prefer higher own actions value higher opponent actions less. In the linear-quadratic model, $a^0(\theta_i) = \alpha + \phi \theta_i$ and $h(\theta_i, a^0(\theta_i)) = \beta + \rho \alpha + V_{\theta A} \theta_i$, so the condition reduces exactly to $\phi V_{\theta A} < 0$.


\section{Conclusion} \label{sec: conclusion}
This paper studies when informative mediated communication is possible in a class of linear-quadratic games with strategic complements. Under misaligned differences, informative communication is impossible regardless of the type distribution: the opponent behavior required for truth-telling is incompatible with obedience. Under aligned differences, informative communication is possible if and only if the extreme-type separation condition holds, which determines whether obedient recommendations can generate sufficient information value to offset manipulation incentives. Our comparison with ex-ante information sharing shows that players may be unwilling to reveal their information once informed, even when they would prefer to commit to sharing it ex ante.

Our characterization also provides a benchmark for understanding how intermediaries facilitate information sharing among competing firms. 
In practice, intermediaries may possess information of their own, verify firms' reports, or implement decisions delegated by firms. They may also design mechanisms to maximize fee revenue rather than firms' joint profit. By abstracting from these additional instruments and incentives, our analysis isolates what can be achieved through the mediated exchange of unverifiable information alone. 
A complementary direction is to sharpen this benchmark by characterizing the set of communication equilibria in terms of primitives and identifying the outcomes that maximize aggregate player payoffs. This remains an open challenge even in the linear-quadratic setting because of the interaction between obedience and truth-telling constraints.

\begingroup
\begin{spacing}{1.0}
\setlength{\bibsep}{2.6pt plus 0.2ex}
\bibliographystyle{ecta}
\bibliography{Reference}
\end{spacing}
\endgroup

\clearpage
\begin{appendix}
\begin{appendices}


\crefalias{section}{appendix}
\crefname{appendix}{Appendix}{Appendices}
\Crefname{appendix}{Appendix}{Appendices}

\section{Preliminaries}\label{app:preliminaries}
In all appendices, for any outcome $\pi \in \D(\Theta\times \mathcal{A})$, we write $\E_\pi[\cdot],\Var_\pi(\cdot), \Cov_\pi(\cdot,\cdot)$ for expectation, variance, and covariance under $\pi$.
When a proof fixes an outcome $\pi$, we suppress the subscript $\pi$ whenever no confusion can arise.

As in the main text, $\bar a$ denotes the no-information benchmark mean action, $\mu$ denotes the mean type. We use $\Var_\theta$ to denote $\Var_\pi(\theta_i)$ for every $i\in I$.


We use $\Sigma^\pi$ to denote the covariance matrix of $(a,\theta)$ under $\pi$, which can be written in block form
\[
\Sigma^\pi
=
\begin{pmatrix}
\Sigma_{aa}^\pi & \Sigma_{a\theta}^\pi\\
\Sigma_{\theta a}^\pi & \Sigma_{\theta\theta}^\pi
\end{pmatrix}.
\]
$I_n$ denotes the $n\times n$ identity matrix, and $J_n$ is the $n\times n$ matrix of ones.

\section{Incentive Compatibility}
\label{sec:TT}
\begin{proof}[Proof of \cref{prop:IC-characterization}]
Fix a consistent and obedient outcome $\pi$. Given the linear-quadratic payoff, for player $i$ with true type $\o_i$ who reports $\hat{\theta}_i$, the unique optimal action for $i$ after observing recommendation $a_i$ is 
\[\delta_i^*(a_i; \theta_i, \hat{\theta}_i)= \alpha + \phi\theta_i + \rho \bar A_{-i}(a_i;\hat{\theta}_i).\]
Therefore, $\pi$ is incentive compatible if and only if, 
for every $i\in I$ and $\theta_i, \hat{\theta}_i \in \Theta_i$,
\begin{align*}
        U_i^\pi(\theta_i; \theta_i, \textrm{id}) \geq U_i^\pi(\theta_i;\hat{\theta}_i,{\delta}^*_i). 
\end{align*}

Substituting the optimal deviation function $\delta^*_i(a_i; \theta_i,\hat{\theta}_i)$ above, we calculate the interim expected payoff net of $g(\theta_i)$ as
\begin{align*}
    U_i^\pi(\theta_i;\hat{\theta}_i,{\delta}^*_i) - g(\theta_i) =& \; \E_{\pi_i^{\hat\theta_i}}\left[ \frac{1}{2}(\alpha + \phi\theta_i + \rho \bar A_{-i}(a_i;\hat{\theta}_i))^2 + (\beta + \psi\theta_i) \bar A_{-i}(a_i;\hat{\theta}_i) \right] \\
    =& \; \frac{1}{2}\Var_{\pi_i^{\hat\theta_i}}(\delta_i^*(a_i)) + \frac{1}{2}(\alpha + \phi\theta_i + \rho \bar A_{-i}(\hat{\theta}_i))^2  + (\beta + \psi\theta_i) \bar A_{-i}(\hat{\theta}_i)\\
    =& \; \frac{1}{2}\sigma^2_i(\hat{\theta}_i) + V(\theta_i, \bar A_{-i}(\hat{\theta}_i)),
\end{align*}
where the first equality follows from iterated expectation, and the second one follows from variance decomposition. The last equality holds because by \eqref{eq: OB},
\[
\delta_i^*(a_i; \theta_i, \hat{\theta}_i) = a_i + \phi(\theta_i - \hat\theta_i) \qquad \pi_i^{\hat\theta_i}-\text{a.s.},
\]
so $\Var_{\pi_i^{\hat\theta_i}}(\delta_i^*(a_i)) = \Var_{\pi_i^{\hat\theta_i}}(a_i) = \sigma^2_i(\hat{\theta}_i)$.

The same expression with $\hat\theta_i = \theta_i$ gives
\[
U_i^\pi(\theta_i;\theta_i,\textrm{id}) - g(\theta_i) = \frac{1}{2}\sigma^2_i(\theta_i) + V(\theta_i, \bar A_{-i}(\theta_i)).
\]
Subtracting the two identities shows that incentive compatibility is equivalent to 
\[
\frac12 \sigma_i^2(\theta_i)-\frac12 \sigma_i^2(\hat\theta_i)
\ge
V\bigl(\theta_i,\bar A_{-i}(\hat\theta_i)\bigr)-V\bigl(\theta_i,\bar A_{-i}(\theta_i)\bigr),
\]
which is exactly \eqref{eq: IC simplified}.
\end{proof}
\subsection{Monotonicity and Envelope Formula}
Recall that in \cref{sec:IC}, given a consistent and obedient outcome $\pi\in\D(\Theta \times \mathcal{A})$, if player $i$'s true type is $\theta_i$ but reports $\hat\theta_i$ and responds optimally to recommendations, the interim expected payoff net of $g(\theta_i)$ is denoted as $U_i^\pi(\theta_i;\hat\theta_i)$:
\[
U_i^{\pi}(\theta_i; \hat{\theta}_i) = \frac{1}{2}\sigma_i^2(\hat{\theta}_i)+ \frac{1}{2}(\alpha + \phi\theta_i + \rho \bar A_{-i}(\hat{\theta}_i))^2 +(\beta + \psi \theta_i) \bar A_{-i}(\hat\theta_i).
\]
We denote the corresponding truthful interim payoff as $U_i^{\pi}(\theta_i) \coloneqq U_i^{\pi}(\theta_i; \theta_i)$. In the proofs below, we omit the dependencies on $\pi$ when there is no confusion.

\begin{lemma}
\label{lem:gamma-positive}
Suppose $\pi\in\D(\Theta\times \mathcal{A})$ is consistent and obedient. If $V_{\theta A}\neq 0$, then $\pi$ is a CE outcome if and only if for every $i \in I$
\begin{enumerate}
\item[(i)] $V_{\theta A} \bar A_{-i}(\cdot)$ is weakly increasing on $[\theta_L,\theta_H]$; and
\item[(ii)] for every $\theta_i \in [\theta_L, \theta_H]$,
\begin{equation}
\label{eq:envelope-gamma-positive}
U^{\pi}_i(\theta_i)
=
U^{\pi}_i(\theta_L) + \frac{1}{2} \phi^2(\theta^2_i - \theta_L^2) + \alpha\phi(\theta_i-\theta_L) +V_{\theta A} \int_{\theta_L}^{\theta_i} \bar A_{-i}(s) \de s.
\end{equation}
\end{enumerate}
\end{lemma}
\begin{proof}
Suppose $\pi$ is a CE outcome. For every $i\in I$ and any pair of types $\theta_i$ and $\hat{\theta}_i$, \eqref{eq: IC simplified} implies that $U_i(\theta_i) \geq U_i(\theta_i; \hat{\theta}_i)$ and $U_i(\hat{\theta}_i; \theta_i) \leq U_i(\hat{\theta}_i)$. Summing up those two inequalities, we have 
    \begin{align*}
        0 \leq & \, U_i(\theta_i) - U_i(\theta_i;\hat{\theta}_i) + U_i(\hat{\theta}_i) - U_i(\hat{\theta}_i;\theta_i)\\
        = & \, V(\theta_i,\bar A_{-i}(\theta_i)) - V(\theta_i, \bar A_{-i}(\hat{\theta}_i)) + V(\hat{\theta}_i, \bar A_{-i}(\hat{\theta}_i)) - V(\hat{\theta}_i, \bar A_{-i}(\theta_i)) \\
        = &\, V_{\theta A} (\theta_i - \hat{\theta}_i) (\bar A_{-i}(\theta_i) - \bar A_{-i}(\hat{\theta}_i)).
    \end{align*}
    Hence $V_{\theta A} \bar A_{-i}(\cdot)$ is weakly increasing.

    By the envelope theorem, for almost every $\theta_i$
    \begin{align*}
        U_i'(\theta_i) = \phi(\alpha+\phi\theta_i+\rho \bar A_{-i}(\theta_i))+ \psi \bar A_{-i}(\theta_i) = \alpha\phi + \phi^2 \theta_i + V_{\theta A} \bar A_{-i}(\theta_i).
    \end{align*}
    Integrating from $\theta_L$ to $\theta_i$ gives (ii).

   Conversely, suppose $\pi$ satisfies (i) and (ii). For any $\theta_i$ and $\hat{\theta}_i$,
    \begin{align*}
        U_i(\theta_i;\hat{\theta}_i) =&\, U_i(\hat{\theta}_i) + V(\theta_i,\bar A_{-i}(\hat{\theta}_i)) - V(\hat{\theta}_i,\bar A_{-i}(\hat{\theta}_i)) \\ 
        =&\, U_i(\hat{\theta}_i) + \frac{1}{2}\phi^2(\theta_i^2 - \hat{\theta}_i^2) + \alpha \phi(\theta_i - \hat{\theta}_i) + V_{\theta A} (\theta_i - \hat{\theta}_i) \bar A_{-i}(\hat{\theta}_i).
    \end{align*}
    Hence, 
    \begin{align*}
        U_i(\theta_i) - U_i(\theta_i; \hat{\theta}_i) =&\, U_i(\theta_i) - U_i(\hat{\theta}_i) - \frac{1}{2}\phi^2(\theta^2_i - \hat{\theta}^2_i) - \alpha \phi(\theta_i - \hat{\theta}_i) - V_{\theta A} (\theta_i - \hat{\theta}_i) \bar A_{-i}(\hat{\theta}_i) \\
        =&\,V_{\theta A} \int_{\hat{\theta}_i}^{\theta_i} \bar A_{-i}(s) \de s - V_{\theta A} (\theta_i - \hat{\theta}_i) \bar A_{-i}(\hat{\theta}_i) \geq 0,
    \end{align*}
    where the second equality follows from (ii), and the inequality holds by the monotonicity of $V_{\theta A} \bar A_{-i}(\cdot)$.
\end{proof}

\begin{lemma}
\label{lem:gamma-zero}
Suppose $\pi\in\D(\Theta\times \mathcal{A})$ is consistent and obedient. If $V_{\theta A}=0$, then $\pi$ is a CE outcome if and only if for every $i\in I$
\begin{equation}
\label{eq:gamma-zero-characterization}
\sigma^2_i(\theta_i) + \rho^2\bar A_{-i}(\theta_i)^2 + 2(\alpha\rho+\beta)\bar A_{-i}(\theta_i) \text{ is constant  on }[\theta_L,\theta_H].
\end{equation}
\end{lemma}
\begin{proof}
    \eqref{eq: IC simplified} implies that for every $\theta_i, \hat{\theta}_i\in\Theta_i$
    \begin{align*}
        U_i(\theta_i) \geq U_i(\theta_i;\hat{\theta}_i)  =&\; U_i(\hat{\theta}_i)  + \frac{1}{2}\phi^2(\theta^2_i - \hat{\theta}^2_i) + \alpha\phi(\theta_i - \hat{\theta}_i) + V_{\theta A} (\theta_i - \hat{\theta}_i) \bar A_{-i}(\hat{\theta_i}).\\
        =&\; U_i(\hat{\theta}_i) + \frac{1}{2}\phi^2 (\theta^2_i - \hat{\theta}^2_i) + \alpha\phi(\theta_i - \hat{\theta}_i),
    \end{align*}
    where the last equality follows from $V_{\theta A} = 0$. Therefore, $U_i(\theta_i) - \frac{1}{2}\phi^2 \theta^2_i - \alpha\phi\theta_i$ is constant in $\theta_i$.

    Since $U_i(\theta_i) = \frac12 \sigma^2_i(\theta_i)+V(\theta_i,\bar A_{-i}(\theta_i))$, we have $U_i(\theta_i) - \frac{1}{2}\phi^2\theta^2_i - \alpha\phi\theta_i  = \frac{1}{2}\sigma^2_i(\theta_i) + \frac{1}{2}\rho^2 \bar A_{-i}(\theta_i)^2 + (\alpha\rho + \beta +V_{\theta A}\theta_i) \bar A_{-i}(\theta_i) + \frac{1}{2}\alpha^2$ is constant in $\theta_i$.
\end{proof}

\subsection{Symmetry Lemmas}
\begin{lemma}\label{lem:action-sign-flip}
Fix a game $G$ with payoff parameters $(\alpha,\phi,\beta,\psi,\rho,g)$, and let $\hat G$ be the game with payoff parameters $(-\alpha,-\phi,-\beta,-\psi,\rho,g)$.
The set of players, type spaces, and type distributions are the same in $G$ and $\hat G$.

For any outcome $\pi\in\D(\Theta\times \mathcal{A})$, let $\hat\pi$ denote the distribution of $(\theta,-a)$ when $(\theta,a)\sim\pi$. Then $\pi$ is a BCE outcome, CE outcome, or BNE outcome in $G$ if and only if $\hat\pi$ is a BCE outcome, CE outcome, or BNE outcome in $\hat G$, respectively.
\end{lemma}

\begin{proof}
Fix an outcome $\pi\in\D(\Theta\times \mathcal{A})$. Let $\hat\pi$ be the distribution of $(\theta,-a)$ when $(\theta,a)\sim\pi$. Note that for every $(\theta,a)\in \Theta\times \mathcal{A}$ and every player $i$,
\begin{align*}
\hat u_i(\theta_i,-a)
=
\Bigl(-\alpha-\phi\theta_i-\tfrac12(-a_i)\Bigr)(-a_i)
+\sum_{j\neq i}\Bigl(-\beta-\psi\theta_i+\rho(-a_i)\Bigr)(-a_j)
+g(\theta_i) 
=
u_i(\theta_i,a).
\end{align*}

Since the type marginal is unchanged, $\pi$ is consistent in game $G$ if and only if $\hat \pi$ is consistent in $\hat G$.

Next, for every $i\in I$, $\theta_i, \hat\theta_i \in \Theta_i$, and measurable $\delta_i:\mathcal{A}_i\to \mathcal{A}_i$, we have 
\[
U_i^{\pi}(\theta_i;\hat\theta_i,\delta_i)
=
U_i^{\hat\pi}(\theta_i;\hat\theta_i,\hat\delta_i),
\]
where $\hat\delta_i(a_i) \coloneqq -\delta_i(-a_i)$. This holds because $\hat u_i(\theta_i,-a) = u_i(\theta_i, a)$ and $\hat\pi_i^{\hat\theta_i}$ is exactly the image of $\pi_i^{\hat\theta_i}$ under the map $a\mapsto -a$. Therefore, BCE outcomes and CE outcomes are preserved by the transformation.

Finally, suppose $\pi$ is a BNE outcome in $G$, induced by strategies $a_i(\theta_i)$. Then $\hat\pi$ is induced by the strategy $\hat a_i(\theta_i)\coloneqq -a_i(\theta_i)$. Since $\hat u_i(\theta_i,-a) = u_i(\theta_i, a)$, player $i$'s payoff from $\hat a_i$ against $\hat a_{-i}$ in $\hat G$ is exactly the same as the payoff from $a_i$ against $a_{-i}$ in $G$. So $\hat\pi$ is a BNE outcome in $\hat G$. The converse implication is immediate.
\end{proof}

\begin{lemma}\label{lem:joint-reflection}
Fix a game $G=(n,\Theta,p,u)$ with payoff parameters $(\alpha,\phi,\beta,\psi,\rho,g)$. Let $\check G=(n,\check\Theta,\check p,\check u)$ be the game with $\check\Theta_i=[-\theta_H,-\theta_L]$, prior $\check p$ given by the distribution of $-\theta_i$ when $\theta_i \sim p$, and payoff parameters $(-\alpha,\phi,-\beta,\psi,\rho,\check g)$, where $\check g(\check \theta_i) = g(- \check\theta_i)$.

For any outcome $\pi\in\D(\Theta\times \mathcal{A})$, let $\check\pi$ denote the distribution of $(-\theta,-a)$ when $(\theta,a)\sim\pi$. Then $\pi$ is a BCE outcome, CE outcome, or BNE outcome in $G$ if and only if $\check\pi$ is a BCE outcome, CE outcome, or BNE outcome in $\check G$, respectively.
\end{lemma}
\begin{proof}
Fix an outcome $\pi\in\D(\Theta\times \mathcal{A})$. Let $\check\pi$ be the distribution of $(-\theta,-a)$ when $(\theta,a)\sim\pi$.

Note that for every $(\check\theta,\check a)\in \check\Theta\times \mathcal{A}$ and every player $i$, 
\begin{align*}
\check u_i(\check\theta_i,\check a)
=
\Bigl(-\alpha+\phi\check\theta_i-\tfrac12\check a_i\Bigr)\check a_i
+\sum_{j\neq i}\Bigl(-\beta+\psi\check\theta_i+\rho\check a_i\Bigr)\check a_j
+g(-\check\theta_i) = u_i(-\check\theta_i,-\check a).
\end{align*}

Consistency is immediate: $\marg_{\check\Theta_i} \check\pi$ is exactly the image of $\marg_{\Theta_i} \pi$ under the map $\theta \mapsto -\theta$.

For every $i\in I$, a true type $\theta_i, \hat{\theta}_i \in \Theta_i$, and  measurable $\delta_i:\mathcal{A}_i\to \mathcal{A}_i$, we have 
\[
U_i^\pi(\theta_i; \hat{\theta}_i, \delta_i) = U_i^{\check \pi}(-\theta_i; -\hat\theta_i, \check{\delta}_i), 
\]
where $\check\delta_i(a_i) \coloneqq -\delta_i(-a_i)$. This holds because $u_i(\theta_i,a) = \check u_i(-\theta_i, -a)$ and $\check\pi_i^{-\hat\theta_i}$ is the image of $\pi_i^{\hat\theta_i}$ under the map $a\mapsto -a$.

The continuation law under $\check\pi$ after report $\check{\hat\theta}_i$ is the image of the continuation law under $\pi$ after report $-\check{\hat\theta}_i$ under the map $a\mapsto -a$. Therefore obedience and incentive compatibility hold under $\check\pi$ if and only if they hold under $\pi$. BCE outcomes and CE outcomes are preserved.

Finally, suppose $\pi$ is a BNE outcome in $G$, induced by strategies $a_i(\theta_i)$. Then $\check\pi$ is induced by the strategy $\check a_i(\check\theta_i)\coloneqq -a_i(-\check\theta_i)$. Since $\check u_i(\check\theta_i,\check a) = u_i(-\check\theta_i,-\check a)$, player $i$'s payoff from $\check a_i$ against $\check a_{-i}$ in $\check G$ is exactly the same as the payoff from $a_i$ against $a_{-i}$ in $G$. So $\check\pi$ is a BNE outcome in $\check G$. The converse implication is immediate.
\end{proof}

\section{Moment Restrictions under Obedience}
\label{app:BCE-moments}
In this section, we provide some necessary moment conditions for symmetric BCE outcomes. \cref{lem:PSD} provides a pair of inequalities in covariances that apply to any symmetric outcome, which follows from the positive semidefiniteness of covariance matrices. \cref{lem:bounded-actions-BCE} shows that the actions are bounded in any BCE outcome, and hence the moments are finite. \cref{lem:OB-moments} presents some identities in first and second-order moments in any BCE outcome. Combining those lemmas, \cref{prop:boundSD_OB} shows that in any symmetric BCE outcome $\pi$, the covariance $\Cov_\pi(a_i,\theta_i)$ between a player's action and their own type is bounded between the no-information and full-information level. Moreover, the variance in actions is upper and lower bounded by two quadratic functions in $\Cov_\pi(a_i,\theta_i)$. 

Similar results have been used to study BCE in linear-quadratic games with Gaussian priors \cite{miyashita2023lqg}. In those settings, it is without loss to study Gaussian BCE outcomes, which are fully characterized by the first and second moments, and the necessary moment requirements are also sufficient. \citet{bonatti2025information} also used similar results to study communication with transfers in linear-quadratic games with Gaussian priors. We provide a self-contained proof here for completeness.

\begin{lemma}
\label{lem:PSD}
Let $\pi$ be a symmetric, consistent outcome with finite second moments. Then for every $j \neq i$
\[
\begin{cases}
    \bigl(\Var_\pi(a_i)-\Cov_\pi(a_i,a_j)\bigr)\Var_\theta
\ge
\bigl(\Cov_\pi(a_i,\theta_i)-\Cov_\pi(a_i,\theta_j)\bigr)^2\\
\bigl(\Var_\pi(a_i)+(n-1)\Cov_\pi(a_i,a_j)\bigr)\Var_\theta
\ge
\bigl(\Cov_\pi(a_i,\theta_i)+(n-1)\Cov_\pi(a_i,\theta_j)\bigr)^2
\end{cases}.
\]
\end{lemma}

\begin{proof}
Since types are i.i.d., $\Sigma_{\theta\theta}= \Var_\theta I_n.$ Because $\Sigma$ is a covariance matrix, it is positive semidefinite. Since $\Sigma_{\theta\theta}$ is positive definite, $\Sigma$ is positive semidefinite if and only if its Schur complement
\[
S = \Sigma_{aa}-\Sigma_{a\theta}\Sigma_{\theta\theta}^{-1}\Sigma_{a \theta}
\]
is positive semidefinite. Fix a pair $i \neq j$. By symmetry, $\Sigma_{aa} =(\Var(a_i)- \Cov(a_i,a_j)) I_n + \Cov (a_i, a_j) J_n$ and $\Sigma_{a\theta} =( \Cov(a_i,\theta_i) -\Cov(a_i,\theta_j) ) I_n + \Cov (a_i, \theta_j) J_n$. Since $\Sigma_{\theta\theta}^{-1} = \tfrac{1}{\Var_\theta} I_n$ and $J_n^2 = n J_n$, direct calculation implies that $S = (a-b) I_n + b J_n$, where 
\[
\begin{cases}
    a = \Var(a_i) - \tfrac{1}{\Var_\theta} (\Cov(a_i,\theta_i)^2 + (n-1)\Cov(a_i, \theta_j)^2)\\ b = \Cov(a_i, a_j) - \tfrac{1}{\Var_\theta} (2 \Cov(a_i,\theta_i) \Cov(a_i, \theta_j) + (n-2)\Cov(a_i, \theta_j)^2)
\end{cases}
\]
A matrix of the form $(a-b) I_n + b J_n$ has eigenvalues $a+(n-1)b$ and $a-b$, so the positive semidefiniteness of $S$ implies the stated inequalities.
\end{proof}

\begin{lemma}
\label{lem:bounded-actions-BCE}
Let $\pi\in \D(\Theta\times \mathcal{A})$ be a BCE outcome. 
Define
\[
a^H
\coloneqq
\bar a+
\frac{\max\{\phi(\theta_L-\mu),\phi(\theta_H-\mu)\}}
{1-\rho(n-1)},
\qquad
a^L
\coloneqq
\bar a+
\frac{\min\{\phi(\theta_L-\mu),\phi(\theta_H-\mu)\}}
{1-\rho(n-1)}.
\]
Then for every player $i\in I$, $a^L \leq a_i \leq a^H$ $\pi$-almost surely.
In particular, all moments of $a_i$ are finite.
\end{lemma}

\begin{proof}
By obedience \eqref{eq: OB} and the definition of $a^H$, for every $i\in I$
\[
a_i-a^H
\leq
\rho\E_\pi\left[
\sum_{j\neq i}(a_j-a^H)
\,\middle|\,
a_i,\theta_i
\right].
\]
Taking positive parts, we have
\[
(a_i-a^H)_+
\leq
\rho\E_\pi\left[
\sum_{j\neq i}(a_j-a^H)_+
\,\middle|\,
a_i,\theta_i
\right].
\]
Taking expectations and summing over $i$ then yields
\[
\sum_{i\in I}\E_\pi[(a_i-a^H)_+]
\leq
\rho(n-1)
\sum_{i\in I}\E_\pi[(a_i-a^H)_+].
\]
Those expectations are finite since actions are integrable. Since $\rho(n-1) < 1$, the sum on the left must be zero, and hence $a_i \leq a^H$ almost surely for every $i$. The analogous argument shows that $a_i \geq a^L$ almost surely.
\end{proof}

\begin{lemma}
\label{lem:OB-moments}
Let $\pi$ be a BCE outcome. Then for every $i \in I$,
\[
\E_{\pi}[a_i] = \bar a = \frac{\alpha + \phi \mu}{1 - \rho(n-1)} .
\]
For every $i \in I$, 
\[
\begin{cases}
    \Cov(a_i,\theta_i) = \phi \Var_\theta+\rho\sum_{j\neq i}\Cov(a_j,\theta_i)\\
    \Var(a_i) = \phi \Cov(a_i,\theta_i)+\rho\sum_{j\neq i}\Cov(a_i,a_j).
\end{cases}
\]
\end{lemma}

\begin{proof}
Because $\pi$ is obedient,
\[
a_i
=
\alpha+\phi\theta_i+\rho\,\E_{\pi^{\theta_i}_i}\!\left[\sum_{j\neq i}a_j \,\middle|\, a_i\right]
\qquad \pi\text{-a.s.}
\]
Taking expectation with respect to $\pi$ on both sides implies that $\E_\pi[a_i] = \alpha + \phi\mu + \rho \sum_{j\neq i}\E_\pi[a_j]$ for every $i\in I$. This system has a unique solution $\E_\pi[a_i] = \bar a$ since $\rho (n-1) < 1$. Taking covariance with $\theta_i$ on both sides implies the first covariance identity, and taking covariance with $a_i$ on both sides implies the second.
\end{proof}

For $\xi\in\{\rho,-\rho(n-1)\}$, define
\[
f_{\xi}(\lambda,\rho,n,\phi)
\coloneqq
\frac{1}{\xi}(\lambda-\phi)\left(\lambda-\frac{\phi}{1+\xi}\right).
\]
\begin{proposition}
\label{prop:boundSD_OB}
Suppose $\rho\in(0,\frac{1}{n-1})$, and let $\pi$ be a symmetric BCE outcome. Define $\lambda \coloneqq \Cov_\pi(a_i,\theta_i) / \Var_\theta$, we have:
\begin{enumerate}
    \item[(i)] Covariance of action and own type lies between $\pi^{NI}$ and $\pi^{FI}$:
    \[
    (\lambda - \phi) (\lambda - k(n,\rho)\phi) \leq 0
    \qquad
    k(n,\rho)\coloneqq \frac{1-\rho(n-2)}{(1+\rho)(1-\rho(n-1))}\geq 1.
    \]
    In particular, $\lambda$ has the same sign as $\phi$, and $|\phi| \leq |\lambda| \leq k(n,\rho) |\phi|$. Moreover, $\lambda = \phi$ if and only if $\pi = \pi^{NI}$, and $\lambda = k(n,\rho)\phi$ if and only if $\pi = \pi^{FI}$.

    \item[(ii)] The residual variation in actions satisfies:
    \[
    \rho(\Var(a_i) - \lambda^2 \Var_\theta) \in [\rho f_\rho(\lambda,\rho,n,\phi) \Var_\theta,\; \rho f_{-\rho(n-1)}(\lambda,\rho,n,\phi) \Var_\theta].
    \]
\end{enumerate}
\end{proposition}

\begin{proof}
Fix $\pi$ and a pair $i\neq j$.

By \cref{lem:OB-moments},
\[
\Cov(a_i,\theta_j)
=
\frac{\Cov(a_i,\theta_i)-\phi \Var_\theta}{\rho(n-1)}, 
\qquad 
\Cov(a_i,a_j)
=
\frac{\Var(a_i)-\phi\Cov(a_i,\theta_i)}{\rho(n-1)}.
\]
Substituting these identities into the two inequalities in \cref{lem:PSD}, dividing by $\Var_\theta^2$, we obtain
\[
\frac{1+\xi}{\xi}\Bigl(\frac{\Var(a_i)}{\Var_\theta}-\lambda^2-f_{\xi}(\lambda,\rho,n,\phi)\Bigr)\ge 0
\]
for $\xi \in \{\rho, -\rho(n-1)\}$. Since $1+\xi > 0$ for both values of $\xi$, the coefficient $\frac{1+\xi}{\xi}$ is positive for $\xi = \rho$ and negative for $\xi = -\rho(n-1)$. Hence,
\[
    f_\rho(\lambda,\rho,n,\phi) \leq \frac{\Var(a_i)}{\Var_\theta}-\lambda^2\leq f_{-\rho(n-1)}(\lambda,\rho,n,\phi)
\]
which proves part (ii).

The intervals in part (ii) are nonempty if and only if 
\[
\rho\Bigl(f_\rho(\lambda,\rho,n,\phi)-f_{-\rho(n-1)}(\lambda,\rho,n,\phi)\Bigr) = \frac{n}{n-1}(\lambda-\phi)\bigl(\lambda-k(n,\rho)\phi\bigr)\le 0,
\]
which is exactly the inequality in part (i). To see the attainment, note that if $\lambda=\phi$ or $\lambda=k(n,\rho)\phi$, then the two bounds in part (ii) coincide. By the proof of \cref{lem:PSD}, this happens if and only if the Schur complement $S = \Sigma_{aa}-\Sigma_{a\theta}\Sigma_{\theta\theta}^{-1}\Sigma_{a \theta}$ of the covariance matrix is a zero matrix. It follows that the random vectors $a$ and $\theta$ satisfy
\[
a = \E[a] + \Sigma_{a\theta} \Sigma_{\theta\theta}^{-1}(\theta - \E[\theta]) \qquad \pi-\text{a.s.},
\]
because $S$ is exactly the covariance matrix of the linear-projection residual $a - \E[a] - \Sigma_{a\theta} \Sigma_{\theta\theta}^{-1}(\theta - \E[\theta])$. Since $\Sigma_{\theta\theta} = \Var_\theta I_n$, $\E[a_i] = \bar a$ and $\E[\theta_i] = \mu$ for every $i\in I$, we have 
\begin{align*}
    a_i =& \;\bar a + \frac{\Cov(a_i,\theta_i)}{\Var_\theta}(\theta_i - \mu) + \sum_{j\neq i} \frac{\Cov(a_i,\theta_j)}{\Var_\theta}(\theta_j - \mu) \\
    =&\; \bar a + \lambda(\theta_i - \mu) + \frac{\lambda - \phi}{\rho(n-1)}\sum_{j\neq i}(\theta_j - \mu),
\end{align*}
where the second equality holds because $\Cov(a_i,\theta_i) = \lambda \Var_\theta$ and $\Cov(a_i, \theta_j) = \frac{\lambda - \phi}{\rho(n-1)}\Var_\theta$ by symmetry and \cref{lem:OB-moments}. 

Therefore, if $\l = \phi$, then the preceding expression reduces to the no-information action \eqref{eq: NI}, so $\pi = \pi^{NI}$. If $\l = k(n,\rho)\phi$, substituting $k(n,\rho)$ into the preceding expression leads to \eqref{eq: FI}, so $\pi = \pi^{FI}$.  The converse is immediate.
\end{proof}

\section{Characterizing Informative Communication}
In this section, we prove \cref{Thm: misaligned,Thm: aligned}. 
It suffices to prove the theorems for symmetric CE outcomes, because if there exists an informative CE outcome, then there exists a symmetric informative CE outcome by randomizing over permutations of players.

Fix a symmetric outcome $\pi \in \D(\Theta\times \mathcal{A})$ and a player $i \in I$. By symmetry, the player-wise objects $\bar A_{-i}^\pi(\cdot)$, $\sigma_i^{2}(\cdot)$, and $U_i^\pi(\cdot)$ do not depend on $i$. Throughout the proof we therefore write $\bar A(\cdot)$, $\sigma^2(\cdot)$, and $U(\cdot)$, suppressing both $i$ and $\pi$.

\paragraph{Normalize the sign of $\phi$.} We show that it suffices to consider the case $\phi > 0$. 

Suppose $\phi < 0$, consider the alternative game $\hat G = (n,\Theta,p,\hat u)$ with payoff parameters $(-\alpha, -\phi, -\beta, -\psi, \rho, g)$. By \cref{lem:action-sign-flip}, there is a bijection between CE and BNE outcomes in $G$ and $\hat G$. The corresponding $\hat V_{\theta A} = - V_{\theta A}$ and $\hat {\overline{V}}_{A} = - \overline{V}_{A}$. 
So the signs of $\phi V_{\theta A}$ and $\phi\overline{V}_{A}$, as well as the cutoff in the extreme-type separation condition \eqref{eq: main cutoff} are invariant under this transformation. 

\begin{proof}[Proof of \cref{Thm: misaligned}]
    Suppose preferences have misaligned differences. Under normalization $\phi>0$, this is the case $V_{\theta A}<0$. Suppose $\pi$ is a CE outcome. 
    By \cref{lem:OB-moments} and \cref{prop:boundSD_OB}, $\rho \Cov(\bar A(\theta_i), \theta_i) = \Cov(a_i, \theta_i) - \phi \Var_\theta \geq 0$ as $\phi > 0$. Since $V_{\theta A} <0$, \cref{lem:gamma-positive} implies that $\bar A$ is weakly decreasing, so $\Cov(\bar A(\theta_i), \theta_i) \leq 0$. Therefore, $\Cov(\bar A(\theta_i), \theta_i) = 0$ and it follows that $\Cov(a_i,\theta_i) = \phi \Var_\theta$. By \cref{prop:boundSD_OB}, $\pi = \pi^{NI}$.

    For the converse, suppose preferences have aligned differences. We show that some type distribution admits an informative CE outcome. Consider a uniform distribution over $[-L, L]$ for some large enough $L > 0$. Under this prior, $\overline{V}_{A} = \frac{\rho \alpha}{1-\rho(n-1)}+\beta$, which is independent of $L$, and hence so is the cutoff in \eqref{eq: main cutoff}. The extreme-type separation condition in \cref{Thm: aligned} holds for sufficiently large $L$ and hence an informative CE outcome exists. 
\end{proof}

\begin{proof}[Proof of \cref{Thm: aligned}]
Suppose preferences have aligned differences.
Under normalization $\phi>0$, this is the case $V_{\theta A}\geq 0$.  We focus on the case $\overline{V}_{A}\geq 0$ and show that an informative symmetric CE outcome exists if and only if  the extreme-type separation condition holds, that is,
\[
\mu-\theta_L>
\frac{\overline{V}_{A}}{
\frac{\rho\phi}{2(1-\rho(n-1))}+V_{\theta A}}.
\]
The opposite case $\overline{V}_{A}  < 0$ will be addressed at the end.

\noindent\textbf{Necessity:} If the  extreme-type separation condition  fails, the only CE outcome is $\pi^{NI}$. Suppose the symmetric outcome $\pi$ we fixed is a CE outcome. Let 
\[
x\coloneqq \frac{\Cov(a_i,\theta_i)}{\Var_\theta}-\phi\ge 0,
\qquad
\bar A^{NI}\coloneqq (n-1)\bar a,
\qquad
\Delta(\theta_i)\coloneqq \bar A(\theta_i)-\bar A^{NI}.
\]
where $x \geq 0$ follows from $\phi \geq 0$ and \cref{prop:boundSD_OB}. By obedience (\cref{lem:OB-moments}),
\[
x
=
\frac{\rho\,\Cov(\bar A(\theta_i),\theta_i)}{\Var_\theta}
=
\frac{\rho\,\Cov(\Delta(\theta_i),\theta_i)}{\Var_\theta},
\qquad
\E_\pi[\Delta(\theta_i)]=0.
\]
Let $z \coloneqq \esssup \Delta \geq 0$. Using $\E[\Delta(\theta_i)]=0$,
\[
\frac{x\Var_\theta}{\rho}
=
\E_\pi[\Delta(\theta_i)(\theta_i-\theta_L)]
\le
z(\mu-\theta_L),
\]
so
\begin{equation}
\label{eq:z-lower}
z\ge \frac{x\Var_\theta}{\rho(\mu-\theta_L)}.
\end{equation}

Next, define
\[
R(\theta_i)
\coloneqq
2U(\theta_i)-(\alpha + \phi\theta_i)^2 
= \sigma^2(\theta_i) + \rho^2 \bar A(\theta_i)^2+2(\rho\alpha+\beta+V_{\theta A}\theta_i)\bar A(\theta_i).
\]

If $V_{\theta A}>0$, \cref{lem:gamma-positive} implies that $\bar A(\cdot)$ is weakly increasing and
\[
R(\theta_i)=R(\theta_H)-2V_{\theta A}\int_{\theta_i}^{\theta_H} \bar A(s)\,\de s.
\]
Taking expectations and changing the order of integration yields
\[
\E[R(\theta_i)]
=
R(\theta_H)-2V_{\theta A}\int_{\theta_L}^{\theta_H} \bar A(s)F(s)\,\de s.
\]
Recall that by \eqref{eq: OB}
\[
U(\theta_i)=\frac12 \E[a_i^2\mid \theta_i]+(\beta+\psi\theta_i)\bar A(\theta_i),
\]
rearranging, and subtracting $\E[a_i]^2=\bar a^2$, we obtain
\begin{align}
\Var_\pi(a_i)
=&\,
R(\theta_H) -2V_{\theta A}\int_{\theta_L}^{\theta_H}\bar A(s)F(s)\de s.
 \notag\\
&\,
-2 \beta \bar A^{NI}
-2\psi\E[\theta_i\bar A(\theta_i)]
+\phi^2\Var_\theta +(\alpha + \phi\mu)^2
-\bar a^2
\label{eq:TT-lower-bound-P}
\end{align}

Since $\bar A(\cdot)$ is weakly increasing, $\D(\theta_H) \geq z$, $\bar A(\theta_H) \geq z + \bar A^{NI}$.
Also,
\[
\E_\pi[\theta_i\bar A(\theta_i)]
=
\mu \bar A^{NI}+\frac{x\Var_\theta}{\rho}, \qquad
\bar a = \alpha + \phi\mu + \rho \bar A^{NI},
\qquad
\int_{\theta_L}^{\theta_H}\bar A(s)F(s)\de s
\le
\bar A(\theta_H)(\theta_H-\mu).
\]
Substituting these bounds into \eqref{eq:TT-lower-bound-P} and simplifying yields
\begin{align*}
    \Var_\pi(a_i) \geq&\, \sigma^2(\theta_H)  + \rho^2 \bar A(\theta_H)^2 + 2(\alpha \rho+\beta+V_{\theta A}\theta_H)\bar A(\theta_H) - 2 (\rho\phi +\psi)\bar A(\theta_H)(\theta_H -\mu) \\
    &-2\beta \bar A^{NI}
-2\psi(\mu \bar A^{NI} + \frac{x\Var_\theta}{\rho})
+\phi^2\Var_\theta +(\alpha + \phi\mu)^2
-\bar a^2  \\
\geq &\, \rho^2z^2+2\overline{V}_{A}z-2\frac{\psi}{\rho}x\Var_\theta+\phi^2\Var_\theta.
\end{align*}

If $V_{\theta A} = 0$, by \cref{lem:gamma-zero}, there exists a constant $K$ such that
\[
R(\theta_i) = \sigma^2(\theta_i)+\rho^2\bar A(\theta_i)^2+2(\alpha\rho+\beta)\bar A(\theta_i)=K
\qquad \forall \theta_i\in \Theta_i.
\]
Since conditional variances are non-negative,
\[
K\ge \rho^2(\bar A^{NI}+z)^2+2(\alpha\rho+\beta)(\bar A^{NI}+z).
\]
Substituting this into \eqref{eq:TT-lower-bound-P} yields the same lower bound
\[
\Var_\pi(a_i)
\ge
\rho^2z^2+2\overline{V}_{A}z-2\frac{\psi}{\rho}x\Var_\theta+\phi^2\Var_\theta.
\]

Thus, in both cases $V_{\theta A}=0$ and $V_{\theta A}>0$,
\begin{equation}
\Var_\pi(a_i)\ge L_{CE}(x),
\qquad
L_{CE}(x)
:=
\phi^2\Var_\theta-2\frac{\psi}{\rho}x\Var_\theta
+
2\overline{V}_{A}\frac{x\Var_\theta}{\rho(\mu-\theta_L)}
+
\Bigl(\frac{x\Var_\theta}{\mu-\theta_L}\Bigr)^2.
\label{eq:LGamma-signrobust}
\end{equation}
We compare this with the upper bound of $\Var_\pi(a_i)$ in \cref{prop:boundSD_OB} with $\rho>0$, which is 
\begin{equation}
\Var_\pi(a_i)\le U_{BCE}(x),
\qquad
U_{BCE}(x)
=
\phi^2\Var_\theta+2\phi x\Var_\theta+\frac{\phi}{1-\rho(n-1)}x\Var_\theta-\Bigl(\frac{1}{\rho(n-1)}-1\Bigr)x^2\Var_\theta.
\label{eq:UGamma-signrobust}
\end{equation}
Subtracting \eqref{eq:UGamma-signrobust} from \eqref{eq:LGamma-signrobust}, we obtain
\begin{equation}
L_{CE}(x)-U_{BCE}(x)
=
x \Var_\theta
\Biggl[
x\Bigl(\frac{\Var_\theta}{(\mu-\theta_L)^2}+\frac{1}{\rho(n-1)}-1\Bigr)
+
\frac{2\overline{V}_{A}}{\rho(\mu-\theta_L)}
-\frac{2V_{\theta A}}{\rho}
-\frac{\phi}{1-\rho(n-1)}
\Biggr].
\label{eq:gap-signrobust}
\end{equation}
The bracket is strictly increasing in $x\geq 0$ as $\rho(n-1) \in (0,1)$. Therefore, if the extreme-type separation condition fails, which is equivalent to
\begin{equation*}
\frac{2\overline{V}_{A}}{\rho(\mu-\theta_L)}
\ge
\frac{2V_{\theta A}}{\rho}+\frac{\phi}{1-\rho(n-1)},
\label{eq:key-ineq-signrobust}
\end{equation*}
then \eqref{eq:gap-signrobust} implies $L_{CE}(x) > U_{BCE}(x)$ for every $x > 0$, and hence necessarily $x = 0$. Part (i) of \cref{prop:boundSD_OB} then implies $\pi = \pi^{NI}$.

\noindent\textbf{Sufficiency.} Assume the extreme-type separation condition holds. We construct an informative symmetric CE outcome as follows. 

Fix $\varepsilon\in(0,1)$; in the knife-edge case $V_{\theta A}=\overline{V}_{A}=0$, we will later choose $\varepsilon$ small enough. For a cutoff $\omega\in(\theta_L,\theta_H)$, let
\[
\chi\coloneqq \rho(n-1),\qquad
\nu\coloneqq 1-F(\omega),
\qquad
\mu_L\coloneqq \E[\theta_i\mid \theta_i\le \omega],
\qquad
\zeta\coloneqq (1-\varepsilon)F(\omega)^{n-1}.
\]
The mediator only uses whether each report is below or above $\omega$. If at least one player reports above $\omega$, everyone receives the public signal $s^0$. If all reports are below $\omega$, everyone receives $s^1$ with probability $1-\varepsilon$ and $s^0$ with probability $\varepsilon$.

Let $z_H$, $z_L^0$, and $z_L^1$ denote $i$'s posterior means of $\sum_{j\neq i}a_j$ after a high report $\hat\theta_i>\omega$, a low report $\hat\theta_i\leq \omega$ followed by $s^0$, and a low report followed by $s^1$, respectively. The corresponding action recommendations are
\[
a_i=
\begin{cases}
\alpha+\phi\hat\theta_i+\rho z_H, & \hat\theta_i>\omega,\\
\alpha+\phi\hat\theta_i+\rho z_L^0, & \hat\theta_i\le \omega \text{ and } s^0,\\
\alpha+\phi\hat\theta_i+\rho z_L^1, & \hat\theta_i\le \omega \text{ and } s^1.
\end{cases}
\]
Given the construction, the corresponding posterior means are
\begin{align*}
    z_L^1 &= \frac{n-1}{1-\chi}(\alpha+\phi \mu_L),\\
    z_L^0 &=\frac{n-1}{1-\chi} \left[\alpha +\frac{\phi}{1-\zeta(1-\chi\nu)} \bigl[\mu-(1-\chi\nu)\zeta \mu_L\bigr] \right], \\
    z_H &=\frac{n-1}{1-\chi} \left[\alpha +\frac{\phi}{1-\zeta(1-\chi\nu)} \bigl[(1-\zeta(1-\chi))\mu-\chi(1-\nu)\zeta \mu_L\bigr] \right].
\end{align*}

Let $\bar z_L\coloneqq \zeta z_L^1+(1-\zeta)z_L^0$ be the expected opponent action after reporting low. Using the formulas above,
\[
z_H-\bar z_L
=
\frac{(n-1)\chi\phi\,\zeta}{(1-\chi)\bigl(1-\zeta(1-\chi\nu)\bigr)}(\mu-\mu_L)>0.
\]
For a true type $\theta_i \in \Theta_i$, let $G_\omega(\theta_i)$ be the gain from reporting low rather than high:
\[
G_\omega(\theta)
=
\zeta V(\theta,z_L^1)+(1-\zeta)V(\theta,z_L^0)-V(\theta,z_H).
\]
Therefore,
\[
G_\omega'(\theta)=V_{\theta A}(\bar z_L-z_H)\le 0.
\]
So the constructed outcome is incentive-compatible once the cutoff type is indifferent: $G_\omega(\omega)=0$.

To show that such a cutoff exists, substituting the values of $z_H, z_L^1,z_L^0$ yields
\[
G_\w(\w) = \frac{(\mu-\mu_L)\chi^2\phi^2\zeta}{2(1-\chi)^2\bigl(1-\zeta(1-\chi\nu)\bigr)^2} J(\w),
\]
where 
\[
J(\omega)
=
\bigl[1-\zeta\bigl(1+\chi^2(1-2\nu)\bigr)\bigr](\mu-\mu_L)
-
\frac{2(1-\chi)(1-\zeta(1-\chi\nu))}{\rho\phi}\bigl(\overline{V}_{A}+V_{\theta A}(\omega-\mu)\bigr).
\]
Since $\nu = 1- F(\omega)$ and $\mu_L(\omega)$ are continuous, so is $J(\omega)$. As $\w \downarrow \theta_L$, we have $\nu \to 1$, $\zeta \to 0$ and $\mu_L \to \theta_L$, so
\[
\lim_{\omega\downarrow\theta_L}J(\omega)
=
\mu-\theta_L - \frac{2(1-\chi)}{\rho\phi}(\overline{V}_{A} + V_{\theta A}(\theta_L -\mu))>0,
\]
where the sign follows from the extreme-type separation condition.

As $\w \uparrow \theta_H$, we have $\nu \to 0$, $\zeta \to 1-\varepsilon$ and $\mu_L \to \mu$. If $(V_{\theta A},\overline{V}_{A})\neq (0,0)$, then
\[
\lim_{\omega\uparrow\theta_H}J(\omega)
=
-\frac{2\varepsilon(1-\chi)}{\rho\phi}\bigl(\overline{V}_{A}+V_{\theta A}(\theta_H-\mu)\bigr)<0.
\]
Hence continuity implies that $J(\omega)=0$, equivalently $G_\omega(\omega)=0$, for some $\omega\in(\theta_L,\theta_H)$.

It remains to consider the knife-edge case $V_{\theta A}=\overline{V}_{A}=0$. In that case,
\[
J(\omega)=\bigl[1-\zeta\bigl(1+\chi^2(1-2\nu)\bigr)\bigr](\mu-\mu_L).
\]
Choose $\varepsilon<\chi^2/(1+\chi^2)$. Then, as $\omega\uparrow\theta_H$, we have $\nu\to 0$ and $\zeta\to 1-\varepsilon$, so the bracket converges to $1-(1-\varepsilon)(1+\chi^2)<0$. Thus $J(\omega)<0$ for $\omega$ close enough to $\theta_H$, while $\lim_{\omega\downarrow\theta_L}J(\omega)=\mu-\theta_L>0$. Continuity again yields a cutoff with $J(\omega)=0$.

Finally, consider the case $\overline{V}_{A}<0$. Apply \cref{lem:joint-reflection}. In the reflected game, $\check\theta_i=-\theta_i$ and $\check a_i=-a_i$, the payoff parameters are $(-\alpha,\phi,-\beta,\psi,\rho,\check g)$, and
\[
\check V_{\theta A}=V_{\theta A},
\qquad
\check{\overline V}_A=-\overline{V}_{A}>0.
\]
The lower endpoint of the reflected type space is $\check\theta_L=-\theta_H$, and the separation condition there becomes
\[
-\mu-(-\theta_H)>
\frac{\check{\overline V}_A}{\frac{\rho\phi}{2(1-\rho(n-1))}+\check V_{\theta A}} \Longleftrightarrow
\theta_H-\mu>
\frac{|\overline{V}_{A}|}{\frac{\rho\phi}{2(1-\rho(n-1))}+V_{\theta A}}.
\]
This is exactly the original extreme-type separation condition when $\theta_{ext}=\theta_H$. By the case already proved, the reflected game admits an informative symmetric CE if and only if this inequality holds. \cref{lem:joint-reflection} then transfers the result back to the original game.
\end{proof}

\section{Welfare Implications}

\begin{proof}[Proof of \cref{Prop: welfare CE}]
Fix an arbitrary CE outcome $\pi$, a player $i$, and type $\theta_i$. 
Since $\pi$ is a CE outcome, truthful reporting weakly dominates a randomized report $\hat\theta_i$ drawn from the prior. Recall that $U_i^\pi(\theta_i) = U_i^\pi(\theta_i;\theta_i)$ denotes type $\theta_i$'s interim expected payoff net of $g(\theta_i)$ under truthful report. By the information--manipulation decomposition in \cref{prop:IC-characterization},
\[
U_i^\pi(\theta_i)
\geq
\E_{\hat\theta_i\sim p}\!\left[U_i^\pi(\theta_i;\hat\theta_i)\right]
=
\frac12\E_{\hat\theta_i\sim p}\!\left[\sigma_i^2(\hat\theta_i)\right]
+
\E_{\hat\theta_i\sim p}\!\left[V\bigl(\theta_i,\bar A_{-i}(\hat\theta_i)\bigr)\right].
\]
By consistency and obedience (\cref{lem:OB-moments}), $\E_{\hat\theta_i\sim p}[\bar A_{-i}(\hat\theta_i)]=(n-1)\bar a$. Since $V(\theta_i,\cdot)$ is strictly convex, Jensen's inequality implies
\[
\E_{\hat\theta_i\sim p}\!\left[V\bigl(\theta_i,\bar A_{-i}(\hat\theta_i)\bigr)\right]
\geq
V\bigl(\theta_i,(n-1)\bar a\bigr)
=
U_i^{\pi^{NI}}(\theta_i).
\]
This inequality is strict unless $\bar A_{-i}(\hat\theta_i)$ is constant $p$-almost surely.
Therefore,
\[
U_i^\pi(\theta_i)-U_i^{\pi^{NI}}(\theta_i)
\geq
\frac12\E_{\pi}\!\left[\sigma_i^2(\theta_i)\right]
\geq 0.
\]
Since $i$ and $\theta_i$ were arbitrary, every type of every player weakly prefers $\pi$ to the no-information benchmark. Moreover, the gain is strict for every type of player $i$ if either $\E_\pi[\sigma_i^2(\theta_i)] > 0$ or $\bar A_{-i}(\hat\theta_i)$ is not constant $p$-almost surely. 

If neither condition holds, then almost every type obtains exactly the no-information payoff. Assume to the contrary that there is a type that is strictly better off. Other types would misreport as this type, which violates the incentive compatibility.
\end{proof}

As in the necessity proof of \cref{Thm: aligned}, write 
\[
x = 
\frac{\Cov_\pi(a_i,\theta_i)}{\Var_\theta}-\phi
=
\frac{\rho\,\Cov_\pi(\theta_i,\bar A_{-i}(\theta_i))}{\Var_\theta}.
\] 
By \cref{prop:boundSD_OB}, $x$ has the same sign as $\phi$, and $x=0$ iff $\pi=\pi^{NI}$.

\begin{lemma}\label{lem: NI benchmark}
Let $\pi$ be a symmetric BCE outcome. Then, for every $i\in I$,
\begin{align}
\E_\pi\!\left[
\bigl(a_i-a_i^{NI}(\theta_i)\bigr)^2
\right]
&=
\E_\pi[\sigma_i^2(\theta_i)]
+\rho^2\Var_\pi(\bar A_{-i}(\theta_i))
\leq
x\Var_\theta
\left[
\frac{\phi}{1-\rho(n-1)}
-\left(\frac{1}{\rho(n-1)}-1\right)x
\right],
\label{eq: NI action bound}\\
\E_\pi[u_i]-\E_{\pi^{NI}}[u_i]
&=
\frac12\E_\pi[\sigma_i^2(\theta_i)]
+V_{\theta A}\Cov_\pi(\theta_i,\bar A_{-i}(\theta_i))
+\frac12\rho^2\Var_\pi(\bar A_{-i}(\theta_i)).\label{eq:BCE-welfare-decomp}
\end{align}
\end{lemma}
\begin{proof}
Conditional on $\theta_i$, $a_i^{NI}(\theta_i)$ is deterministic, so 
\[
\E_\pi\!\left[\bigl(a_i-a_i^{NI}(\theta_i)\bigr)^2 \mid \theta_i \right] = \sigma_i^2(\theta_i) + (\E_\pi[a_i\mid\theta_i]-a_i^{NI}(\theta_i))^2.
\]
By \eqref{eq: OB}, \eqref{eq: NI}, and \cref{lem:OB-moments}, we have 
$\E_\pi[a_i\mid\theta_i]-a_i^{NI}(\theta_i)
=\rho\bigl(\bar A_{-i}(\theta_i)-\E_\pi[\bar A_{-i}(\theta_i)]\bigr)$ because $\E_\pi[\bar A_{-i}(\theta_i)]=\bar A^{NI}=(n-1)\bar a$. 
Taking expectations over $\theta_i$ gives the first equality in \eqref{eq: NI action bound}.

By the law of total variance, \eqref{eq: OB} and \cref{lem:OB-moments}, we have
\[
    \Var_\pi(a_i)
    =
    \E_\pi[\sigma^2_i(\theta_i)]
    +\Var_\pi\bigl(\alpha+\phi\theta_i+\rho \bar A_{-i}(\theta_i)\bigr)
    =
    \E_\pi[\sigma_i^2(\theta_i)]
    +\phi^2\Var_\theta+2\phi x\Var_\theta+\rho^2\Var_\pi(\bar A_{-i}(\theta_i)).
\]
Substituting this expression into the upper variance bound in
\cref{prop:boundSD_OB}:
\[
    \Var_\pi(a_i)
    \le
    \phi^2\Var_\theta
    + 2\phi x\Var_\theta
    +\frac{\phi}{1-\rho(n-1)}x\Var_\theta
    -\Bigl(\frac{1}{\rho(n-1)}-1\Bigr)x^2\Var_\theta,
\]
yields the inequality in \eqref{eq: NI action bound}.

For \eqref{eq:BCE-welfare-decomp}, following the same steps as in the proof of \cref{Prop: welfare CE}, for any type $\theta_i$, the difference in interim payoff between $\pi$ and the non-informative outcome is:
    \begin{align*}
        U_i^\pi(\theta_i;\theta_i) - U^{\pi^{NI}}_i(\theta_i) =&\; \frac{1}{2}\sigma^2_i(\theta_i) + V(\o_i,\bar A_{-i}(\o_i)) - V(\o_i,\bar A^{NI}) \\
        =&\; \frac{1}{2}\sigma^2_i(\theta_i) +V_A(\theta_i)(\bar A_{-i}(\theta_i) - \bar A^{NI}) + \frac{1}{2} \rho^2 (\bar A_{-i}(\theta_i) - \bar A^{NI}) ^2,
    \end{align*}
    where $V_A(\theta_i) = \overline{V}_{A} + V_{\theta A} (\theta_i - \mu)$ is the $A_{-i}$-partial derivative of $V$ evaluated at $(\theta_i, \bar A^{NI})$. The second equality here holds because $V$ is quadratic. Taking expectations and using $\E_\pi[\bar A_{-i}(\theta_i)]=\bar A^{NI}$ gives 
\eqref{eq:BCE-welfare-decomp}.
\end{proof}

\section{Proofs for \cref{Sec: comparison}}

\subsection{The Costs of Truth-telling}
\begin{lemma} \label{Lem: FI opt}
    If preferences have aligned differences, then $\pi^{FI}$ uniquely maximizes expected aggregate player payoff among all BCE outcomes.
\end{lemma}
\begin{proof}
    By \cref{lem:action-sign-flip}, it is without loss to assume $\phi > 0$. Since preferences have aligned differences, $V_{\theta A} \geq 0$. Applying the same symmetrization argument as before, it suffices to consider symmetric BCE outcomes. Fix a symmetric BCE outcome $\pi$ and let $x = \Cov_\pi(\theta_i, a_i) / \Var_\theta - \phi$. By \cref{prop:boundSD_OB}, 
    \[
    0 \leq x \leq \frac{\rho^2 (n-1) \phi}{( 1 + \rho )(1- \rho (n-1))},
    \]
    where the upper bound is attained if and only if $\pi = \pi^{FI}$. 

    By \cref{lem: NI benchmark} and using $\Cov_\pi(\theta_i, \bar A_{-i}(\theta_i)) = x\Var_\theta/\rho$ (by \cref{lem:OB-moments}), we have 
    \begin{equation} \label{eq: welfare gain in x}
        \E_\pi[u_i] - \E_{\pi^{NI}}[u_i] \leq x \Var_\theta \left[ \frac{\phi}{2(1-\rho(n-1))} + \frac{V_{\theta A}}{\rho}  - \frac{1}{2}\left( \frac{1}{\rho(n-1)} - 1 \right)x \right].
    \end{equation}
    The right-hand side is strictly increasing over the feasible range for $x$: Its derivative is minimized at the upper endpoint, which equals 
    \[
    \Var_\theta \frac{V_{\theta A}}{\rho} + \Var_\theta \phi \left[\frac{1}{2(1-\rho(n-1))} - \frac{\rho}{1+\rho} \right] > 0.
    \]
    The strict inequality here follows from $V_{\theta A} \geq 0$ and $\frac{1}{2(1-\rho(n-1))} > 1/2 > \rho/(1+\rho)$.

    Recall that the welfare upper bound in \eqref{eq: welfare gain in x} is derived from the upper variance bound in \cref{prop:boundSD_OB}, which is attained at the maximum feasible $x$. Since the bound in \eqref{eq: welfare gain in x} is strictly increasing in $x$,  $\pi^{FI}$ uniquely maximizes the aggregate player payoff among symmetric BCE outcomes. 
\end{proof}

\begin{proof}[Proof of \cref{Prop: CE BCE EA}]
    We first show that BCE and EA attain the same maximum aggregate payoff. It suffices to show that the maximum aggregate payoff under BCE can be attained under EA. It is without loss to consider a symmetric BCE $\pi$ that maximizes the aggregate player payoff among BCE outcomes. Since $\pi^{NI}$ is a BCE outcome, optimality and symmetry of $\pi$ imply that, under $\pi$, every player obtains at least the no-information payoff. This outcome $\pi$ can be implemented under EA as follows: the mediator commits to implement $\pi$ when all players participate, and otherwise switches to $\pi^{NI}$. 

    Next, we show the equivalence between (i) and (ii). If $\pi^{NI}$ maximizes aggregate player payoff among BCE outcomes, then CE also attains the common maximum since $\pi^{NI}$ is a CE outcome. Conversely, assume to the contrary that a CE outcome $\pi \neq \pi^{NI}$ attains the BCE maximum. Since $\pi$ is informative, \cref{Thm: misaligned} implies that preferences have aligned differences. By \cref{Lem: FI opt}, $\pi^{FI}$ is then the unique welfare-maximizing BCE outcome, so $\pi = \pi^{FI}$. However, $\pi^{FI}$ is not a CE outcome: Under \eqref{eq: FI}, $\sigma_i^2(\theta_i)$ is constant in $\theta_i$, whereas 
    \[
    \bar A_{-i}^{FI}(\theta_i) = (n-1)\bar a + \frac{\rho\phi (n-1)}{(1+\rho)(1-\rho(n-1))} (\theta_i - \mu)
    \]
    is strictly monotone in $\theta_i$. Since $V(\theta_i, \cdot)$ is strictly convex, with $V_{AA} = \rho^2 > 0$, any interior type $\theta_i \in (\theta_L, \theta_H)$ would find reporting either $\theta_L$ or $\theta_H$ strictly more profitable than reporting truthfully. Hence, \eqref{eq: IC simplified} fails for those interior types, contradicting the fact that $\pi = \pi^{FI}$ is a CE outcome. By \cref{lem: CE cpt}, the set of CE outcomes is compact. Therefore, the maximum aggregate player payoff can be attained over CE outcomes. 
    The equivalence between (i) and (ii) then implies a strict welfare loss whenever (ii) fails.

    Finally, the equivalence between (ii) and (iii) follows from \cref{lem: NI welfare}.
\end{proof}

\begin{lemma}\label{lem: NI welfare}
The no-information benchmark $\pi^{NI}$ uniquely maximizes aggregate player payoff among BCE outcomes if and only if preferences have strongly misaligned differences.
\end{lemma}
\begin{proof}
Suppose preferences have strongly misaligned differences. By \cref{lem:action-sign-flip}, it is without loss to assume $\phi>0$. Fix a symmetric BCE outcome $\pi$ and let $x = \Cov_\pi(\theta_i, a_i) / \Var_\theta - \phi$.
By \cref{prop:boundSD_OB}, $x\geq0$, with equality if and only if $\pi=\pi^{NI}$.
Under the stated condition, the linear coefficient in the right-hand side of \eqref{eq: welfare gain in x} is nonpositive, while the quadratic term is strictly negative whenever $x>0$. Hence every symmetric BCE outcome other than $\pi^{NI}$ yields strictly lower aggregate player payoff. If an arbitrary BCE maximizer symmetrized to $\pi^{NI}$, the original outcome is itself $\pi^{NI}$. Thus $\pi^{NI}$ is the unique BCE maximizer.

For the converse, suppose instead that $V_{\theta A}
+
\frac{\rho\phi}{2(1-\rho(n-1))}
>0.$
We construct symmetric BCE outcomes that yield strictly higher aggregate player payoff than $\pi^{NI}$.

\noindent\underline{\emph{Construction.}} Let $Y_i=\theta_i-\mu$ and $\chi = \rho (n-1)$. For sufficiently small $\varepsilon>0$, let
all players observe a public signal $S\in\{-1,1\}$ with
\[
\Pr(S=s\mid\theta)
=
\frac12
\left(
1+s\varepsilon\sum_{i\in I}Y_i
\right),
\qquad s\in\{-1,1\}.
\]
This is a valid information structure because the type space is bounded.
Since the $Y_i$'s are independent and have mean zero,
$\Pr(S=s)=1/2$ and
$\Pr(S=s\mid Y_i=y_i)=\frac12(1+s\varepsilon y_i)$.
Hence, by Bayes' rule, the conditional law of $Y_i$ given $S=s$ has Radon-Nikodym derivative $1+ s\e Y_i$ with respect to the prior of $Y_i$. 
Therefore, for any integrable function $f$,
\begin{align} \label{eq: local-signal-bayes}
    \E[f(Y_i)\mid S=s]
=
\E_p[f(Y_i)(1+s\varepsilon Y_i)],
\end{align}
where the expectation is taken with respect to the prior of $Y_i$ unless otherwise stated. 

Similarly, for $j\neq i$,
$\Pr(S=s\mid Y_i=y_i,Y_j=y_j)
=
\frac12(1+s\varepsilon(y_i+y_j))$. Therefore, conditional on $Y_i = y_i$ and $S= s$, the law of $Y_j$ has Radon-Nikodym derivative 
\[
\frac{1+s\e (y_i+Y_j)}{1+s\e y_i}
\]
with respect to the prior law of $Y_j$.
It follows that
\begin{align}
\E[f(Y_j)\mid Y_i=y_i,S=s]
&=
\E[f(Y_j)]
+
\frac{s\varepsilon}{1+s\varepsilon y_i}
\E[f(Y_j)Y_j]
\nonumber\\
&=
\E[f(Y_j)\mid S=s]
-
\frac{\varepsilon^2y_i}{1+s\varepsilon y_i}
\E[f(Y_j)Y_j],
\label{eq: local-signal-bayes-1}
\end{align}
where the second equality follows from \eqref{eq: local-signal-bayes}.

For each $s\in\{-1,1\}$, define
\[
C_s^\varepsilon
=
\frac{\phi \Var_\theta}{
1+\chi\varepsilon^2
\E\left[
\frac{Y_i^2}{1+s\varepsilon Y_i}
\right]
},
\]
and consider the symmetric continuation strategy
\begin{equation}
a_s^\varepsilon(y_i)
=
a_i^{NI}(\mu+y_i)
+
s\varepsilon
\frac{\chi\phi \Var_\theta}
{1-\chi}
-
\chi\varepsilon^2
\frac{y_i}{1+s\varepsilon y_i}
C_s^\varepsilon.
\label{eq: local-BCE-construction}
\end{equation}

We first verify that this is a continuation BNE with the given public information structure. Since $a_i^{NI}(\mu+Y_i)=\bar a+\phi Y_i$ and $\E[Y_i]=0$, the definition of $a_s^\e$ yields
\begin{align*}
\E[a_s^\varepsilon(Y_i)\mid S=s]
&=
\E\left[
a_s^\varepsilon(Y_i)(1+s\varepsilon Y_i)
\right]
=
\bar a
+s\varepsilon\phi \Var_\theta
+s\varepsilon
\frac{\chi\phi \Var_\theta}
{1-\chi}
=
\bar a
+
s\varepsilon
\frac{\phi \Var_\theta}
{1-\chi},
\end{align*}
where the first equation follows from \eqref{eq: local-signal-bayes} and the second follows from the definition of $a_s^\e$.

Applying the other conditional expectation identity \eqref{eq: local-signal-bayes-1} gives 
\begin{align*}
\E[a_s^\varepsilon(Y_j)\mid Y_i=y_i,S=s]
&=
\E[a_s^\varepsilon(Y_j)\mid S=s]
-
\frac{\varepsilon^2 y_i}{1+s\varepsilon y_i}
\E[a_s^\varepsilon(Y_j)Y_j]
\\
&=
\bar a
+
s\varepsilon
\frac{\phi \Var_\theta}
{1-\rho(n-1)}
-
\frac{\varepsilon^2 y_i}{1+s\varepsilon y_i}
C_s^\varepsilon,
\end{align*}
where the second equation follows from the definition of $a_s^\e$ and $C_s^\e$ because 
\begin{align*}
\E[a_s^\varepsilon(Y_i)Y_i]
&=
\phi \Var_\theta
-
\chi\varepsilon^2 C_s^\varepsilon
\E\left[
\frac{Y_i^2}{1+s\varepsilon Y_i}
\right]=
C_s^\varepsilon.
\end{align*}

Since $a_i^{NI}(\mu + y_i) = \alpha + \phi(\mu+y_i) + \chi \bar a$, the best response given $Y_i = y_i$ and $S = s$ is thus
\begin{align*}
\alpha+\phi(\mu+y_i)
+\chi\E[a_s^\varepsilon(Y_j)\mid Y_i=y_i,S=s]
=
a_i^{NI}(\mu+y_i)
+s\varepsilon\frac{\chi\phi \Var_\theta}{1-\chi}
-\chi\varepsilon^2
\frac{y_i}{1+s\varepsilon y_i}C_s^\varepsilon
=
a_s^\varepsilon(y_i).
\end{align*}
Thus \eqref{eq: local-BCE-construction} is a symmetric continuation
BNE. Let $\pi^\varepsilon$ denote the induced BCE outcome.

\noindent\underline{\emph{Welfare.}} We now evaluate the welfare gain relative to $\pi^{NI}$ as $\varepsilon\downarrow0$. Since the type space is bounded, $C_s^\e \to \phi \Var_\theta$ for each $s\in\{-1,1\}$. 

\eqref{eq: local-BCE-construction} therefore implies, uniformly
in $y_i$,
\[
\frac{
a_s^\varepsilon(y_i)-a_i^{NI}(\mu+y_i)
}{\varepsilon} = s\frac{\chi\phi \Var_\theta}{1-\chi}
-\chi\varepsilon
\frac{y_i}{1+s\varepsilon y_i}C_s^\varepsilon
\longrightarrow
s\frac{\chi\phi \Var_\theta}
{1-\chi}.
\]
The second term converges uniformly to zero because the type space is bounded and $C_s^\e \to \phi \Var_\theta$. Squaring and taking expectation yields
\begin{equation}
\frac{1}{\varepsilon^2}
\E_{\pi^\varepsilon}
\left[
\bigl(a_i-a_i^{NI}(\theta_i)\bigr)^2
\right]
\longrightarrow
\frac{
\chi^2\phi^2\Var_\theta^2
}{
(1-\chi)^2
}.
\label{eq: local-BCE-square}
\end{equation}

Since $\Pr(S=s) = 1/2$, by iterated expectation and \eqref{eq: local-signal-bayes},
\begin{align*}
\Cov_{\pi^\varepsilon}(\theta_i,a_i)
&=
\frac12\sum_{s\in\{-1,1\}}
\E\left[
Y_i a_s^\varepsilon(Y_i)(1+s\varepsilon Y_i)
\right]
\\
&=
\frac12\sum_{s\in\{-1,1\}}
\E\left[
Y_i(\bar a+\phi Y_i)(1+s\varepsilon Y_i)
\right]
\\
&\quad
+
\frac12\sum_{s\in\{-1,1\}}
s\varepsilon\frac{\chi\phi \Var_\theta}{1-\chi}
\E\left[Y_i(1+s\varepsilon Y_i)\right]
-
\frac{\chi\varepsilon^2\Var_\theta}{2}
\sum_{s\in\{-1,1\}}C_s^\varepsilon
\\
&=
\phi \Var_\theta
+
\varepsilon^2\frac{\chi\phi \Var_\theta^2}{1-\chi}
-
\frac{\chi\varepsilon^2\Var_\theta}{2}
\sum_{s\in\{-1,1\}}C_s^\varepsilon.
\end{align*}
The last equality uses $\E[Y_i]=0$,
$\E[Y_i^2]=\Var_\theta$, and
$\E[Y_i(1+s\varepsilon Y_i)]=s\varepsilon \Var_\theta$; in the first
term, the terms proportional to $s$ cancel when summed over
$s\in\{-1,1\}$.
It follows that
\begin{align*}
\frac{1}{\varepsilon^2}
\left(
\frac{\Cov_{\pi^\varepsilon}(\theta_i,a_i)}{\Var_\theta}
-\phi
\right)
=
\frac{\chi\phi \Var_\theta}{1-\chi}
-\frac{\chi}{2}
\sum_{s\in\{-1,1\}}C_s^\varepsilon
\longrightarrow
\frac{\chi^2\phi \Var_\theta}{1-\chi}.
\end{align*}
By \cref{lem:OB-moments}, this implies
\[
\frac{
\Cov_{\pi^\varepsilon}(\theta_i,\bar A_{-i}(\theta_i))
}{
\varepsilon^2
}
\longrightarrow
\frac{\chi(n-1)\phi \Var_\theta^2}{1-\chi}.
\]

Finally, combining this limit with
\eqref{eq: local-BCE-square} in \cref{lem: NI benchmark}, and
substituting back $\chi=\rho(n-1)$, gives
\[
\frac{
\E_{\pi^\varepsilon}[u_i]
-
\E_{\pi^{NI}}[u_i]
}{
\varepsilon^2
}
\longrightarrow
\frac{
\rho(n-1)^2\phi \Var_\theta^2
}{
1-\rho(n-1)
}
\left[
V_{\theta A}
+
\frac{\rho\phi}{2(1-\rho(n-1))}
\right]
>0.
\]
Thus every sufficiently small $\varepsilon>0$ yields a BCE outcome
that strictly improves on $\pi^{NI}$, so no information is not
BCE-optimal.
\end{proof}

\begin{lemma}\label{lem: CE cpt}
    A consistent outcome $\pi \in \D(\Theta\times \mathcal{A})$ is a CE outcome if and only if, for every player $i\in I$ and every bounded continuous $\lambda: \Theta_i \times \mathcal{A}_i \to \R$, 
    \begin{align}\label{eq: OB-int}
        \int_{\Theta\times \mathcal{A}}  \lambda(\theta_i, a_i)[a_i - \alpha - \phi \theta_i - \rho A_{-i}] \de \pi(\theta, a) = 0  \tag{OB-int}
    \end{align}
    and for every nonnegative continuous function $\xi: \Theta_i^2 \to \R_+$,
    \begin{align}\label{eq: TT-int}
        \int_{\Theta_i^2} \xi(\theta_i, \hat\theta_i) 
    [ \int u_i(\theta_i,a) \de \pi_i^{\theta_i}(a) 
    - \int u_i(\theta_i,a_i+\phi(\theta_i-\hat\theta_i), a_{-i}) \de \pi_i^{\hat\theta_i}(a)] 
    \de p(\theta_i) \de p(\hat\theta_i) \geq 0. \tag{TT-int}
    \end{align}
    The integral conditions depend only on $\pi$, independently of the choice of regular conditional distributions. Therefore, the set of CE outcomes is closed under the weak topology and hence is compact.
\end{lemma}
\begin{proof}
    The necessity follows from the definition and linear-quadratic structure. For sufficiency, suppose \eqref{eq: OB-int} and \eqref{eq: TT-int} hold. Applying similar arguments as in \cref{lem:bounded-actions-BCE}, action recommendations are in $B = [a^L, a^H]^n$ almost surely. We can therefore fix a version of the regular conditional distribution $\pi(\cdot\mid \theta)$ supported on $B$ for every $\theta$. We modify it into a version that satisfies \eqref{eq: OB} and \eqref{eq: IC simplified} for every type and report. 

    Fix $i\in I$, consider the pushforward of $p$ under the map $\theta_i \mapsto (\theta_i, \pi_i^{\theta_i}) \in \Theta_i \times \Delta(B)$. We equip  $\Delta(B)$ with the topology of weak convergence. Let $\Gamma_i$ denote the support of this pushforward, which does not rely on the choice of conditional distribution. Since $\Theta_i$ and $B$ are compact, $\Gamma_i$ is a closed subset of a compact set, which is compact. By definition, $(\theta_i, \pi_i^{\theta_i}) \in \Gamma_i$ for $p$-almost all $\theta_i$. Moreover, for every $\theta_i \in \Theta_i$, there exists at least one $Q\in \Delta(B)$ such that $(\theta_i, Q) \in \Gamma_i$ since $B$ is compact and $\Gamma_i$ is closed.

    Next, \eqref{eq: OB-int} implies that every $(\theta_i, Q)\in \Gamma_i$ satisfies \eqref{eq: OB}: for every continuous $h: [a^L, a^H] \to \R$, 
    \[
    \int_B h(a_i) (a_i - \alpha - \phi\theta_i - \rho A_{-i})\de Q(a) = 0.
    \]
    This holds because the integral is continuous in $(\theta_i, Q)$ and  \eqref{eq: OB-int} implies that this equation holds for $(\theta_i, \pi_i^{\theta_i})$ $p$-almost surely. Similarly,  \eqref{eq: TT-int} implies that every pair $(\theta_i, Q), (\theta_i', Q') \in \Gamma_i$ satisfies \eqref{eq: IC simplified}: 
    \[
    \int_B u_i(\theta_i, a) \de Q(a) - \int_B u_i(\theta_i, a_i + \phi(\theta_i - \theta_i'), a_{-i}) \de Q'(a) \geq 0. 
    \] 
    The difference in integrals is continuous in $(\theta_i, Q, \theta_i', Q')$. If the inequality fails strictly, then there are open neighborhoods of $(\theta_i, Q)$ and $(\theta_i', Q')$ such that truth-telling fails at every pair of types drawn from those neighborhoods with their corresponding conditional distribution, which violates \eqref{eq: TT-int}.

    We now construct a version of the conditional distribution that ensures incentive compatibility everywhere. Let $E_i \coloneqq \{\theta_i \in \Theta_i : (\theta_i, \pi_i^{\theta_i}) \notin \Gamma_i\}$, which has $p$-measure zero. Since $\Gamma_i$ is compact and has a nonempty section at every type, by Theorem 18.13 of \cite{aliprantis06}, we can measurably select $\tilde\pi_i^{\theta_i}$ such that $(\theta_i, \tilde\pi_i^{\theta_i})\in \Gamma_i$ for every $\theta_i$, while retaining $\tilde\pi_i^{\theta_i} = \pi_i^{\theta_i}$ outside $E_i$.
    
    Given a profile $\theta$ where some $\theta_j\in E_j$, we modify the conditional kernel as follows: let $i = \min\{j: \theta_j \in E_j\}$ and set $\tilde\pi(\cdot \mid \theta) = \tilde\pi_i^{\theta_i}$. Otherwise, we retain the original kernel $\pi(\cdot \mid \theta)$. This only modifies the kernel at type profiles with measure zero, so $\tilde \pi(\cdot\mid \theta)$ is another version of the regular conditional distribution of outcome $\pi$. Moreover, for every type $\theta_i$, their opponent's type lies outside the set that requires modification $P_{-i}$-almost surely, so the corresponding distribution of recommendations after reporting $\theta_i$ is exactly $ \tilde\pi_i^{\theta_i}$. By the properties of $\Gamma_i$ established above, obedience and truth-telling are satisfied for every type and report. Hence $\pi$ is a CE outcome.

    Finally, given this representation of obedience and truth-telling, those constraints are linear and weakly continuous in the outcome distribution, so the set of CE outcomes is weakly closed. Since the types and actions are bounded (\cref{lem:bounded-actions-BCE}), the set of CE outcomes is compact.
\end{proof}

\subsection{Public Cheap talk}
Next, we formalize the public cheap talk game. For each player $i$, fix a compact metrizable space $M_i$ that is rich enough, and let $M \coloneqq \prod_{i\in I} M_i$. After observing $\theta_i$, player $i$ sends a costless public message $m_i \in M_i$. The profile $m \in M$ is observed by every player, after which the players choose actions simultaneously.

\begin{lemma}
\label{lem:ct_binary}
In any public cheap-talk equilibrium, the posterior mean induced by a player's message takes at most two values almost surely. If player $i$ uses two posterior means $\m_L<\m_H$, then the gain from inducing $\m_H$ rather than $\m_L$ is
\begin{equation} \label{eq: CT payoff diff}
    \Lambda^{CT}(\m_H-\m_L)
\left[
\overline{V}_{A}+V_{\theta A}(\theta_i-\mu)
+\frac{\rho^2\Lambda^{CT}}{2}(\m_H+\m_L-2\mu)
\right],
\end{equation}
where $\Lambda^{CT} \coloneqq \frac{\phi\rho(n-1)}{(1+\rho)(1-\rho(n-1))} \neq 0$.
This payoff difference is affine in $\theta_i$, with slope having the sign of $\phi V_{\theta A}$.
\end{lemma}
\begin{proof}
    Fix an arbitrary PBE of the public cheap talk game. 
    
    \emph{Continuation play.} Fix an on-path public message profile $m$. All expectations in this
paragraph are conditional on $m$, under the posterior and continuation
strategies induced by the equilibrium. Since types are independent and
messages are sent separately, player $i$'s belief about player $j$'s
continuation action conditional on $(\theta_i,m)$ is $\E[a_j\mid m]$. Abusing the notation, we use $\bar A_{-i}(m)=\sum_{j\neq i}\E[a_j\mid m]$ to denote $i$'s expected aggregate opponent action given $m$, sequential rationality and strict
concavity imply that the equilibrium continuation action is degenerate at
\[
a_i(\theta_i,m)=\alpha+\phi\theta_i+\rho \bar A_{-i}(m)
\]
for almost every type. Taking expectations conditional on the full public message profile $m$
gives
\[
\E[a_i\mid m]=\alpha+\phi \E[\theta_i\mid m]+\rho \bar A_{-i}(m).
\]
Solving this linear system across players implies that
\begin{equation}\label{eq:CT-continuation}
\bar A_{-i}(m)=\bar A^{NI}+\frac{\phi}{(1+\rho)(1-\rho(n-1))}\sum_{j\neq i}\bigl(\E[\theta_j\mid m]-\mu\bigr)+\frac{\phi\rho(n-1)}{(1+\rho)(1-\rho(n-1))}\bigl(\E[\theta_i\mid m]-\mu\bigr),
\end{equation}

\emph{Message payoffs.} Since types are independent and the players' messages only depend on their types, $\E[\theta_j \mid m] = \E[\theta_j \mid m_j]$ for every on-path profile $m$ and every $j \in I$. Fix a player $i$ and an on-path message $m_i$. Conditional on $(\theta_i, m_i)$, taking expectation over opponent messages and types, $i$'s expected aggregate opponent action is
\[
\bar A_{-i}(m_i) = \bar A^{NI} + \Lambda^{CT}\bigl(\E[\theta_i\mid m_i]-\mu\bigr)
\]
since the opponents' posterior mean averages back to the prior mean. Applying the same decomposition argument as in \cref{sec:IC}, type $\theta_i$'s expected payoff from sending $m_i$ is
\[
\frac{\rho^2}{2}\Var_{m_{-i}}(\bar A_{-i}(m_i,m_{-i})) + V(\theta_i, \bar A_{-i}(m_i)) + g(\theta_i),
\]
where the first term does not depend on $\theta_i$ by \eqref{eq:CT-continuation} and the independence of types. Since $V$ is quadratic, a similar decomposition as in the proof of \cref{lem: NI benchmark} implies that this expected payoff can be rewritten as
\begin{align*}
     (\overline{V}_{A} + V_{\theta A}(\theta_i - \mu))&\Lambda^{CT}\bigl(\E[\theta_i\mid m_i]-\mu\bigr) + \frac{1}{2}\rho^2 (\Lambda^{CT})^2\bigl(\E[\theta_i\mid m_i]-\mu\bigr)^2\\
     +&\frac{\rho^2}{2}\Var_{m_{-i}}(\bar A_{-i}(m_i,m_{-i})) + V(\theta_i, \bar A^{NI}) + g(\theta_i),
\end{align*}
which is strictly convex in the posterior mean $\E[\theta_i\mid m_i]$ since $\rho^2 (\Lambda^{CT})^2 > 0$. 

\emph{At most two on-path posterior means.} Let $S_i$ be the support of the distribution of the on-path posterior means $\E[\theta_i\mid m_i]$ for player $i$. Since $S_i$ is closed and $S_i\subseteq [\theta_L, \theta_H]$, it is compact. Therefore, $\max S_i$ and $\min S_i$ can be attained or approximated by on-path messages. Since player $i$'s expected payoff is strictly convex in the induced posterior mean, any posterior mean $s \in (\min S_i, \max S_i)$ is strictly dominated by either $\min S_i$ or $\max S_i$ (or on-path approximations of those endpoints). Therefore, $S_i \subseteq \{\min S_i, \max S_i\}$, so there are at most two posterior means. Suppose $i$ induces two posteriors $\mu_L < \mu_H$ on-path; direct calculation leads to \eqref{eq: CT payoff diff}.
\end{proof}

\begin{proof}[Proof of \cref{prop: public_cheap_talk}]
    Part (i) follows from \cref{Thm: misaligned} since any public cheap-talk equilibrium outcome is also a CE outcome. We focus on part (ii). Suppose throughout that $\phi V_{\theta A} \geq 0$.

    \emph{Necessity.} Suppose there exists an informative public cheap-talk outcome. There must be $i \in I$ that induces two distinct posterior means on-path, denoted as $\mu_L < \mu_H$. 

    First, suppose $V_{\theta A} \neq  0$. Since $\phi \neq 0$ and $\phi V_{\theta A} \geq 0$, in this case $\phi V_{\theta A} > 0$. By \eqref{eq: CT payoff diff}, the gain from the high-posterior message is strictly increasing in $\theta_i$. In this case, there exists a cutoff $\w \in (\theta_L, \theta_H)$ such that $\mu_L = \nu_L(\w)\coloneqq  \E[\theta_i \mid \theta_i \leq \w]$ and $\mu_H = \nu_H(\w)\coloneqq \E[\theta_i \mid \theta_i > \w]$, and the cutoff type is exactly indifferent between those two posteriors by \eqref{eq: CT payoff diff}
    \[
    J(\w) \coloneqq \phi \left( \overline{V}_{A}+V_{\theta A}(\w-\mu) +\frac{\rho^2\Lambda^{CT}}{2}(\nu_H(\w)+\nu_L(\w)-2\mu)\right) = 0.
    \]
    Since the conditional means $\nu_L(\w)$ and $\nu_H(\w)$ are continuous and strictly increasing on $(\theta_L, \theta_H)$, so is $J(\w)$. Moreover, $\phi V_{\theta A} > 0$ and $\phi \Lambda^{CT} > 0$, so $J$ is strictly increasing. The limits at the endpoints are
    \[
    \lim_{\omega\downarrow\theta_L}J(\omega) = \phi\overline{V}_{A} + \phi (V_{\theta A} + \frac{\rho^2 \Lambda^{CT}}{2})(\theta_L - \mu),\quad \lim_{\omega\uparrow\theta_H}J(\omega) = \phi\overline{V}_{A} + \phi (V_{\theta A} + \frac{\rho^2 \Lambda^{CT}}{2})(\theta_H - \mu).
    \]
    An interior solution therefore requires the first limit to be negative and the second to be positive. If $\phi \overline{V}_{A} \geq 0$, only the first inequality binds and it is equivalent to $(\mu - \theta_L) (|V_{\theta A}| + \frac{1}{2}\rho^2 |\Lambda^{CT}|) > |\overline{V}_{A}|$ after normalizing the signs. If $\phi \overline{V}_{A} < 0$, only the second inequality binds and it is equivalent to $(\theta_H - \mu) (|V_{\theta A}| + \frac{1}{2}\rho^2 |\Lambda^{CT}|) > |\overline{V}_{A}|$. This is exactly the cheap-talk separation condition.

    Now suppose $V_{\theta A} = 0$, all types are indifferent between the two posterior means, and \eqref{eq: CT payoff diff} implies that 
    \[
    \phi \left( \overline{V}_{A} +\frac{\rho^2\Lambda^{CT}}{2}(\m_H+\m_L-2\mu)\right) = 0.
    \]
    Note that $\theta_L - \mu < \m_H+\m_L-2\mu < \theta_H - \mu$ by Bayes plausibility and the fact that the two posteriors are distinct. The preceding indifference equation consequently implies the same strict endpoint inequalities, hence the cheap-talk separation condition.
    
    \emph{Sufficiency.} Suppose the cheap-talk separation condition holds. The preceding arguments imply that $J$ has a unique zero $\w^* \in (\theta_L, \theta_H)$. We construct an informative public cheap-talk equilibrium as follows: For every $i\in I$, choose two distinct messages $L_i, H_i\in M_i$, and $i$ sends $L_i$ when $\theta_i \leq \w^*$ and $H_i$ when $\theta_i > \w^*$. The on-path beliefs are consistent with Bayes' rule, and the off-path messages for $i$ are interpreted as $L_i$. At every public history, the players take the unique best response characterized by \eqref{eq:CT-continuation}. By construction, type $\w^*$ is indifferent. If $V_{\theta A} \neq 0$, the monotonicity of the payoff difference \eqref{eq: CT payoff diff} ensures that no player type would strictly prefer another message. If $V_{\theta A} = 0$, every type is indifferent between $L_i$ and $H_i$ so no one strictly prefers another message. Therefore, this construction yields an informative public cheap-talk equilibrium.
\end{proof}

\subsection{Restricted Access}
\begin{proof}[Proof of \cref{Prop: comm size}]
    Fix a subset $J\subseteq I$ of participants with $|J|=k<n$ and a CE outcome $\pi\in \D(\Theta \times \mathcal{A})$ with $k$ participants. 

    Note that $\pi$ is also a BCE outcome in the $n$-player communication game, so \cref{lem:OB-moments} implies that $\E_{\pi}[a_i] = \bar a$ for every $i\in I$, regardless of whether $i$ is a participant or not.

    For a nonparticipant $\ell \in I \setminus J$, since $\ell$ only observes the realized private type $\theta_\ell$, which is independent of others' types, $\ell$'s conditional expectation over others' actions is precisely the unconditional one. Thus, type $\theta_\ell$ will take action $a_\ell(\theta_\ell) = \alpha + \phi\theta_\ell + \rho(n-1)\bar a$, which is exactly the strategy in the no-information benchmark $\pi^{NI}$.

    For the participants, their reports do not affect the non-participants' actions, and the action recommendations they observe are not informative about non-participants. For a participant $j \in J$ with type $\theta_j \in \Theta_j$, report $\hat{\theta}_j \in \Theta_j$, and measurable $\delta_j : \mathcal{A}_j \to \mathcal{A}_j$, the corresponding interim payoff is 
    \begin{align*}
        U_j^{\pi}(\theta_j; \hat{\theta}_j,\delta_j) =& \,\E_{\pi_j^{\hat{\theta}_j}}[(\alpha + \phi \theta_j - \tfrac{1}{2}\delta_j(a_j))\delta_j(a_j) + \sum_{s\neq j}(\beta + \psi \theta_j + \rho \delta_j(a_j)) a_s ] + g(\theta_j) \\
        =& \,\E_{\pi_j^{\hat{\theta}_j}}[(\tilde{\alpha} + \phi \theta_j - \tfrac{1}{2}\delta_j(a_j))\delta_j(a_j) + \sum_{s\in J \setminus\{j\}}(\beta + \psi \theta_j + \rho \delta_j(a_j)) a_s ] + \tilde{g}(\theta_j)
    \end{align*}
    where $\tilde\alpha = \alpha + \rho(n-k) \bar a$ and $\tilde{g}(\theta_j) = g(\theta_j) + (\beta + \psi\theta_j) (n-k)\bar a$. This is exactly the interim payoff of player $j$ in a $k$-player communication game with parameters $(\tilde\alpha,\phi,\beta,\psi,\rho,\tilde g)$.

    Therefore, the marginal of $\pi$ over the participants' types and actions $(\theta_j, a_j)_{j\in J}$ is a CE outcome of this reduced $k$-player game. Conversely, given any CE outcome of this $k$-player game, one can construct a CE outcome of the original game by letting each nonparticipant $\ell\notin J$ play $a_\ell(\theta_\ell)=\alpha+\phi\theta_\ell+\rho(n-1)\bar a$.

    Given this one-to-one correspondence, there exists an informative CE outcome with $k$ participants if and only if there exists an informative CE outcome in the reduced $k$-player game. Note that $\phi$ and $V_{\theta A} = \rho\phi + \psi$ are common across those two games, and in the $k$-player game
    \[
    \rho\frac{\tilde \alpha + \phi\mu}{1-\rho (k-1)} + \beta + \psi \mu = \rho\frac{ \alpha + \phi\mu}{1-\rho (n-1)} + \beta + \psi \mu = \overline{V}_{A},
    \]
    so $\overline{V}_{A}$ is also unchanged. Therefore, the only thing that changed in the extreme-type separation condition is the total number of players $k$, and the corresponding cutoff is decreasing in $k$.
\end{proof}

\section{Proofs for \cref{sec:general_LQ}}\label{sec:proof_general_LQ}
\begin{proof}[Proof of \cref{prop: second-order_effect}]
As in the proof of \cref{Thm: misaligned,Thm: aligned}, it suffices to prove the result for symmetric CE. Fix a symmetric obedient outcome $\pi\in \D(\Theta\times \mathcal{A})$ and a player $i \in I$. We suppress $i$ and $\pi$ when there is no confusion. Recall that $\bar A^{NI} = (n-1)\bar a$ and $\Delta(\theta_i) = \bar A(\theta_i) - \bar A^{NI}$. Define
\[
Q(\hat\theta_i)\coloneqq
\E_{\pi_i^{\hat\theta_i}}\bigl[(A_{-i}-\bar A^{NI})^2\bigr].
\]

The second-order term does not involve $a_i$, so the obedience constraint \eqref{eq: OB} is unchanged. The information--manipulation decomposition in \cref{prop:IC-characterization} still applies to the baseline part of the payoff. In particular, the optimized payoff of a true type $\theta_i$ from reporting $\hat\theta_i$, net of $g(\theta_i)$, is
\[
\frac12\sigma^2(\hat\theta_i)
+V\bigl(\theta_i,\bar A^{NI}+\Delta(\hat\theta_i)\bigr)
+\kappa\E_{\pi_i^{\hat\theta_i}}[A_{-i}^2].
\]
Since $V(\theta_i,\cdot)$ is quadratic,
\[
V\bigl(\theta_i,\bar A^{NI}+\Delta(\hat\theta_i)\bigr)
=
V(\theta_i,\bar A^{NI})
+\bigl[\overline{V}_{A}+V_{\theta A}(\theta_i-\mu)\bigr]\Delta(\hat\theta_i)
+\frac12\rho^2\Delta(\hat\theta_i)^2.
\]
Moreover, by the definition of $Q$,
\[
\E_{\pi_i^{\hat\theta_i}}[A_{-i}^2]
=
Q(\hat\theta_i)
+2\bar A^{NI}\Delta(\hat\theta_i)
+(\bar A^{NI})^2.
\]
Dropping the terms independent of the report and using
$\overline{V}_{A}^\kappa=\overline{V}_{A}+2\kappa \bar A^{NI}$, we obtain
\begin{equation}\label{eq:second-order-report-payoff}
\frac12\sigma^2(\hat\theta_i)
+\frac12\rho^2\Delta(\hat\theta_i)^2
+\kappa Q(\hat\theta_i)
+\bigl[\overline{V}_{A}^\kappa
+V_{\theta A}(\theta_i-\mu)\bigr]\Delta(\hat\theta_i).
\end{equation}

Adding the incentive constraints of two types $\theta_i<\theta_i'$ against each other's reports gives
\begin{equation}\label{eq:second-order-monotonicity}
V_{\theta A}(\theta_i'-\theta_i)
\bigl[\Delta(\theta_i')-\Delta(\theta_i)\bigr]\ge0.
\end{equation}

\noindent\textbf{Part (i).}
By \cref{lem:action-sign-flip}, we may assume $\phi>0$. If preferences have misaligned differences, then $V_{\theta A}<0$, so \eqref{eq:second-order-monotonicity} implies that $\Delta$ is weakly decreasing. Hence
$\Cov(\theta_i,\Delta(\theta_i))\le0$. On the other hand, by obedience, \cref{lem:OB-moments,prop:boundSD_OB} imply
$\rho\Cov\bigl(\theta_i,\Delta(\theta_i)\bigr)
=\Cov(a_i,\theta_i)-\phi \Var_\theta\ge0$,
with equality only at $\pi^{NI}$. Thus the outcome is uninformative. Conversely, suppose preferences have aligned differences. As in the proof of \cref{Thm: misaligned}, a uniform prior over $[-L,L]$ for large enough $L > 0$ satisfies \eqref{eq: cutoff kappa}, and part (ii) below provides an informative CE outcome. This proves part (i).

\noindent\textbf{Part (ii): sufficiency.}
The symmetry lemmas \cref{lem:action-sign-flip,lem:joint-reflection} used in the proof of \cref{Thm: aligned} also preserve $\kappa A_{-i}^2$. So it is without loss to assume $\phi > 0, V_{\theta A} \geq 0$ and $\overline{V}_{A}^\kappa\geq 0$, and thereby the relevant endpoint is $\theta_L$. Suppose the modified extreme-type separation condition \eqref{eq: cutoff kappa} holds. We construct an informative symmetric CE as follows.

Fix $\varepsilon\in(0,1)$; in the knife-edge case $V_{\theta A}=\overline{V}_{A}^\kappa=0$, we will later choose $\varepsilon$ small enough. For a cutoff $\omega\in(\theta_L,\theta_H)$, let
\[
\chi\coloneqq \rho(n-1),\qquad
\nu\coloneqq 1-F(\omega),
\qquad
\mu_L\coloneqq \E[\theta_i\mid \theta_i\le \omega],
\qquad
\zeta\coloneqq (1-\varepsilon)F(\omega)^{n-1}.
\]

Use the threshold construction from the sufficiency proof of \cref{Thm: aligned}. The mediator uses only whether each report is below or above $\omega$. If all reports are low, the mediator sends $s^1$ with probability $1-\varepsilon$ and $s^0$ otherwise; if at least one report is high, it sends $s^0$. The mediator then recommends the obedient action.
Let $z_H,z_L^0,z_L^1$ be the posterior mean total opponent actions after a high report, a low report followed by $s^0$, and a low report followed by $s^1$, respectively, and put
\[
d_\omega\coloneqq z_L^0-z_L^1,
\qquad
\bar a_L^0\coloneqq\alpha+\phi\mu_L+\rho z_L^0.
\]
Write $\Delta_L,\Delta_H$ and $\sigma_L^2,\sigma_H^2$ for the values induced by a low and a high report, respectively. The Bayes-consistency calculation in the proof of \cref{Thm: aligned} gives
\begin{align}
 d_\omega
 &=\frac{(n-1)\phi(\mu-\mu_L)}
 {(1-\chi)\bigl[1-\zeta(1-\chi\nu)\bigr]},
 \label{eq:second-order-d}\\
 \Delta_H&=\chi F(\omega)\zeta d_\omega,
 &\Delta_L&=-\chi\nu\zeta d_\omega,
 \notag\\
 \Delta_L-\Delta_H&=-\chi\zeta d_\omega,
 \label{eq:second-order-deltas}\\
 \sigma_H^2&=0,
 &\sigma_L^2&=\rho^2\zeta(1-\zeta)d_\omega^2,
 \notag\\
 \bar a_L^0-\bar a
 &=\frac{d_\omega}{n-1}
 \bigl[\zeta(1-\chi\nu)-(1-\chi)\bigr].
 \label{eq:second-order-moments}
\end{align}
Write $Q_L,Q_H$ for the corresponding values of $Q$. A low report differs from a high report only when every opponent reports low and the mediator sends $s^1$. On that event, every opponent's action falls by $\rho d_\omega$, so total opponent action falls by $\chi d_\omega$. Therefore,
\begin{equation}\label{eq:second-order-Q-difference}
Q_L-Q_H
=-\zeta\chi d_\omega
\bigl[2(n-1)(\bar a_L^0-\bar a)-\chi d_\omega\bigr].
\end{equation}

Let $G_\omega(\theta_i)$ be the gain of type $\theta_i$ from reporting low rather than high. Substituting \eqref{eq:second-order-deltas}--\eqref{eq:second-order-Q-difference} into \eqref{eq:second-order-report-payoff} yields
\begin{equation}\label{eq:second-order-gain}
\frac{G_\omega(\theta_i)}{\zeta\chi d_\omega}
=
\frac{\rho^2d_\omega}{2\chi}
\bigl[1-\zeta+\chi^2\zeta(2\nu-1)\bigr]
-\bigl[\overline{V}_{A}^\kappa+V_{\theta A}(\theta_i-\mu)\bigr]
-\kappa\bigl[2(n-1)(\bar a_L^0-\bar a)-\chi d_\omega\bigr].
\end{equation}
This expression is weakly decreasing in $\theta_i$. Hence the threshold strategy is report incentive compatible whenever $G_\omega(\omega)=0$.

As $\omega\downarrow\theta_L$, we have $\nu\to1$, $\zeta \to 0$, $\mu_L \to \theta_L$, $ d_\omega\to\frac{(n-1)\phi }{1-\chi}(\mu - \theta_L)$, and $\bar a_L^0-\bar a\to-\phi (\mu - \theta_L)$. Hence, \eqref{eq:second-order-gain} implies
\[
\lim_{\omega\downarrow\theta_L}
\frac{G_\omega(\omega)}{\zeta\chi d_\omega}
=
(\mu - \theta_L)\left[
V_{\theta A}
+\frac{\rho\phi}{2(1-\chi)}
+\kappa(n-1)\phi\frac{2-\chi}{1-\chi}
\right]
-\overline{V}_{A}^\kappa>0,
\]
where the sign follows from the modified extreme-type separation condition \eqref{eq: cutoff kappa}. 

As $\omega\uparrow\theta_H$,
$d_\omega\to0$, $\bar a_L^0-\bar a\to0$, and $\zeta\to1-\varepsilon$. If
$(\overline{V}_{A}^\kappa,V_{\theta A})\ne(0,0)$, then
\[
\lim_{\omega\uparrow\theta_H}
\frac{G_\omega(\omega)}{\zeta\chi d_\omega}
=-\bigl[\overline{V}_{A}^\kappa+V_{\theta A}(\theta_H-\mu)\bigr]<0.
\]
If $\overline{V}_{A}^\kappa=V_{\theta A}=0$, then
\[
\lim_{\omega\uparrow\theta_H}
\frac{G_\omega(\omega)}{\zeta\chi d_\omega^2}
=
\frac{\rho^2}{2\chi}
\bigl[\varepsilon-\chi^2(1-\varepsilon)\bigr]
-\kappa(\chi-2\varepsilon),
\]
which is strictly negative at $\varepsilon=0$, and hence for every sufficiently small $\varepsilon>0$. Continuity gives an interior cutoff $\omega^*$ such that $G_{\omega^*}(\omega^*)=0$. At this cutoff $\omega^*$, the corresponding outcome is obedient by construction, incentive compatible by the monotonicity of $G_{\omega^*}$, and is informative because $\zeta,d_{\omega^*}>0$.

\noindent\textbf{Part (ii): necessity when $n=2$.}
Continue with the preceding normalization, so $\chi=\rho$. Define
\[
c\coloneqq
(\mu - \theta_L)\left[
\frac{\rho\phi}{2(1-\rho)}
+\frac{\kappa\phi(2-\rho)}{1-\rho}
\right]
\]
and
\begin{equation}\label{eq:second-order-Gamma}
\Gamma(\theta_i)
\coloneqq
\frac12\sigma^2(\theta_i)
+\frac12\rho^2\Delta(\theta_i)^2
+\kappa Q(\theta_i)
+c\Delta(\theta_i).
\end{equation}

\smallskip
\noindent\underline{\emph{Obedience identities.}}
For the two players $i\ne j$, let
\[
r_i\coloneqq\E[a_j-\bar a\mid a_i,\theta_i],
\qquad
r_j\coloneqq\E[a_i-\bar a\mid a_j,\theta_j].
\]
Rewriting $\eqref{eq: OB}$ using $r_i, r_j$, since $\bar a = (\alpha + \phi\mu)/(1-\rho)$,
\begin{equation}\label{eq:second-order-centered-obedience}
a_i-\bar a=\phi(\theta_i-\mu)+\rho r_i.
\end{equation}
By iterated expectations, $\Delta(\theta_i) =\E[r_i\mid\theta_i].$ Moreover, conditional on $\theta_i$, \eqref{eq:second-order-centered-obedience} implies $\sigma^2(\theta_i)
=\rho^2\Var(r_i\mid\theta_i)$,
and hence
\begin{equation}\label{eq:second-order-conditional-second-moment}
\sigma^2(\theta_i)+\rho^2\Delta(\theta_i)^2
=\rho^2\E[r_i^2\mid\theta_i].
\end{equation}

Taking conditional expectations of player $j$'s obedience constraint \eqref{eq:second-order-centered-obedience}
with respect to $(a_i,\theta_i)$ gives
\begin{equation}\label{eq:second-order-cross-obedience}
\phi\E[\theta_j-\mu\mid a_i,\theta_i]
=r_i-\rho\E[r_j\mid a_i,\theta_i].
\end{equation}
Multiplying \eqref{eq:second-order-cross-obedience} by
$\theta_i-\mu$ and taking expectations, independence of the types implies
\[
\E[(\theta_i-\mu)r_i]
=\rho\E[(\theta_i-\mu)r_j].
\]
On the other hand, multiplying \eqref{eq:second-order-centered-obedience}
by $r_j$ and taking expectations gives
\[
\phi\E[(\theta_i-\mu)r_j]
=
\E[(a_i-\bar a)r_j]-\rho\E[r_ir_j].
\]
Since $r_j=\E[a_i-\bar a\mid a_j,\theta_j]$, $\E[(a_i-\bar a)r_j]=\E[r_j^2]=\E[r_i^2],$
where the last equality follows from symmetry. Therefore,
\begin{equation}\label{eq:second-order-first-moment-r}
\E[(\theta_i-\mu)r_i]
=
\frac{\rho}{\phi}
\left(\E[r_i^2]-\rho\E[r_ir_j]\right) = \frac{\rho}{\phi} \left( (1-\rho)\E[r_i^2]
+\frac{\rho}{2}\E[(r_i-r_j)^2] \right),
\end{equation}
where the second equality follows from symmetry.

\smallskip
\noindent\underline{\emph{A covariance bound from obedience.}}
Since $n=2$, $Q(\theta_i)=\E[(a_j-\bar a)^2\mid\theta_i]$.
Substituting \eqref{eq:second-order-conditional-second-moment} into
\eqref{eq:second-order-Gamma} therefore gives
\begin{align}
\Cov\bigl(\theta_i,\Gamma(\theta_i)\bigr)
={}&
\frac{\rho^2}{2}
\E[(\theta_i-\mu)r_i^2]
+\kappa
\E[(\theta_i-\mu)(a_j-\bar a)^2]
+c\E[(\theta_i-\mu)r_i].
\label{eq:second-order-Gamma-covariance}
\end{align}

We first rewrite the second term. Expanding player $j$'s  obedience constraint \eqref{eq:second-order-centered-obedience} and using independence of types implies
\begin{align*}
\E[(\theta_i-\mu)(a_j-\bar a)^2]
={}&
2\rho\phi
\E[(\theta_i-\mu)(\theta_j-\mu)r_j]
+\rho^2\E[(\theta_i-\mu)r_j^2].
\end{align*}
To rewrite the first expectation, multiply
\eqref{eq:second-order-cross-obedience} by
$(\theta_i-\mu)r_i$ and take expectations. This gives
\[
\phi\E[(\theta_i-\mu)(\theta_j-\mu)r_i]
=
\E[(\theta_i-\mu)r_i^2]
-\rho\E[(\theta_i-\mu)r_ir_j].
\]
By symmetry, the left-hand side is unchanged if $r_i$ is replaced by
$r_j$. Hence
\begin{align}
\E[(\theta_i-\mu)(a_j-\bar a)^2]
={}&
2\rho\E[(\theta_i-\mu)r_i^2]
-2\rho^2\E[(\theta_i-\mu)r_ir_j]+\rho^2\E[(\theta_i-\mu)r_j^2] \notag \\
={}&\rho(2-\rho)\E[(\theta_i-\mu)r_i^2]+\rho^2\E[(\theta_i-\mu)(r_i-r_j)^2] \notag\\
={}&\rho(2-\rho)\E[(\theta_i-\mu)r_i^2]
+\frac{\rho^2}{2}
\E\bigl[(\theta_i+\theta_j-2\mu)(r_i-r_j)^2\bigr], \label{eq:second-order-Q-covariance}
\end{align}
where the last equality follows from symmetry.

Substituting \eqref{eq:second-order-Q-covariance} and
\eqref{eq:second-order-first-moment-r} into
\eqref{eq:second-order-Gamma-covariance} gives
\begin{align}
\Cov\bigl(\theta_i,\Gamma(\theta_i)\bigr)
={}&
\left[
\frac{\rho^2}{2}
+2\kappa\rho\left(1-\frac\rho2\right)
\right]
\E\bigl[(\theta_i-\theta_L)r_i^2\bigr]
\notag\\
&+\frac{\rho^3 (\mu - \theta_L)}{4(1-\rho)}
\E\bigl[(r_i-r_j)^2\bigr]
\notag\\
&+\frac{\kappa\rho^2}{2}
\E\left[
\left(
\theta_i+\theta_j-2\theta_L
+\frac{\rho (\mu-\theta_L)}{1-\rho}
\right)(r_i-r_j)^2
\right]
\ge0.
\label{eq:second-order-obedience-bound}
\end{align}
Because the prior is atomless, $\theta_i>\theta_L$ almost surely.
Equality in \eqref{eq:second-order-obedience-bound} therefore forces
$r_i=0$ almost surely, and thereby
$\pi=\pi^{NI}$ by \cref{prop:boundSD_OB}.

\smallskip
\noindent\underline{\emph{Impossibility.}}
Suppose that the modified extreme-type separation condition \eqref{eq: cutoff kappa} fails. Under the normalization
above,
\begin{equation}\label{eq:second-order-cutoff-failure}
\overline{V}_{A}^\kappa\ge (\mu - \theta_L)V_{\theta A}+c.
\end{equation}
If $V_{\theta A}>0$, \eqref{eq:second-order-monotonicity} makes
$\Delta$ weakly increasing. For $\theta_i<\hat\theta_i$, the incentive
constraint of type $\theta_i$ against report $\hat\theta_i$, together with
\eqref{eq:second-order-report-payoff}, implies
\[
\Gamma(\hat\theta_i)-\Gamma(\theta_i)
\le
\bigl[
c-\overline{V}_{A}^\kappa-V_{\theta A}(\theta_i-\mu)
\bigr]
\bigl[\Delta(\hat\theta_i)-\Delta(\theta_i)\bigr] \leq 0,
\]
where the last inequality follows from $\theta_i \geq \theta_L$ and \eqref{eq:second-order-cutoff-failure}.
Therefore, $\Gamma$ is weakly decreasing and $\Cov(\theta_i,\Gamma(\theta_i))\le0$.

If $V_{\theta A}=0$, report incentive compatibility makes
\[
\frac12\sigma^2(\theta_i)
+\frac12\rho^2\Delta(\theta_i)^2
+\kappa Q(\theta_i)
+\overline{V}_{A}^\kappa\Delta(\theta_i)
\]
constant for almost every $\theta_i$. Hence, $\Gamma(\theta_i)$ is a constant plus
$(c-\overline{V}_{A}^\kappa)\Delta(\theta_i)$. By
\eqref{eq:second-order-cutoff-failure}, this coefficient is nonpositive,
while obedience implies $\Cov(\theta_i,\Delta(\theta_i))\ge0$. Therefore, $\Cov(\theta_i,\Gamma(\theta_i))\le0$.

Combining either conclusion with
\eqref{eq:second-order-obedience-bound} forces equality, so
$\pi=\pi^{NI}$. This proves necessity for $n=2$. 
\end{proof}

\begin{proof}[Proof of \cref{prop:beyond_lq}]
    By \cref{ass:linearBR}, the maximizer in
$v(\theta_i,A_{-i})=\max_{a_i}\{f(\theta_i,a_i)+h(\theta_i,a_i)A_{-i}\}$ is
$a^0(\theta_i)+\rho A_{-i}$. The envelope theorem and \cref{ass:value_learning} therefore imply $v_{A_{-i}}(\theta_i, A_{-i}) = h(\theta_i, a^0(\theta_i) + \rho A_{-i})$, hence
\begin{equation}\label{eq:beyond-lq-indirect}
v(\theta_i,A_{-i})
=f\bigl(\theta_i,a^0(\theta_i)\bigr)+h(\theta_i, a^0(\theta_i))A_{-i}+K(A_{-i}),
\end{equation}
where $K(A_{-i})
\coloneqq
\int_0^{A_{-i}} H(\rho t)\,\de t$
is common across types.

Fix a CE outcome $\pi \in \D(\Theta \times \mathcal{A})$. Retain the notations $\pi_i^{\hat{\theta}_i}$, $\bar A_{-i}(\hat{\theta}_i)$ and $\bar A_{-i}(a_i;\hat\theta_i)$ from \cref{sec:IC}. For every $i\in I$ and type $\theta_i \in \Theta_i$, by \cref{ass:linearBR}, the obedience constraint is of the form
\[
a_i=a^0(\theta_i)+\rho \bar A_{-i}(a_i;\theta_i)
\qquad
\pi_i^{\theta_i}\text{-almost surely}.
\]
A true type $\theta_i$ who reports $\hat\theta_i$ can therefore choose the optimal action
$a^0(\theta_i)+\rho \bar A_{-i}(a_i;\hat\theta_i)$ after observing the recommendation. Let $h^0(\theta_i) \coloneqq h(\theta_i, a^0(\theta_i))$.  Using \eqref{eq:beyond-lq-indirect} and the law of iterated expectations, the corresponding payoff is
\begin{equation}\label{eq:beyond-lq-report-payoff}
U_i^\pi(\theta_i;\hat\theta_i)
=f\bigl(\theta_i,a^0(\theta_i)\bigr)
+h^0(\theta_i)\bar A_{-i}(\hat\theta_i)
+I_i(\hat\theta_i), \qquad I_i(\hat\theta_i)
\coloneqq
\E_{\pi_i^{\hat\theta_i}}
\bigl[K\bigl(\bar A_{-i}(a_i;\hat\theta_i)\bigr)\bigr],
\end{equation}
where $I_i$ does not depend on the true type because $K$ does not.

Adding the incentive constraints $U_i^\pi(\theta_i; \theta_i) \geq U_i^\pi(\theta_i; \theta_i')$ and $U_i^\pi(\theta_i'; \theta_i') \geq U_i^\pi(\theta_i'; \theta_i)$ implies that
\[
\bigl[h^0(\theta_i)-h^0(\theta_i')\bigr]
\bigl[\bar A_{-i}(\theta_i)-\bar A_{-i}(\theta_i')\bigr]\ge0.
\]
for every pair of $(\theta_i, \theta_i')$. Under the hypothesis of \cref{prop:beyond_lq},
\[
\bigl[a^0(\theta_i)-a^0(\theta_i')\bigr]
\bigl[\bar A_{-i}(\theta_i)-\bar A_{-i}(\theta_i')\bigr]\le0.
\]
If $\theta_i'$ is an independent copy of $\theta_i$, it follows that
\begin{equation}\label{eq:beyond-lq-IC-covariance}
\Cov\bigl(a^0(\theta_i),\bar A_{-i}(\theta_i)\bigr)
=\frac12\E\left[
\bigl(a^0(\theta_i)-a^0(\theta_i')\bigr)
\bigl(\bar A_{-i}(\theta_i)-\bar A_{-i}(\theta_i')\bigr)
\right]
\le0.
\end{equation}

We first suppose that $\pi$ is symmetric. Let $X_i\coloneqq a^0(\theta_i)$. 
Because $a_i-X_i$ is measurable with respect to $(a_i,X_i)$, the obedience constraint can be written as 
\[
a_i=X_i+\rho\E[A_{-i}\mid a_i,X_i].
\]
The proof of \cref{prop:boundSD_OB} uses only independence, symmetry, finite second moments, and this linear obedience equation. Applying that argument with $X_i$ in place of $\theta_i$ and coefficient one yields
\[
\Cov(a_i,X_i)\ge\Var(X_i),
\]
with equality only if
\[
a_i=X_i+\rho(n-1)\bar a
\quad\text{almost surely},
\qquad
\bar a=\frac{\E[X_i]}{1-\rho(n-1)}.
\]
Moreover, by obedience
\[
\Cov(a_i,X_i)
=\Var(X_i)+\rho\Cov(A_{-i},X_i)
=\Var(X_i)+\rho\Cov\bigl(\bar A_{-i}(\theta_i),X_i\bigr).
\]
Consequently,
\begin{equation}\label{eq:beyond-lq-obedience-covariance}
\Cov\bigl(a^0(\theta_i),\bar A_{-i}(\theta_i)\bigr)\ge0,
\end{equation}
with equality only at the no-information outcome. Combining \eqref{eq:beyond-lq-IC-covariance} and \eqref{eq:beyond-lq-obedience-covariance} therefore gives $\pi=\pi^{NI}$.

Finally, let $\pi$ be an arbitrary CE outcome and let $\bar \pi$ denote its symmetrization over player permutations. By symmetry of the primitive environment, $\bar \pi$ is a symmetric CE outcome, and thereby $\bar \pi = \pi^{NI}$. Let  
$
G  = \left\{(\theta,a):
 a_i=a^0(\theta_i)+\rho(n-1)\bar a
 \text{ for every }i
\right\}.
$
Since $\bar\pi$ assigns probability one to this graph, every outcome entering the permutation average, including $\pi$, does as well. Consistency then implies $\pi=\pi^{NI}$.
\end{proof}

\section{Positive Spillovers and Strategic Complementarity} \label{app:investment}

In the private-demand case of \cref{eg: Bertrand}, stronger strategic complementarity makes the extreme-type separation condition more demanding. The following investment example shows that this need not hold when firms also benefit directly from others' investments.

\begin{example}[Investment with Positive Spillovers]\label{eg: Investment}
\emph{Firms choose investment levels $a_i\in\mathbb{R}$ in adoption of a new technology. Firm $i$'s private return to investment is $(1-q) \theta_i + \frac{q}{n-1}A_{-i}$, where $\theta_i\in\Theta_i\subseteq\mathbb{R}_+$ is firm $i$'s productivity and $q\in(0,\tfrac12)$ measures strategic complementarity in investment. Investing $a_i$ costs $\frac{1}{2}a_i^2$, and induces a direct spillover $\frac{b}{n-1}a_i$ in others' payoffs with $b > 0$. Thus, firm $i$'s payoff is
\[
u_i(\theta_i,a)
=
\left((1-q)\theta_i+\tfrac{q}{n-1}A_{-i}\right)a_i
-\tfrac{1}{2}a_i^2
+\tfrac{b}{n-1}A_{-i}.
\]
This is a special case of \eqref{eq:general_payoff} with $\alpha = 0$, $\phi = 1-q$, $\beta = \frac{b}{n-1}$, $\psi = 0$, $\rho = \frac{q}{n-1}$, and $g\equiv 0$. This is a variant of the investment application in \cite{angeletos2007efficient} with an extra positive welfare externality term, which is motivated by mechanisms such as aggregate demand spillovers in \cite{murphy1989industrialization} and knowledge externalities in \cite{romer1986increasing}.
}
\end{example}

\begin{proposition}
    In \cref{eg: Investment}, informative communication is possible if and only if 
    \[
    \mu - \theta_L > \frac{2(q\mu + b)}{q(3-2q)} \; \Longleftrightarrow \; b < \frac{1}{2}(\mu-3\theta_L) q -  (\mu-\theta_L) q^2.
    \]
    In particular, if $\mu > 3\theta_L$ and $0 < b < \frac{(\mu-3\theta_L)^2}{16(\mu-\theta_L)}$, informative communication is possible precisely for an intermediate interval of $q \in (0, 1/2)$. 
\end{proposition}

A larger direct spillover $b$ increases firms' incentives to manipulate other firms' investments without affecting their own best responses, which makes the extreme-type separation condition more demanding. 

The effect of stronger strategic complementarity $q$ is more subtle. When $q$ is small, information about others' investments has little value for a firm's own decision, so the informational benefit is insufficient to offset the manipulation incentive induced by the direct spillover. Increasing $q$ makes such information more valuable, which can help sustain truthful reporting. However, stronger complementarity also increases firms' gains from inducing higher investments by others. When $q$ becomes sufficiently large, the manipulation incentive dominates again. Therefore, increasing $q$ can initially enable informative communication but eventually prevent it.


\section{Pure Correlation}

Under pure correlation, the mediator does not receive any report from the players and draws a private signal profile according to some distribution independent of the realized types. Players observe their types and their private signals, and then choose actions. A \emph{pure-correlation outcome} is an outcome induced by a BNE of this game. 

\begin{proposition}\label{prop: pure corr}
    The only pure-correlation outcome is uninformative.
\end{proposition}



Formally, a pure-correlation device consists of compact metrizable signal spaces $(M_i)_{i\in I}$ that are rich enough and a distribution $\l \in \D(\prod_{i \in I}M_i)$ that is independent of the players' types. 
\begin{proof}[Proof of \cref{prop: pure corr}]
    Fix any pure-correlation device $\l \in \D(\prod_{i\in I}M_i)$ and any BNE strategy profile $a_i: \Theta_i \times M_i \to \R$, where we abuse the notation so $a_i$ denotes a measurable function rather than a random variable for every $i\in I$. Since the signals are independent of types and types are independent across players, the expectations are taken with respect to the product probability measure $P \times \l$. The equilibrium condition implies that
    \[
    a_i(\theta_i, m_i) = \alpha + \phi \theta_i + \rho \E[\sum_{j\neq i} a_j(\theta_j, m_j) \mid m_i]
    \]
    almost surely.
    Recall that the BNE strategy in the no-information benchmark is $a_i^{NI}(\theta_i) = \alpha + \phi \theta_i + \rho (n-1)\bar a$. Taking the difference, we have 
    \[
    a_i(\theta_i, m_i) - a_i^{NI}(\theta_i) = \rho\E[\sum_{j\neq i} a_j(\theta_j, m_j)\mid m_i] - \rho (n-1)\bar a = \rho\E[\sum_{j\neq i} (a_j(\theta_j, m_j) - a_j^{NI}(\theta_j))\mid m_i],
    \]
    where the last equality holds because $\bar a = \E[a_j^{NI}(\theta_j) \mid m_i]$ by independence of signals from types. Taking the absolute value and by iterated expectation, we have 
    \[
    \E[|a_i(\theta_i, m_i) - a_i^{NI}(\theta_i)|] \leq \rho \E[\sum_{j\neq i} |a_j(\theta_j, m_j) - a_j^{NI}(\theta_j)|].
    \]
    Let $B\coloneqq \max_{i\in I} \E[|a_i(\theta_i, m_i) - a_i^{NI}(\theta_i)|]$, the preceding equation implies that $B \leq \rho(n-1) B$, and thereby $B = 0$ because $\rho(n-1)\in (0,1)$. Therefore, $a_i(\theta_i,m_i) = a_i^{NI}(\theta_i)$ almost surely, so the induced BNE is $\pi^{NI}$.
\end{proof}

\section{Interim Welfare with Quadratic Externalities}\label{sec: non_lq_improvement}
In the baseline model, \cref{Prop: welfare CE} shows that, under any CE outcome, every type of every player obtains an interim expected payoff that is weakly higher than the no-information benchmark. This welfare result depends on how opponents' actions enter payoffs. We illustrate this point with the following example, where informative communication is possible, but every type is strictly worse off than with no communication.

\begin{example}[Negative Quadratic Externalities]
\emph{Two players have types $\theta_i, \theta_j$ independently and identically uniformly distributed on $[-1,1]$. The payoff of player $i$ equals
\[u_i(\theta_i,a_i,a_j) \;=\; \theta_i a_i + \tfrac{1}{2} a_i a_j - \tfrac{1}{2} a_i^2 - a_j^2,\]
and symmetrically for player $j$.
Agent $i$'s best reply given $j$'s action equals $a_i=\theta_i + \tfrac{1}{2}\,a_j$, implying that best responses are linear and actions are strategic complements. \\
In the no-communication benchmark, the agents' strategies are $a_i^{NI}(\theta_i) = \theta_i$ and payoffs are given by
\[U_i^{NI}(\theta_i) \;=\; \tfrac{1}{2}\theta_i^2 - \E[\theta_j^2] \;=\; \tfrac{1}{2}\theta_i^2 - \tfrac{1}{3}.\]
Let $s(\theta_i)$ equal to $1$ if $\theta_i\ge 0$ and $-1$ if $\theta_i<0$. Suppose that the mediator, after reports $(\hat\theta_i,\hat\theta_j)$, recommends
\[a_i \;=\; \hat\theta_i + \frac{s(\hat\theta_i) + 2 s(\hat\theta_j)}{6}.\]
This mechanism essentially reveals whether the opponent's type is above or below zero. Under truthful reporting,
\[
\mathbb E[a_j\mid\theta_i,a_i] = \mathbb E[a_j\mid\theta_i,s(\theta_j)]
=\frac23s(\theta_j)+\frac13s(\theta_i),
\]
so the recommended action is a best response and obedience holds. Truth-telling also holds: after reporting any type with sign $z \in \{-1,1\}$ and choosing action optimally, type $\theta_i$'s expected payoff is
\[
U_i(\theta_i; z) =
\frac12\mathbb E\left[
\left(\theta_i+\frac z6+\frac{s_j}3\right)^2
\right]-\mathbb E\left[
\left(\theta_j+\frac{s_j}6+\frac z3\right)^2
\right]= \frac12\theta_i^2+\frac16\theta_i z-\frac{41}{72},
\]
so reporting $z = s(\theta_i)$ is optimal. Therefore, the mechanism induces an informative CE outcome $\pi$.
For every $\theta_i \in [-1,1]$, the payoff difference between this outcome and the no-information benchmark is
\[
U_i^\pi(\theta_i)-U_i^{NI}(\theta_i)
=\frac{|\theta_i|}{6}-\frac{17}{72} <0.
\] 
In this example, although communication improves players' ability to respond to their opponents' actions, it also increases the dispersion of actions. The resulting negative quadratic externality dominates the information benefit for every type.}
\end{example}

\section{Consumer Welfare in the Bertrand Illustration}
\cref{Thm: aligned} shows that, in the Bertrand oligopoly of \cref{eg: Bertrand} with private demand, informative communication can be possible, while \cref{Prop: welfare CE} proves that it makes all firms in the oligopoly weakly better off. In this section, we study the consequences of informative communication on consumer welfare.

To evaluate consumer welfare, we use the standard quadratic representative-consumer formulation of linear differentiated-products demand \citep{singh1984price}.\footnote{Note that quasilinearity implies that preferences belong to the Gorman polar form with a common, constant marginal utility of income. Thus, the representative consumer's utility measures aggregate welfare. This also implies, however, that  all distributional questions among consumers are assumed away.} Write the demand system in vector form as
\begin{equation*}
  q(\theta,a) \;=\; b(\theta) - H a,
  \qquad\text{with}\quad
  b(\theta) := D\mathbf{1} + \theta,
  \quad
  H := \Bigl(\tfrac12 + r\Bigr) I_n - r J_n,
  \label{eq:demand}
\end{equation*}
whose $i$-th component is the demand for product $i$.\footnote{For $r > 0$, the assumption $r < 1/(2(n-1))$ is exactly positive definiteness of $H$.
}
Because $H$ is symmetric and positive definite, the demand above is generated by a representative consumer with quasilinear utility
\begin{equation}
  \mathcal{U}(q,x_0;\theta)
  \;=\; x_0 + b(\theta)^\top\, H^{-1} q - \tfrac12\, q^\top\, H^{-1} q ,\tag{R}
  \label{eq:utility}
\end{equation}
where $x_0$ is numeraire consumption. Maximizing the utility in \eqref{eq:utility} subject to
$x_0 + a^\top\, q \le y$, where $y$ is the consumer's income, gives the first-order condition $H^{-1}\bigl(b(\theta) - q\bigr) = a$, which is exactly the demand above. Consumer surplus at $(\theta,a)$ is the maximized value of \eqref{eq:utility} net of income,
\begin{equation}
  CS(\theta,a)
  \;=\; \tfrac12\, q(\theta,a)^\top\, H^{-1} q(\theta,a)
  \;=\; \tfrac12\, b(\theta)^\top\, H^{-1} b(\theta) - \tfrac{1}{2} \, b(\theta)^\top\, a - \tfrac12\, q(\theta,a)^\top\, a .\tag{CS}
  \label{eq:CS}
\end{equation}
For any outcome $\pi$, let
\[
  PS(\pi) \;:=\; \E_\pi\Bigl[\, \sum_{i=1}^{n} (a_i - C) q_i(\theta, a) \Bigr]
\]
denote expected industry profit, call $CS(\pi)$ the expected consumer surplus under the representative consumer, and define total surplus as $TS(\pi) \;:=\; PS(\pi) + CS(\pi)$.\footnote{We sum consumer surplus to profits since both are measured in numeraire units.}

\begin{proposition}\label{prop: consumer_welfare}
Compared to the no-information outcome, every informative BCE outcome, and hence every informative CE outcome, induces strictly higher expected producer surplus and total surplus, but strictly lower expected consumer surplus.
\end{proposition}

Although expected prices and quantities are unchanged across BCE outcomes, informative communication increases price dispersion and the covariance between prices and firm's private demand shocks. Hence, consumers face prices that are more responsive to demand shocks, leading to a lower consumer surplus. 


Full information provides a further benchmark. \cref{Lem: FI opt} shows that full information uniquely maximizes producer surplus in the aligned-differences case. It maximizes both price dispersion and the covariance between prices and demand shocks. Since expected consumer surplus decreases with both, full information also minimizes consumer surplus. Hence, for any informative CE outcome $\pi$,
\[
   PS(\pi^{NI}) \;<\; PS(\pi) \;<\; PS(\pi^{FI}),\quad\text{ while }\quad CS(\pi^{NI}) \;>\; CS(\pi) \;>\; CS(\pi^{FI}).\]

\begin{proof}[Proof of \cref{prop: consumer_welfare}]
Fix an informative BCE outcome $\pi$. Obedience in the private-demand model implies
\[
  a_i \;=\; D + \frac{C}{2} + \theta_i
        + r\, \E_\pi\Bigl[\, \sum_{j \neq i} a_j \;\Big|\; a_i, \theta_i \Bigr].
\]
Therefore,
\begin{align*}
  \E_\pi[q_i \mid a_i, \theta_i]
    &= D + \theta_i - \tfrac{1}{2} a_i
       + r\, \E_\pi\Bigl[\, \sum_{j\neq i} a_j \;\Big|\; a_i,\theta_i \Bigr] 
    = \tfrac{1}{2}\,(a_i - C).
\end{align*}
It follows that $\E_\pi\bigl[(a_i - C) q_i\bigr]
  \;=\; \tfrac{1}{2}\, \E_\pi\bigl[(a_i - C)^2\bigr]$.
Since $\E_{\pi}[a_i] = \bar a$, we have  
\[PS(\pi) - PS(\pi^{NI})
    = \frac{1}{2} \sum_{i=1}^{n} \bigl[\Var_\pi(a_i) - \Var_\theta\bigr].\]
For consumer surplus, recall that $CS(\theta,a) = \tfrac{1}{2}\, b(\theta)^\top\, H^{-1} b(\theta)
    - \tfrac{1}{2}b(\theta)^\top a - \tfrac{1}{2}\, q(\theta,a)^\top a$.
Taking expectations and using $\E_{\pi}[a_i q_i] = \frac{1}{2}\E_{\pi}[a_i(a_i -C)]$ by obedience yields
\[
CS(\pi) = \frac{1}{2}\E_\pi[b(\theta)^\top\, H^{-1} b(\theta)] - \frac{1}{2} \sum_{i=1}^n \E_\pi[(D+\theta_i)a_i] - \frac{1}{4}\sum_{i=1}^n \E_\pi[a_i(a_i -C)].
\]
Since the first term only depends on the prior and $\E_{\pi}[a_i] = \bar a$, changes in consumer surplus depend on price moments and on the covariance between prices and demand:
\[CS(\pi) - CS(\pi^{NI})
    = -\frac{1}{4} \sum_{i=1}^{n} \bigl[\Var_\pi(a_i) - \Var_\theta\bigr]
       - \frac{1}{2} \sum_{i=1}^{n} \bigl[\Cov_\pi(a_i,\theta_i) - \Var_\theta\bigr].\]

Adding the producer and consumer surplus changes then implies
\begin{align*}
  TS(\pi) - TS(\pi^{NI})
    &= \frac{1}{4} \sum_i \Bigl[ \Var_\pi(a_i) - \Var_\theta
        - 2\bigl(\Cov_\pi(a_i,\theta_i) - \Var_\theta\bigr) \Bigr] \\
    &= \frac{1}{4} \sum_i
        \E_\pi\Bigl[\bigl(a_i - a_i^{NI}(\theta_i)\bigr)^2\Bigr],
\end{align*}
where the last equality follows from \eqref{eq: NI} with $\phi =1$ and $\E_\pi[a_i] = \bar a$.

Finally, it suffices to consider symmetric BCE outcomes, since symmetrization preserves aggregate producer surplus, consumer surplus, and total surplus. \cref{prop:boundSD_OB} then implies $\Cov_\pi(a_i,\theta_i) > \Var_\theta$ and, by Cauchy--Schwarz,
\[
  \Var_\pi(a_i) \;\ge\; \frac{\Cov_\pi(a_i,\theta_i)^2}{\Var_\theta} > \Var_\theta ,
\]
Thus, the change in producer and total surplus is strictly positive, while that in consumer surplus is strictly negative.
\end{proof}

\section{Robustness to Rare Extreme Types}
Under aligned differences, \cref{Thm: aligned} shows that informative communication is impossible when the relevant extreme point is too close to the mean. Extending the support to sufficiently more extreme types can, however, make informative communication possible, even if those added types occur with arbitrarily small probability. We examine whether this sensitivity of existence also leads to substantial changes in equilibrium behavior or welfare.

Fix a prior $p$ under which the extreme-type separation condition fails strictly with relevant extreme type $\theta_{ext}$. Choose a more extreme type $\tilde\theta_{ext}$ such that the condition would hold strictly when evaluated at the original mean $\mu$ and $\overline{V}_{A}$. Let $q$ be any full-support atomless distribution on the interval between $\theta_{ext}$ and $\tilde \theta_{ext}$, and let $p^\varepsilon=(1-\varepsilon)p+\varepsilon q$. We use $a_i^{NI,\varepsilon}$ and $\pi^{NI,\varepsilon}$ to denote the no-information action rule and outcome under prior $p^\varepsilon$, respectively.


\begin{proposition}\label{prop: tail perturbation}
For every sufficiently small $\varepsilon>0$, an informative CE outcome
exists under $p^\varepsilon$. 

Moreover, there are constants
$C,\bar\varepsilon>0$ such that, for every
$\varepsilon\in(0,\bar\varepsilon)$ and every CE outcome $\pi$ under
$p^\varepsilon$, the average mean-square deviation of actions from
the no-information benchmark $\pi^{NI,\varepsilon}$ and the average welfare gain relative to
$\pi^{NI,\varepsilon}$ are both bounded above by $C\varepsilon$:
\[
\frac{1}{n}\sum_{i\in I}\E_{\pi} \left[
\bigl(a_i-a_i^{NI,\varepsilon}(\theta_i)\bigr)^2
\right] \leq C\e, \qquad \frac1n\sum_{i\in I}
\left(
\E_\pi[u_i]-\E_{\pi^{NI,\varepsilon}}[u_i]
\right) \leq C\e .
\]
\end{proposition}

Adding rare extreme types makes informative communication feasible, but the average mean-square action deviation and the average welfare gain relative to the corresponding no-information outcome vanish uniformly as those types become rare. 

\begin{proof}[Proof of \cref{prop: tail perturbation}]
By \cref{lem:action-sign-flip,lem:joint-reflection}, it suffices to
consider the case $\phi>0$ and $\overline{V}_{A}>0$. Since preferences have aligned differences, $V_{\theta A}\geq0$, and the relevant extreme type is
$\theta_L$, with $\tilde\theta_L<\theta_L$.

Let $\mu_\varepsilon$, $\Var_{\theta,\varepsilon}$, and
$\overline V_{A,\varepsilon}$ denote the corresponding objects under
$p^\varepsilon$. Under the maintained assumption,
\[
\mu-\tilde\theta_L>
\frac{\overline{V}_{A}}
{\frac{\rho\phi}{2(1-\rho(n-1))}+V_{\theta A}}.
\]
Since $\mu_\varepsilon\to\mu$ and
$\overline V_{A,\varepsilon}\to\overline{V}_{A}$, the strict
inequality holds under $p^\varepsilon$ for all sufficiently small
$\varepsilon>0$. Hence, \cref{Thm: aligned} implies the existence of
an informative CE outcome.

Next, we establish the quantitative bounds. Note that it suffices to consider symmetric CE outcomes, because symmetrizing any CE outcome by uniformly randomizing over permutations of players would not change the average mean-square deviation nor the average welfare gain relative to no information.

Fix a symmetric CE outcome $\pi$ under $p^\e$. All no-information benchmark
objects below are those associated with $p^\varepsilon$. As in the necessity proof of \cref{Thm: aligned}, let $\Delta(\theta_i) = \bar A_{-i}(\theta_i) - \bar A^{NI,\varepsilon}$ and $z = \esssup \Delta$, and recall that $x =
\Cov_\pi(a_i,\theta_i)/\Var_{\theta,\varepsilon}-\phi$.

We first bound the maximal increase $z$ in expected opponent actions.
Recall from the variance calculation in the necessity proof of
\cref{Thm: aligned} that
\[
\Var_\pi(a_i)
\geq
\phi^2\Var_{\theta,\varepsilon}
-\frac{2\psi}{\rho}x\Var_{\theta,\varepsilon}
+2\overline V_{A,\varepsilon}z+\rho^2z^2.
\]
Combining this with the upper variance bound 
\eqref{eq:UGamma-signrobust}, evaluated under $p^\varepsilon$, yields
\begin{align*}
2\overline V_{A,\varepsilon}z+\rho^2z^2
\leq{}&
\left(
\frac{2V_{\theta A}}{\rho}
+\frac{\phi}{1-\rho(n-1)}
\right)x\Var_{\theta,\varepsilon}-
\left(\frac{1}{\rho(n-1)}-1\right)x^2\Var_{\theta,\varepsilon}.
\end{align*}
Since $x,z\geq0$ and $\rho(n-1)<1$, dropping the nonnegative term
$\rho^2z^2$ on the left and the nonpositive quadratic term on the
right gives
\begin{equation}\label{eq: tail z bound}
z
\leq
\frac{x \Var_{\theta,\varepsilon}}{2 \overline{V}_{A,\e}} \left(
\frac{2V_{\theta A}}{\rho}
+\frac{\phi}{1-\rho(n-1)}
\right).
\end{equation}

Using $\E_\pi[\Delta(\theta_i)]=0$ and $p^\e = (1-\e)p + \e q$, we split the covariance at the original lower endpoint $\theta_L$,
\begin{align*}
\frac{x\Var_{\theta,\varepsilon}}{\rho}
&= \Cov(\Delta(\theta_i), \theta_i) = 
\E_\pi[\Delta(\theta_i)(\theta_i-\theta_L)]\\
&\leq
z(1-\varepsilon)(\mu-\theta_L)
+\varepsilon
\sqrt{
\E_q[\Delta(\theta_i)^2]\,
\E_q[(\theta_L-\theta_i)^2]
}.
\end{align*}
Here, the first term uses $\Delta(\theta_i)\leq z$ on the support of
$p$, while the second follows from Cauchy--Schwarz on the added tail $q$.
Moreover,
$\varepsilon\E_q[\Delta(\theta_i)^2]
\leq\E_{p^\varepsilon}[\Delta(\theta_i)^2]
=\Var_\pi(\bar A_{-i}(\theta_i))$. By \eqref{eq: NI action bound} in \cref{lem: NI benchmark}, 
\[
\rho^2\Var_\pi(\bar A_{-i}(\theta_i)) \leq \frac{\phi}{1-\rho(n-1)} x \Var_{\theta,\varepsilon}, 
\]
where we drop the nonpositive quadratic terms in that bound. It follows that 
\begin{align*}
    \frac{x\Var_{\theta,\varepsilon}}{\rho}
\leq 
z(1-\varepsilon)(\mu-\theta_L)
+ \frac{1}{\rho}
\sqrt{
\frac{\phi}{1-\rho(n-1)}
x\Var_{\theta,\varepsilon}\,
\varepsilon\E_q[(\theta_L-\theta_i)^2]
}.
\end{align*}
Substituting
\eqref{eq: tail z bound} and multiplying by $\rho$ yields
\[
\left[
1-
\frac{
V_{\theta A}+\frac{\rho\phi}{2(1-\rho(n-1))}
}{
\overline V_{A,\varepsilon}
}
(1-\varepsilon)(\mu-\theta_L)
\right]x\Var_{\theta,\varepsilon}
\leq
\sqrt{
\frac{\phi}{1-\rho(n-1)}
x\Var_{\theta,\varepsilon}\,
\varepsilon\E_q[(\theta_L-\theta_i)^2]
}.
\]
Because the extreme-type separation condition fails strictly under
$p$, the expression in the square bracket converges, as $\e \to 0$, to 
\[
c_0 \coloneqq 1 - \frac{
V_{\theta A}+\frac{\rho\phi}{2(1-\rho(n-1))}
}{
\overline V_{A}
}(\mu-\theta_L) > 0.
\]
Therefore, there exists $\bar \e > 0$ such that the expression in square brackets is at least $c_0/2$ for every $\e \in (0,\bar\e)$. Note that this bound depends only on the primitives and $q$, so it is uniform over all symmetric CE outcomes.

If $x=0$, the desired bounds follow immediately since $\pi = \pi^{NI,\varepsilon}$.
If $x > 0$, dividing the preceding inequality by $\sqrt{x\Var_{\theta,\varepsilon}}$ and squaring implies
\[
x \Var_{\theta,\varepsilon} \leq \frac{4\phi \E_q[(\theta_L-\theta_i)^2]}{c_0^2 (1-\rho(n-1))} \e.
\]

The conclusions now follow directly from \cref{lem: NI benchmark}.
Indeed, dropping the nonpositive quadratic term in \eqref{eq: NI action bound} bounds the mean-square deviation by $\phi x\Var_{\theta,\varepsilon}/(1-\rho(n-1))$, while  \eqref{eq:BCE-welfare-decomp} bounds the welfare gain by one half of this deviation plus $V_{\theta A} x \Var_{\theta,\varepsilon} / \rho$. Therefore, there exist constants $C_D, C_W > 0$ that rely only on the primitives and $q$, such that for every $\e \in (0,\bar \e)$ and every symmetric CE outcome $\pi$,
\[
\E_\pi[(a_i - a_i^{NI,\varepsilon}(\theta_i))^2]\leq C_D \e, \qquad 0 \leq \E_{\pi}[u_i] - \E_{\pi^{NI,\varepsilon}}[u_i] \leq C_W \e.
\]
Setting $C = \max\{C_D,C_W\}$ leads to the desired result.
\end{proof}

\end{appendices}
\end{appendix}
\end{document}